\documentclass[aps,prc,reprint,superscriptaddress]{revtex4-2} 

\usepackage{graphicx}
\usepackage{dcolumn}
\usepackage{bm}

\usepackage{url}

\usepackage{color}
\usepackage{graphicx}
\usepackage{physics}
\usepackage{lineno}
\usepackage{amsmath}

\begin{document}

\preprint{APS/123-QED}

\title{Measurement of the branching ratio of $\mathrm{^{16}N}$, $\mathrm{^{15}C}$, $\mathrm{^{12}B}$, and $\mathrm{^{13}B}$ isotopes\\ through the nuclear muon capture reaction in the Super-Kamiokande detector}

\newcommand{\AFFicrr}{\affiliation{Kamioka Observatory, Institute for Cosmic Ray Research, University of Tokyo, Kamioka, Gifu 506-1205, Japan}}
\newcommand{\AFFkashiwa}{\affiliation{Research Center for Cosmic Neutrinos, Institute for Cosmic Ray Research, University of Tokyo, Kashiwa, Chiba 277-8582, Japan}}
\newcommand{\AFFipmu}{\affiliation{Kavli Institute for the Physics and
Mathematics of the Universe (WPI), The University of Tokyo Institutes for Advanced Study,
University of Tokyo, Kashiwa, Chiba 277-8583, Japan}}
\newcommand{\AFFmad}{\affiliation{Department of Theoretical Physics, University Autonoma Madrid, 28049 Madrid, Spain}}
\newcommand{\AFFubc}{\affiliation{Department of Physics and Astronomy, University of British Columbia, Vancouver, BC, V6T1Z4, Canada}}
\newcommand{\AFFbu}{\affiliation{Department of Physics, Boston University, Boston, MA 02215, USA}}
\newcommand{\AFFuci}{\affiliation{Department of Physics and Astronomy, University of California, Irvine, Irvine, CA 92697-4575, USA }}
\newcommand{\AFFcsu}{\affiliation{Department of Physics, California State University, Dominguez Hills, Carson, CA 90747, USA}}
\newcommand{\AFFcnm}{\affiliation{Institute for Universe and Elementary Particles, Chonnam National University, Gwangju 61186, Korea}}
\newcommand{\AFFduke}{\affiliation{Department of Physics, Duke University, Durham NC 27708, USA}}
\newcommand{\AFFgifu}{\affiliation{Department of Physics, Gifu University, Gifu, Gifu 501-1193, Japan}}
\newcommand{\AFFgist}{\affiliation{GIST College, Gwangju Institute of Science and Technology, Gwangju 500-712, Korea}}
\newcommand{\AFFuh}{\affiliation{Department of Physics and Astronomy, University of Hawaii, Honolulu, HI 96822, USA}}
\newcommand{\AFFicl}{\affiliation{Department of Physics, Imperial College London , London, SW7 2AZ, United Kingdom }}
\newcommand{\AFFkek}{\affiliation{High Energy Accelerator Research Organization (KEK), Tsukuba, Ibaraki 305-0801, Japan }}
\newcommand{\AFFkobe}{\affiliation{Department of Physics, Kobe University, Kobe, Hyogo 657-8501, Japan}}
\newcommand{\AFFkyoto}{\affiliation{Department of Physics, Kyoto University, Kyoto, Kyoto 606-8502, Japan}}
\newcommand{\AFFliv}{\affiliation{Department of Physics, University of Liverpool, Liverpool, L69 7ZE, United Kingdom}}
\newcommand{\AFFmiyagi}{\affiliation{Department of Physics, Miyagi University of Education, Sendai, Miyagi 980-0845, Japan}}
\newcommand{\AFFnagoya}{\affiliation{Institute for Space-Earth Environmental Research, Nagoya University, Nagoya, Aichi 464-8602, Japan}}
\newcommand{\AFFkmi}{\affiliation{Kobayashi-Maskawa Institute for the Origin of Particles and the Universe, Nagoya University, Nagoya, Aichi 464-8602, Japan}}
\newcommand{\AFFpol}{\affiliation{National Centre For Nuclear Research, 02-093 Warsaw, Poland}}
\newcommand{\AFFsuny}{\affiliation{Department of Physics and Astronomy, State University of New York at Stony Brook, NY 11794-3800, USA}}
\newcommand{\AFFokayama}{\affiliation{Department of Physics, Okayama University, Okayama, Okayama 700-8530, Japan }}
\newcommand{\AFFosaka}{\affiliation{Department of Physics, Osaka University, Toyonaka, Osaka 560-0043, Japan}}
\newcommand{\AFFox}{\affiliation{Department of Physics, Oxford University, Oxford, OX1 3PU, United Kingdom}}
\newcommand{\AFFqmul}{\affiliation{School of Physics and Astronomy, Queen Mary University of London, London, E1 4NS, United Kingdom}}
\newcommand{\AFFregina}{\affiliation{Department of Physics, University of Regina, 3737 Wascana Parkway, Regina, SK, S4SOA2, Canada}}
\newcommand{\AFFseoul}{\affiliation{Department of Physics and Astronomy, Seoul National University, Seoul 151-742, Korea}}
\newcommand{\AFFsheff}{\affiliation{School of Mathematical and Physical Sciences, University of Sheffield, S3 7RH, Sheffield, United Kingdom}}
\newcommand{\AFFshizuokasc}{\affiliation{Department of Informatics in
Social Welfare, Shizuoka University of Welfare, Yaizu, Shizuoka, 425-8611, Japan}}
\newcommand{\AFFstfc}{\affiliation{STFC, Rutherford Appleton Laboratory, Harwell Oxford, and Daresbury Laboratory, Warrington, OX11 0QX, United Kingdom}}
\newcommand{\AFFskk}{\affiliation{Department of Physics, Sungkyunkwan University, Suwon 440-746, Korea}}
\newcommand{\AFFtodai}{\affiliation{Department of Physics, University of Tokyo, Bunkyo, Tokyo 113-0033, Japan }}
\newcommand{\AFFtit}{\affiliation{Department of Physics, Institute of Science Tokyo, Meguro, Tokyo 152-8551, Japan }}
\newcommand{\AFFtus}{\affiliation{Department of Physics, Faculty of Science and Technology, Tokyo University of Science, Noda, Chiba 278-8510, Japan }}
\newcommand{\AFFtriumf}{\affiliation{TRIUMF, 4004 Wesbrook Mall, Vancouver, BC, V6T2A3, Canada }}
\newcommand{\AFFtokai}{\affiliation{Department of Physics, Tokai University, Hiratsuka, Kanagawa 259-1292, Japan}}
\newcommand{\AFFtsinghua}{\affiliation{Department of Engineering Physics, Tsinghua University, Beijing, 100084, China}}
\newcommand{\AFFynu}{\affiliation{Department of Physics, Yokohama National University, Yokohama, Kanagawa, 240-8501, Japan}}
\newcommand{\AFFllr}{\affiliation{Ecole Polytechnique, IN2P3-CNRS, Laboratoire Leprince-Ringuet, F-91120 Palaiseau, France }}
\newcommand{\AFFbari}{\affiliation{ Dipartimento Interuniversitario di Fisica, INFN Sezione di Bari and Universit\`a e Politecnico di Bari, I-70125, Bari, Italy}}
\newcommand{\AFFnapoli}{\affiliation{Dipartimento di Fisica, INFN Sezione di Napoli and Universit\`a di Napoli, I-80126, Napoli, Italy}}
\newcommand{\AFFroma}{\affiliation{INFN Sezione di Roma and Universit\`a di Roma ``La Sapienza'', I-00185, Roma, Italy}}
\newcommand{\AFFpadova}{\affiliation{Dipartimento di Fisica, INFN Sezione di Padova and Universit\`a di Padova, I-35131, Padova, Italy}}
\newcommand{\AFFkeio}{\affiliation{Department of Physics, Keio University, Yokohama, Kanagawa, 223-8522, Japan}}
\newcommand{\AFFwinnipeg}{\affiliation{Department of Physics, University of Winnipeg, MB R3J 3L8, Canada }}
\newcommand{\AFFkcl}{\affiliation{Department of Physics, King's College London, London, WC2R 2LS, UK }}
\newcommand{\AFFwarwick}{\affiliation{Department of Physics, University of Warwick, Coventry, CV4 7AL, UK }}
\newcommand{\AFFral}{\affiliation{Rutherford Appleton Laboratory, Harwell, Oxford, OX11 0QX, UK }}
\newcommand{\AFFwu}{\affiliation{Faculty of Physics, University of Warsaw, Warsaw, 02-093, Poland }}
\newcommand{\AFFbcit}{\affiliation{Department of Physics, British Columbia Institute of Technology, Burnaby, BC, V5G 3H2, Canada }}
\newcommand{\AFFtohoku}{\affiliation{Department of Physics, Faculty of Science, Tohoku University, Sendai, Miyagi, 980-8578, Japan }}
\newcommand{\AFFicise}{\affiliation{Institute For Interdisciplinary Research in Science and Education, ICISE, Quy Nhon, 55121, Vietnam }}
\newcommand{\AFFilance}{\affiliation{ILANCE, CNRS - University of Tokyo International Research Laboratory, Kashiwa, Chiba 277-8582, Japan}}
\newcommand{\AFFibs}{\affiliation{Center for Underground Physics, Institute for Basic Science (IBS), Daejeon, 34126, Korea}}
\newcommand{\AFFglasgow}{\affiliation{School of Physics and Astronomy, University of Glasgow, Glasgow, Scotland, G12 8QQ, United Kingdom}}
\newcommand{\AFFoecu}{\affiliation{Media Communication Center, Osaka Electro-Communication University, Neyagawa, Osaka, 572-8530, Japan}}
\newcommand{\AFFminn}{\affiliation{School of Physics and Astronomy, University of Minnesota, Minneapolis, MN  55455, USA}}
\newcommand{\AFFsilesia}{\affiliation{August Che\l{}kowski Institute of Physics, University of Silesia in Katowice, 75 Pu\l{}ku Piechoty 1, 41-500 Chorz\'{o}w, Poland}}
\newcommand{\AFFtoyama}{\affiliation{Faculty of Science, University of Toyama, Toyama City, Toyama 930-8555, Japan}}
\newcommand{\AFFbmcc}{\affiliation{Science Department, Borough of Manhattan Community College / City University of New York, New York, New York, 1007, USA.}}
\newcommand{\AFFnumazu}{\affiliation{National Institute of Technology, Numazu College, Numazu, Shizuoka  410-8501, Japan}}

\AFFicrr
\AFFkashiwa
\AFFmad
\AFFbmcc
\AFFbu
\AFFbcit
\AFFuci
\AFFcsu
\AFFcnm
\AFFduke
\AFFllr
\AFFgifu
\AFFgist
\AFFglasgow
\AFFuh
\AFFibs
\AFFicise
\AFFicl
\AFFbari
\AFFnapoli
\AFFpadova
\AFFroma
\AFFilance
\AFFkeio
\AFFkek
\AFFkcl
\AFFkobe
\AFFkyoto
\AFFliv
\AFFminn
\AFFmiyagi
\AFFnagoya
\AFFkmi
\AFFpol
\AFFnumazu
\AFFsuny
\AFFokayama
\AFFoecu
\AFFox
\AFFral
\AFFseoul
\AFFsheff
\AFFshizuokasc
\AFFsilesia
\AFFstfc
\AFFskk
\AFFtohoku
\AFFtodai
\AFFipmu
\AFFtit
\AFFtus
\AFFtoyama
\AFFtriumf
\AFFtsinghua
\AFFwu
\AFFwarwick
\AFFwinnipeg
\AFFynu

\author{Y.~Maekawa}
\AFFkeio
\author{K.~Abe}
\AFFicrr
\AFFipmu
\author{S.~Abe}
\AFFicrr
\author{Y.~Asaoka}
\AFFicrr
\AFFipmu
\author{M.~Harada}
\AFFicrr
\author{Y.~Hayato}
\AFFicrr
\AFFipmu
\author{K.~Hiraide}
\AFFicrr
\AFFipmu
\author{K.~Hosokawa}
\AFFicrr
\author{K.~Ieki}
\author{M.~Ikeda}
\AFFicrr
\AFFipmu
% ...


\collaboration{The Super-Kamiokande Collaboration}
\noaffiliation

\date{\today}

\begin{abstract}
The Super-Kamiokande detector has measured solar neutrinos for more than $25$~years. The sensitivity for solar neutrino measurement is limited by the uncertainties of energy scale and background modeling. Decays of unstable isotopes with relatively long half-lives through nuclear muon capture, such as $\mathrm{^{16}N}$, $\mathrm{^{15}C}$, $\mathrm{^{12}B}$ and $\mathrm{^{13}B}$, are detected as background events for solar neutrino observations. In this study, we developed a method to form a pair of stopping muon and decay candidate events and evaluated the production rates of such unstable isotopes. We then measured their branching ratios considering both their production rates and the estimated number of nuclear muon capture processes as $Br(\mathrm{^{16}N})=(9.0 \pm 0.1)\%$, $Br(\mathrm{^{15}C})=(0.6\pm0.1)\%$, $Br(\mathrm{^{12}B})=(0.98 \pm 0.18)\%$, $Br(\mathrm{^{13}B})=(0.14 \pm 0.12)\%$, respectively.  The result for $\mathrm{^{16}N}$ has world-leading precision at present and the results for $\mathrm{^{15}C}$, $\mathrm{^{12}B}$, and $\mathrm{^{13}B}$ are the first branching ratio measurements for those isotopes.
\end{abstract}

  \maketitle


\section{Introduction} \label{sec:intro}

\subsection{Solar neutrinos in the Super-Kamiokande}

In the last $50$~years, solar neutrinos have been measured by several experiments~\cite{Davis:1968cp, Kamiokande-II:1989hkh, Abazov:1991rx, GALLEX:1992gcp, GNO:2000avz, Super-Kamiokande:2001ljr, SNO:2001kpb, SNO:2002tuh, KamLAND:2002uet, Borexino:2017rsf, BOREXINO:2018ohr, KamLAND:2014gul, BOREXINO:2023ygs, SNO:2024vjl, PandaX:2024muv, XENON:2024ijk}. Through solar neutrino measurements by the Super-Kamiokande(SK) and SNO experiments, flavour conversion of solar neutrinos was discovered, leading to the confirmation of neutrino oscillation~\cite{Super-Kamiokande:2001ljr, SNO:2001kpb}. The next targets in furthering our understanding of neutrino oscillation are the searches for the matter effect of solar neutrino oscillation, such as the MSW effect in the Sun predicted by Mikheyev, Smirnov, and Wolfenstein~\cite{Mikheyev:1985zog, Wolfenstein:1977ue}, and the day/night flux asymmetry induced by the matter effect in the Earth~\cite{Baltz:1986hn, Bouchez:1986kb, Carlson:1986ui, Cribier:1986ak, Petcov:1998su, Bakhti:2020tcj}.

The SK detector has precisely measured the flux of $\mathrm{\mathrm{^{8}B}}$ solar neutrinos and its energy spectrum for more than $25$~years since $1996$ by selecting recoiled electron events via elastic scattering of solar neutrinos~\cite{Super-Kamiokande:2005wtt, Super-Kamiokande:2008ecj, Super-Kamiokande:2010tar, Super-Kamiokande:2016yck, Super-Kamiokande:2023jbt}. By analyzing the Cherenkov light pattern from the recoil electron, the SK detector can reconstruct the time of elastic scattering, as well as the direction and the energy of the scattered electron. These measurements allow us to search for the MSW effect and the day/night flux asymmetry of solar neutrinos. Although the SK detector has already accumulated more than $10^{5}$ solar neutrino interaction events, the sensitivity of SK to the matter effect is primarily limited by the statistics of observed signal, the uncertainty of the energy scale for the recoil electron, as well as the uncertainties on background events, such as radon dissolved in water~\cite{Nakano:2019bnr} and spallation products induced by penetrating muons~\cite{Super-Kamiokande:2015xra, Super-Kamiokande:2021snn}. To reduce the uncertainty on the total amount of background events and to increase sensitivity to solar neutrinos, the uncertainty that comes from the production rates of muon-induced background should be measured.

\subsection{Unstable isotopes by nuclear muon capture process}

The negative muon capture rate on nucleons has been studied to theoretically and experimentally understand the nuclear response through the weak interaction~\cite{COHEN1964255, Donnelly:1972ppr, Cannata:1976jw, Koshigiri:1979zc, Towner:1981dcc, Marketin:2008ei}. Such reactions are used to study nuclear structure and dynamics, where the weak structure at low energies determines weak reaction rates with nuclei. 

A negative muon that slows down in the water that fills the SK detector can form a muonic atom with an oxygen through the Coulomb force~\cite{Guichon:1979ga}. The muon in the oxygen's orbit can then either decay or undergo a muon capture on nuclei~\cite{Measday:2001yr, Eramzhian:1977fy}. Table~\ref{tb:branch} summarizes the branching ratios of unstable isotopes produced by nuclear muon capture process on oxygen. Because a limited number of isotope branching ratios have been measured~\cite{2002heisinger, Measday:2001yr}, any further development of the simulation will be insufficient to estimate the branching ratios induced by nuclear muon capture processes in oxygen. Currently, the expected branching ratios from three simulation codes are available; \texttt{Geant4}~\cite{Allison:2016}, \texttt{Fluka}~\cite{Battistoni:2015epi}, and \texttt{PHITS}~\cite{Sato01092013, Abe02012017}. The latter is the Monte Carlo simulation using particle and heavy ions transport with recently developed muon interaction models. Table~\ref{tb:branch} summarizes simulated outputs from three simulations. They are inconsistent with each other. Therefore, additional measurements of the branching ratio are required to improve the predictive capability of nuclear muon capture processes in those simulations.

\begin{table}[h]
    \begin{center}
    \caption{Summary of branching ratios of negative muon capture on oxygen nuclei. A limited number of isotopes have been measured~\cite{2002heisinger, Measday:2001yr}. Some isotopes are not produced based on the simulated values from \texttt{PHITS}~\cite{Sato01092013, Abe02012017}, \texttt{Geant4}~\cite{Allison:2016},  and \texttt{Fluka}~(version 4-5.0)~\cite{Battistoni:2015epi, Nairat:2024upg}. The ``Other" row in \texttt{Fluka} includes the combination of light particle emissions without any production of residual nuclei listed\footnote{Such as $3\alpha + \mathrm{^{3}H}+n$, $3\alpha+\mathrm{^{2}H}+2n$, and $3\alpha+p+3n$.}.}
        \label{tb:branch}
            \begin{tabular}{c|cc|ccc}
                \hline
                        & \multicolumn{2}{c|}{Measurement data} & \multicolumn{3}{c}{MC simulation} \\ \hline
                Isotope & Ref.~\cite{2002heisinger} & Ref.~\cite{Measday:2001yr} & \texttt{PHITS} & \texttt{Geant4} & \texttt{Fluka} \\
                        & [\%] & [\%] & [\%] & [\%] & [\%]\\
                \hline \hline
                $\mathrm{^{16}N}$  & -- & $11\pm1$ & $0.5$ & $12.0$ & $5.52$\\
                $\mathrm{^{15}N}$  & -- & -- & $57$ & $40.5$ & $61.9$\\
                $\mathrm{^{14}N}$  & -- & -- & $21$ & $6.7$ & $2.26$\\
                $\mathrm{^{13}N}$  & -- & -- & $0.4$ & $0.1$ & $0.02$\\
                $\mathrm{^{15}C}$  & -- & -- & $0.19$ & $0.7$ & $0.67$\\
                $\mathrm{^{14}C}$  & $13.7\pm1.1$ & -- & $5.14$ & $4.3$ & $3.54$\\
                $\mathrm{^{13}C}$  & -- & -- & $5.1$ & $5.5$ & $2.76$\\
                $\mathrm{^{12}C}$  & -- & -- & $3.2$ & $2.3$ & $1.24$\\
                $\mathrm{^{11}C}$  & -- & -- & -- & $0.007$ & $0.046$\\
                $\mathrm{^{14}B}$  & -- & -- & -- & $0.002$ & $0.002$\\
                $\mathrm{^{13}B}$  & -- & -- & $0.02$ & $-$ & $0.088$\\
                $\mathrm{^{12}B}$  & -- & -- & $1.16$ & $7.6$ & $4.41$\\
                $\mathrm{^{11}B}$  & -- & -- & $5.4$ & $13.6$ & $7.65$\\
                $\mathrm{^{10}B}$  & -- & -- & $0.27$ & $1.2$ & $0.87$\\
                $\mathrm{^{11}Be}$  & -- & -- & $0.01$ &  $0.005$ & $0.037$\\
                $\mathrm{^{10}Be}$ & $0.43\pm0.03$ & -- & $0.14$ & $0.7$ & $0.91$ \\
                $\mathrm{^{9}Be}$ & -- & -- & $0.05$ & $0.7$ & $1.06$\\
                $\mathrm{^{7}Be}$  & $0.02\pm0.01$ & -- & -- &  $0.03$ & $0.15$ \\
                $\mathrm{^{9}Li}$ & -- & -- & -- & $0.008$ & $0.062$\\
                $\mathrm{^{8}Li}$ & -- & -- & $0.13$ & $1.0$ & $0.52$ \\
                $\mathrm{^{7}Li}$ & -- & -- & $0.53$ & $1.9$ & $1.87$ \\
                $\mathrm{^{6}Li}$ & -- & -- & $0.05$ & $0.9$ & $1.69$ \\
                Other & -- & -- & -- & -- & $2.4$ \\
                \hline \hline
        \end{tabular}
    \end{center}
\end{table}

In the SK detector, products of $\beta$ decays with large $Q$-values of unstable isotopes can be detected by Cherenkov light. If these unstable isotopes have a relatively long half-life, the decays occur long enough after the muon stops that the two events are separately recorded in the analysis sample. In such a case, the production rates of unstable isotopes by the nuclear muon capture process can be measured by forming a pair of parent-stopping muon and isotope decay candidate events. Table~\ref{tb:nucleus} summarizes the $Q$-values of $\beta$ decay and half-lives of four unstable nuclei which can be detected in the SK detector. The $\mathrm{^{16}N}$ and $\mathrm{^{15}C}$ nuclei have relatively long half-lives of several seconds and the $\mathrm{^{13}B}$ and $\mathrm{^{12}B}$ nuclei have half-lives of milliseconds. By observing these types of decay events after a stopping muon event is detected, we can evaluate the fraction of isotopes via both decay time and energy distributions and thereby measure the branching ratio of the nuclear muon capture process on oxygen.

\begin{table}[h]
    \begin{center}
    \caption{Summary of the production reaction, $Q$-values of $\beta$ decay, and the half-lives of the unstable isotopes induced by nuclear muon capture. More details are described in Sec.~\ref{sec:overview}.}
        \label{tb:nucleus}
            \begin{tabular}{c|ccc}
                \hline \hline
                Isotope & Reaction & $Q$-value of $\beta$ decay & Half-life \\ 
                 & & [MeV] & [s] \\ \hline
                $\mathrm{^{16}N}$ & $\mathrm{^{16}O}(\mu^{-}, \nu)\mathrm{^{16}N}$  & $10.42$ & $7.13\phantom{00}$ \\ 
                $\mathrm{^{15}C}$ & $\mathrm{^{16}O}(\mu^{-}, \nu p)\mathrm{^{15}C}$ & $\phantom{0}9.77$ & $2.45\phantom{00}$ \\
                $\mathrm{^{13}B}$ & $\mathrm{^{16}O}(\mu^{-}, \nu n 2p)\mathrm{^{13}B}$  & $13.44$ & $0.0172$ \\
                $\mathrm{^{12}B}$ & $\mathrm{^{16}O}(\mu^{-}, \nu \alpha)\mathrm{^{12}B}$  & $13.37$ & $0.0202$ \\
            \hline \hline
        \end{tabular}
    \end{center}
\end{table}

This article is organized as follows. In Sec.~\ref{sec:detector}, we briefly describe the SK detector, data acquisition system, and energy calibration in the SK experiment. In Sec.~\ref{sec:overview}, we give an overview of the decay processes of four unstable isotopes, $\mathrm{^{16}N}$, $\mathrm{^{15}C}$, $\mathrm{^{12}B}$, and $\mathrm{^{13}B}$, which can be detected in SK, and we also present the theoretical predictions of their production rates at the underground experimental site of the SK detector. In Sec.~\ref{sec:analysis}, we describe the analysis methods, which include the selection of stopping muon, the tagging method of $\mathrm{^{16}N}$ and $\mathrm{^{15}C}$ decay candidate events, the $\chi^{2}$ method to determine the production rates of $\mathrm{^{16}N}$ and $\mathrm{^{15}C}$, the systematic uncertainties on the measurement, and the measurement result of the production rates of $\mathrm{^{16}N}$ and $\mathrm{^{15}C}$. In Sec.~\ref{sec:b12}, we describe the analysis method and result for the measurement of $\mathrm{^{12}B}$ and $\mathrm{^{13}B}$ production rates. In Sec.~\ref{sec:branch}, we describe the results of the branching ratio measurement and compare the experimental results with theoretical prediction models. In Sec.~\ref{sec:summary}, we conclude this study and give prospects.

\section{Super-Kamiokande detector} \label{sec:detector}

\subsection{Detector and data set} \label{sec:dataset}

The Super-Kamiokande detector is an imaging water Cherenkov detector. It is located $1000$~m underground~($2700$~m of water equivalent) in the Kamioka mine in Gifu prefecture, Japan. The detector components, operation, and performance are detailed in detector papers~\cite{Super-Kamiokande:2002weg, Abe:2013gga}. 

The SK detector consists of about $50,000$~tonnes of purified water in a stainless steel cylindrical water tank with photomultiplier tubes~(PMTs).  The detector is divided into two regions by an inner structure that optically separates these regions with sheets. One region is the inner detector~(ID) and the other is the outer detector~(OD). The ID serves as the target volume for neutrino interactions and the OD is used to veto external cosmic-ray muons as well as $\gamma$-rays from the surrounding rock. The ID holds $32,000$~tonnes of water as a physics target and its standard fiducial volume~($2$~m from the inner detector structure) for solar neutrino analysis is $22,500$~tonnes. The OD is optically separated from the ID and is used to veto cosmic-ray muons penetrating the mountain above. The ID is viewed by 11129 20-inch PMTs~\cite{Kume:1983hs} and the OD is viewed by 1885 8-inch PMTs.

The SK data set is separated into eight distinct periods, from SK-I to SK-VIII. For the prior SK-I to SK-V phases, spanning April 1996 to July 2020, the detector operated with ultra-pure water~\cite{Super-Kamiokande:2002weg}. Starting in July 2020, SK-VI, in which gadolinium sulfate was dissolved into the detector water tank, has been in operation~\cite{Super-Kamiokande:2021the}. Then, additional gadolinium loading was conducted in June 2022~\cite{Super-Kamiokande:2024kcb} toward SK-VII. Table~\ref{tb:phase} summarizes the period of operation, the livetime used for this analysis~(up to and including SK-VI), the number of ID PMTs, electronics, and the water status.
\begin{table}[h]
    \begin{center}
    \caption{The summary of data sets used in this analysis. The livetime is the total duration of data samples used for this analysis after removing calibration and bad condition runs.}
        \label{tb:phase}
            \begin{tabular}{c|c|c|c}
                \hline \hline
                SK phase & SK-IV & SK-V & SK-VI  \\ \hline
                Period &  Sep. 2008  & Jan. 2019  & Aug. 2020 \\
                       &  -- May 2018 & -- Jul. 2020 & -- Jun. 2022  \\ \hline
                ID PMTs &  \multicolumn{3}{c}{11129} \\ \hline
                Livetime~[days] & $2970.1$ & $379.2$ & $552.2$ \\ \hline
                Electronic & \multicolumn{3}{c} {QBEE~\cite{Nishino:2009zu, Super-Kamiokande:2010kjr}} \\ \hline
                Water &  \multicolumn{2}{c|}{Ultra-pure water} & Gd loaded water~\cite{Super-Kamiokande:2021the} \\
            \hline \hline
        \end{tabular}
    \end{center}
\end{table}

\subsection{Tagging the stopping muon} \label{sec:daq}

In the measurement of unstable isotopes, the tag of the stopping muon is critical to determine the production rate and the branching ratio of nuclear muon capture on oxygen in the SK detector. In this analysis, we developed a method to form pairs of parent stopping muon and decay products from unstable isotopes. In this section, we briefly explain how to tag stopping muons.

\subsubsection{Electronics up to and including SK-III} \label{sec:daq123}

From SK-I to SK-III, Analog Timing Modules~(ATMs), based on the TKO~(Tristan KEK Online) standards, were used as front-end electronics~\cite{Tanimori:1988qi, Ikeda:1990}. It generates four types of hardware triggers, called Super Low Energy~(SLE), Low Energy~(LE), High Energy~(HE), and OD. The first three triggers are generated depending on the pulse height of ID PMT signals while the last one is generated when the pulse height of OD PMT signals exceeds a certain threshold. Once those triggers are generated, a single event is recorded within a $1.3~\mu$s timing window. 

The observed event is recorded in a $1.3~\mu$s window up to and including SK-III. Hence, the parent muon and the decay electron event are recorded separately. However, some charge in the PMT could leak during charge integration after triggering a cosmic-ray muon, such that the number of hit PMTs and hit times are not always correct for subsequent decay electrons, resulting in an inaccurate reconstruction of their energy and timing. For this reason, we have difficulties in reconstructing the stopping muon positions and in forming the pair of the parent muon and the associated decay electron up to and including SK-III. In order to reduce the systematic uncertainties of the analysis presented in this article, we did not use the observed data from SK-I to SK-III.

\subsubsection{Electronics from SK-IV onwards} 

For SK-IV, starting in September 2008, new front-end electronics denoted QBEEs~\cite{Nishino:2009zu} were installed. These are capable of very high-speed signal processing, enabling the integration and recording of charge and time for every PMT signal. Since all PMT signals are digitized and recorded, there is no detector deadtime. Furthermore, a new online data acquisition~(DAQ) system was implemented which generates multiple software triggers depending on the number of hit PMTs within $200$~ns~\cite{Super-Kamiokande:2010kjr}. All PMT signals within a trigger time window are recorded, where the duration of the time window changes depending on the trigger types. Table~\ref{tb:trigger} summarizes the trigger types and the duration of an associated single event.

\begin{table}[h]
    \begin{center}
    \caption{Summary of software~(SK-IV and later) triggers with their timing window to record a single event. AFT trigger is generated when the SHE trigger is issued without the OD trigger in SK-IV~\cite{Super-Kamiokande:2013ufi}. From SK-V onwards, AFT trigger is generated whenever SHE trigger is issued~\cite{Super-Kamiokande:2022cvw}.}
        \label{tb:trigger}
            \begin{tabular}{c|c}
                \hline \hline
                Trigger & \multicolumn{1}{c}{Duration~[$\mu$s]} \\ 
                &  SK-IV and later \\ \hline 
                SLE  & $[-0.5, +0.8]$ \\
                LE and HE  & $[-5, +35] $ \\
                SHE  & $[-5, +35] $ \\
                AFT  &$[+35, +535]$ \\
            \hline \hline
        \end{tabular}
    \end{center}
\end{table}

\begin{figure}[h]
    \includegraphics[width=1.0\linewidth]{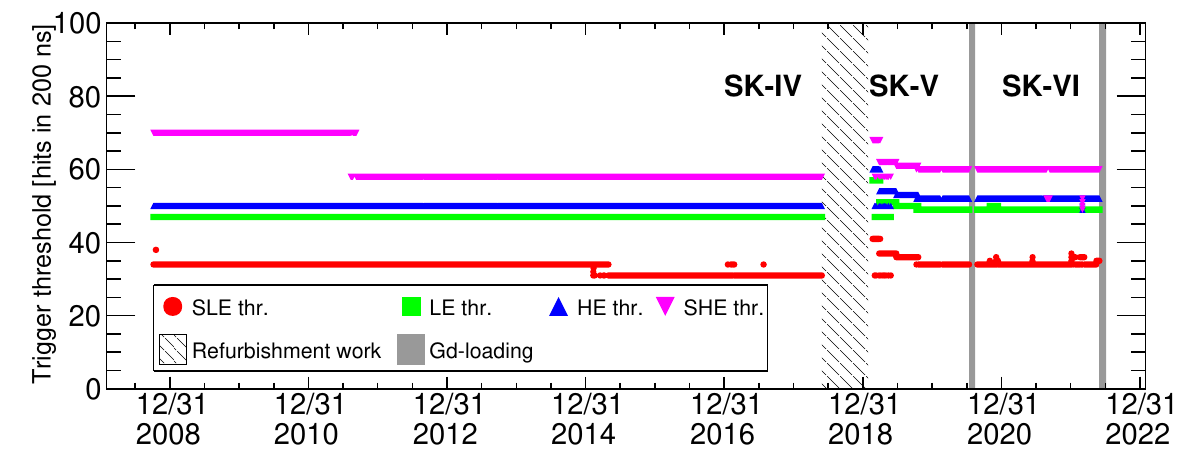}
    \caption{The software trigger thresholds with the QBEE electronics as a function of time from October 2008 to June 2022. Red circles, light-green squares, blue upper triangles, and pink lower triangles are the SLE, LE, HE, and SHE triggers, respectively~\cite{Nishino:2009zu}. The black left slanting region shows the period of the refurbishment work and the light-gray filled region shows the period of first and second Gd-loading~\cite{Super-Kamiokande:2021the, Super-Kamiokande:2024kcb}.}
    \label{fig:software}
\end{figure}

The new DAQ system generates Super High Energy~(SHE) and AFT triggers; the latter records all hits from $[+35, +535]~\mu$s to detect delayed events such as thermal neutron capture on hydrogen~\cite{Super-Kamiokande:2008mmn, Super-Kamiokande:2022hxq} and gadolinium~\cite{Super-Kamiokande:2021the}. In the SK-IV phase, this AFT trigger is generated after the SHE trigger with an OD veto to search for $2.2$~MeV $\gamma$-ray from neutron capture by hydrogen after an inverse $\beta$ decay of anti-electron neutrinos~\cite{Super-Kamiokande:2013ufi}. From SK-V, we changed the configuration of generating the AFT trigger to record spallation neutrons along with through-going cosmic-ray muons~\cite{Super-Kamiokande:2022cvw}. Figure~\ref{fig:software} shows the software trigger thresholds during the SK-IV, SK-V, and SK-VI phases. The thresholds have been changed depending on the detector conditions, such as background rate, water circulation pattern, accumulated data transfer speed, and the recording data size, etc.

In this analysis, the selection of the decay electron is critical to find the decays of unstable isotopes due to the nuclear muon capture on oxygen. Except for the SLE trigger, the duration of time windows, as summarized in Table~\ref{tb:trigger}, is long enough to capture the vast majority of electrons from muon decay; the corresponding decay electron identification procedure is detailed in Ref.~\cite{Kitagawa:2024ipa}. This long timing window to record a single event is helpful to increase the signal-to-noise ratio to tag the nuclear muon capture events as mentioned in the latter section.

\subsection{Energy calibrations in the MeV region}

The energy response of the SK detector is evaluated by several calibration sources. Table~\ref{tb:calib} summarizes the calibration sources used in the SK detector, in particular for the energy region below $50$~MeV. 

\begin{table}[!h]
    \begin{center}
    \caption{Summary of calibration sources used in the SK detector for the MeV energy range.}
        \label{tb:calib}
            \begin{tabular}{ll}
                \hline \hline
                Calibration source & Purpose \\ \hline
                LINAC~\cite{Super-Kamiokande:1998hbb} & Absolute energy scale \\
                DT generator~\cite{Super-Kamiokande:2000kzn} & Position and direction dependence \\
                 & of energy scale \\
                Decay electron~\cite{Super-Kamiokande:2016yck} & Monitoring water transparency \\
            \hline \hline
        \end{tabular}
    \end{center}
\end{table}

An absolute energy scale is determined by an electron linear accelerator~(LINAC), which was installed at the SK detector to vertically inject mono-energetic electrons from $4.4$~MeV to $18.9$~MeV with an accuracy on the energy of less than $\pm0.2\%$~\cite{Super-Kamiokande:1998hbb, Super-Kamiokande:2023jbt}. By comparing the calibration data against the MC simulation, the absolute energy scale is tuned. However the electron is only injected in the vertical direction and this situation limits the evaluation of the position and direction dependence of the energy scale. To compensate for this situation, another calibration source, a deuterium-tritium neutron~(DT) generator, is also used for the evaluation of energy scale~\cite{Super-Kamiokande:2000kzn}. This system generates neutrons via the reaction of $\mathrm{^{2}H}+\mathrm{^{3}H}\to \mathrm{^{4}He}+n$, and the produced neutrons are then captured on oxygen in water. This reaction creates $\mathrm{^{16}N}$ whose final decay products are an electron and a single $\gamma$-ray. The DT calibration data has been taken at several positions where the LINAC calibration cannot reach. This allows us to evaluate the systematic uncertainties for the position and direction dependences of the energy scale~\cite{Super-Kamiokande:2023jbt}. Although DT generator can accumulate a large number of $\beta$~decays, the broad shape of the energy distribution of $\beta$~decay limits the accuracy for determining the absolute energy scale.

Finally, water transparency is critical for the energy scale because photon propagation depends on its properties, such as the probabilities of absorption, scattering, and reflection~\cite{Super-Kamiokande:2002weg}. Decay electrons, which are naturally produced by a stopping cosmic-ray muon, are also sampled to evaluate the stability of the detector response. The total amount of observed charge by the PMTs from the decay electrons depends on the quality of water transparency inside the SK detector. In the MC simulation, the parameters for photon propagation are scaled by the observed charge in stopping muon data from decay electrons to reproduce the daily change of real detector response~\cite{Abe:2013gga}.

\section{Overview of $\bm{\mathrm{^{16}N}}$, $\bm{\mathrm{^{15}C}}$, $\bm{\mathrm{^{12}B}}$, and $\bm{\mathrm{^{13}B}}$ decays} \label{sec:overview}

In this section, we briefly explain the targets of unstable isotopes due to nuclear muon capture in this analysis. We also summarize the expected production rate of $\mathrm{^{16}N}$ at the SK experimental site, which was computed based on theoretical models by other groups.

\subsection{Production processes and their decay}

\begin{figure}[h]
    \includegraphics[width=1.0\linewidth]{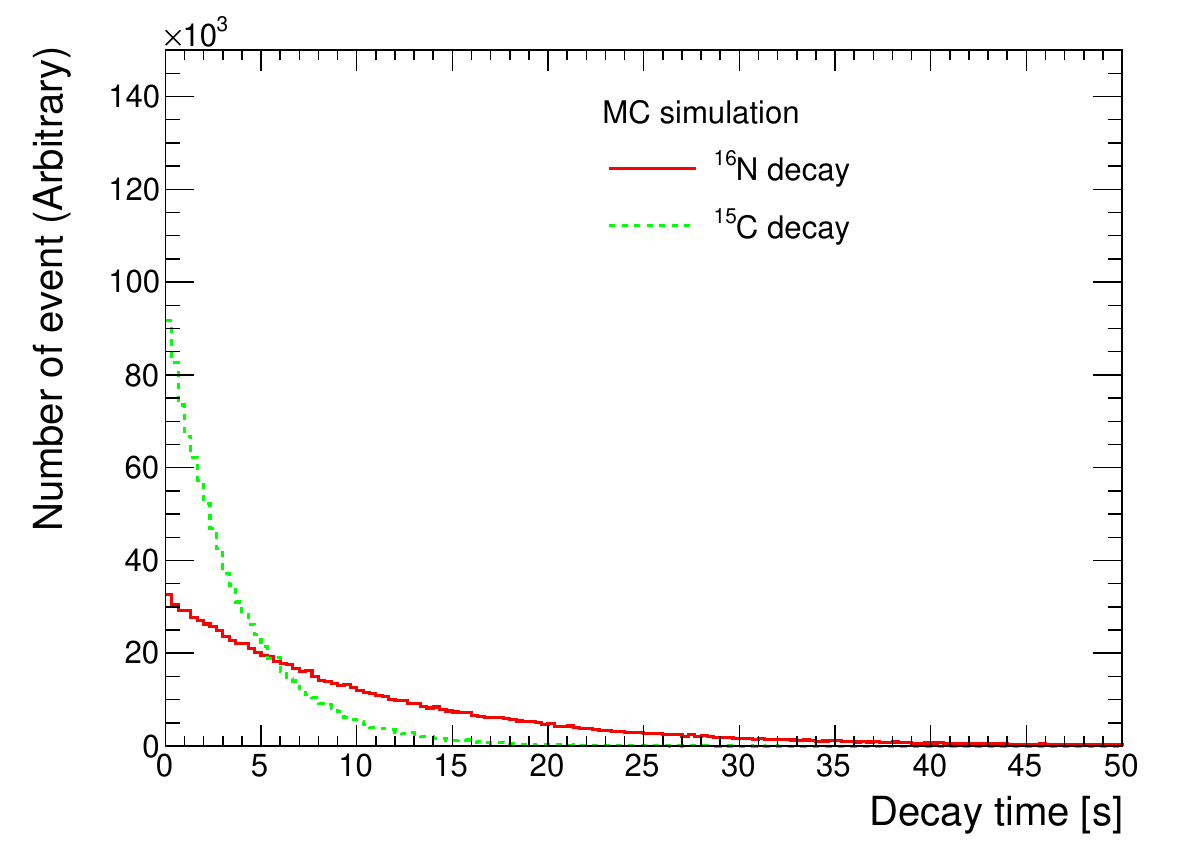} 
    \caption{Examples of the decay curves of $\mathrm{^{16}N}$ and $\mathrm{^{15}C}$. The red solid line and green dashed lines show the decay curve of $\mathrm{^{16}N}$ and $\mathrm{^{15}C}$ decays, respectively. For comparison purposes, $10^{6}$ decays are simulated for each of the two decays.}
    \label{fig:mc_expect-decay}
\end{figure}

\begin{figure}[h]
  \includegraphics[width=1.0\linewidth]{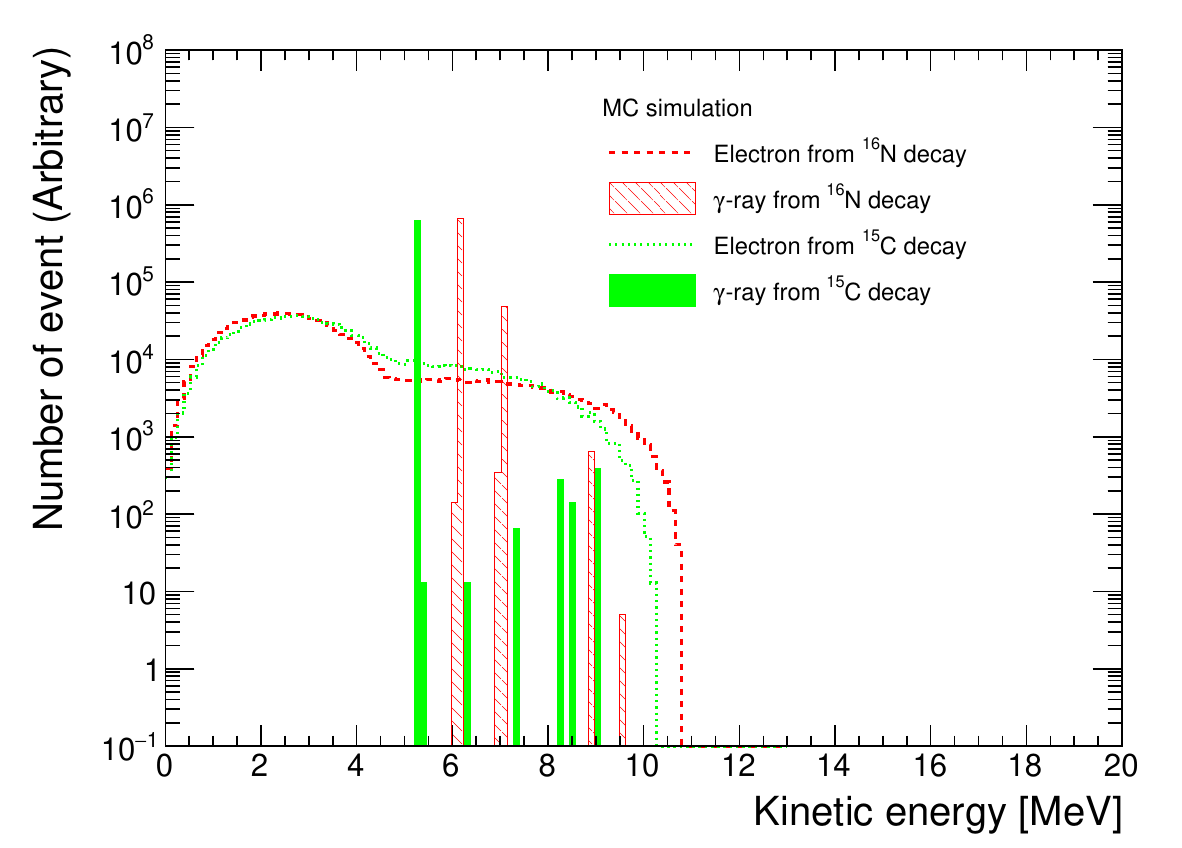}
    \caption{Kinetic energies of emitted electrons and $\gamma$-rays from decays of $\mathrm{^{16}N}$ and $\mathrm{^{15}C}$. The red dashed and green dotted lines show the energies of electrons from $\mathrm{^{16}N}$ and $\mathrm{^{15}C}$ decays, respectively. The red left-slanting-line histogram and green filled histogram show the energies of $\gamma$-rays from $\mathrm{^{16}N}$ and $\mathrm{^{15}C}$ decays, respectively. For comparison purposes, $10^{6}$ decays are simulated for each of the two decays with \texttt{Geant4} as detailed in Sec.~\ref{sec:mc-intro}.}
    \label{fig:mc_expect-energy}
\end{figure}

\begin{figure}[h]
    \includegraphics[width=1.0\linewidth]{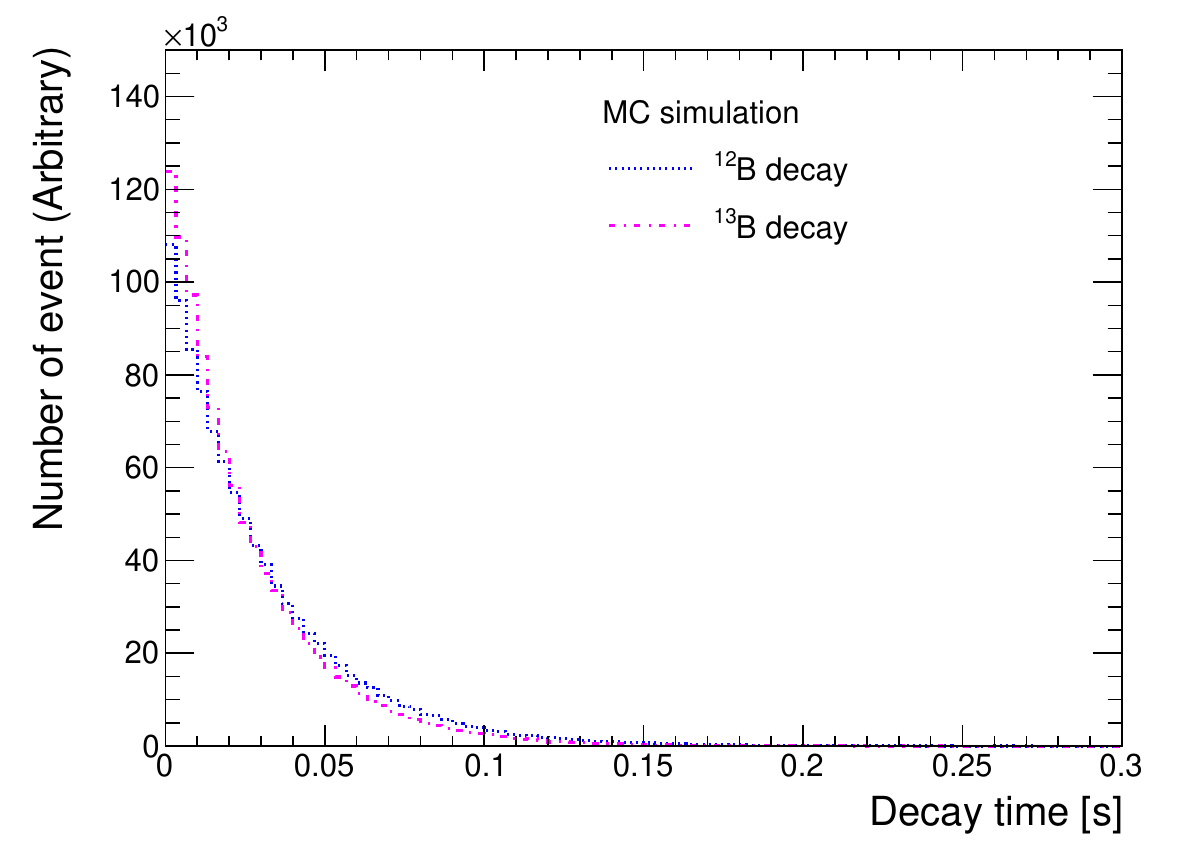}
    \caption{Examples of the decay curve of $\mathrm{^{12}B}$ and $\mathrm{^{13}B}$. The blue dotted line and magenta dot-dashed lines show the decay curve of $\mathrm{^{12}B}$ and $\mathrm{^{13}B}$ decays. Comparing with the decay curves of $\mathrm{^{16}N}$ and $\mathrm{^{15}C}$ described in Fig.~\ref{fig:mc_expect-decay}, the decay of $\mathrm{^{12}B}$ quickly occurs and fully completes within $0.3$~s. For comparison purposes, $10^{6}$ decays are simulated for each of the two decays.}
    \label{fig:mc_expect-decay-b12b13}
\end{figure}

\begin{figure}[h]
    \includegraphics[width=1.0\linewidth]{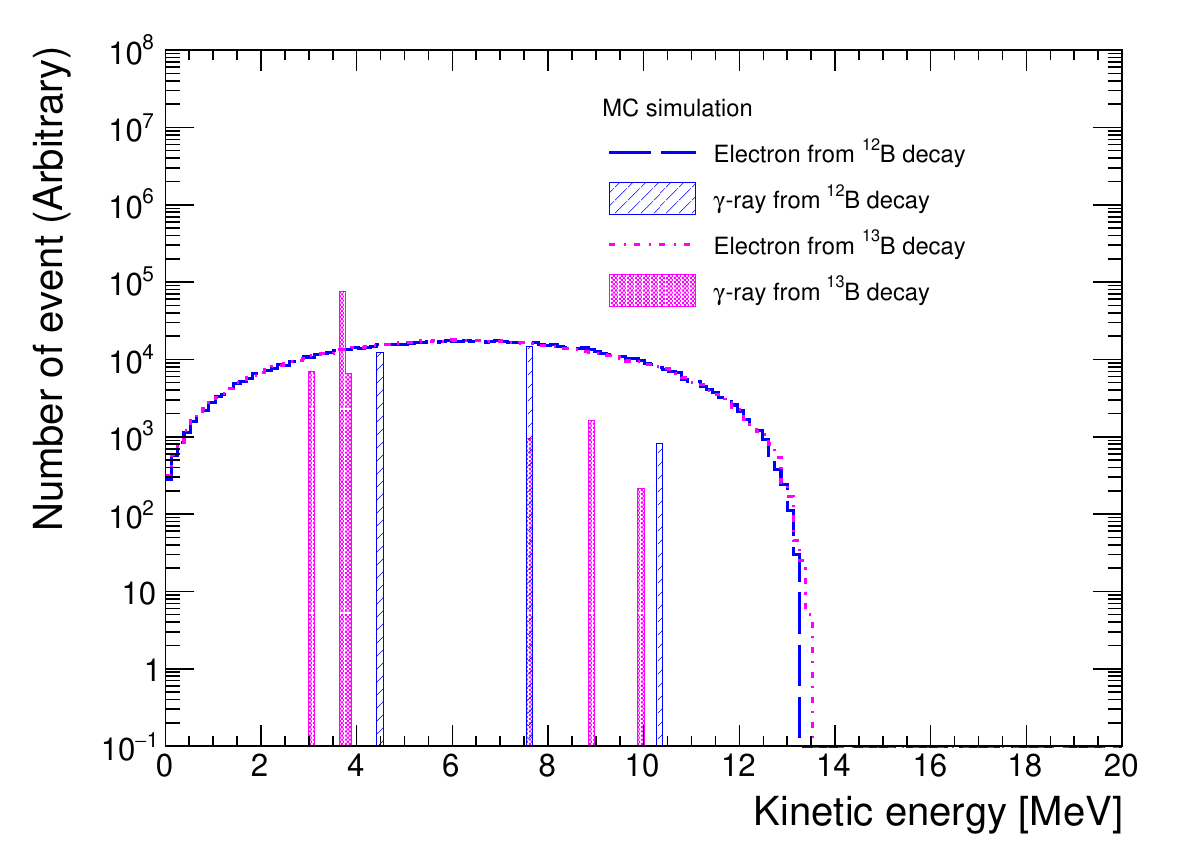}
    \caption{Kinetic energies of emitted electrons and $\gamma$-rays from decays of $\mathrm{^{12}B}$ and $\mathrm{^{13}B}$. The blue dashed and magenta dot-dashed lines show the energies of electrons from $\mathrm{^{12}B}$ and $\mathrm{^{13}B}$ decays, respectively. The blue right-slanting-line histogram and magenta checkered histogram show the energies of $\gamma$-rays from $\mathrm{^{12}B}$ and $\mathrm{^{13}B}$ decays, respectively. For comparison purposes, $10^{6}$ decays are simulated for each of the two decays with \texttt{Geant4} as detailed in Sec.~\ref{sec:mc-intro}.}
    \label{fig:mc_expect-energy-b12b13}
\end{figure}

Soon after the nuclear muon capture process, an excited state of nitrogen, $\mathrm{^{16}N}^{*}$, is produced in water. Due to the large energy transfer within the multi-body system, energy below $100$~MeV is distributed to the compound nucleus. In general, the excited $\mathrm{^{16}N}^{*}$ emits particles, resulting in the production of various residual nuclei. The evaporation of neutron(s) leads to the production of $\mathrm{^{15}N}$ or $\mathrm{^{14}N}$~\cite{vanderSchaaf:1983ah, Measday:2001yr, Jokiniemi:2024zdl, Super-Kamiokande:2025vvn,Kaplan:1969bd}, while in some cases no neutrons are emitted and, for example, $\mathrm{^{16}N}$ may remain after emitting $\gamma$-rays. Proton emission results in the production of $\mathrm{^{15}C}$, and the emission of an $\alpha$~particle results in the production of $\mathrm{^{12}B}$. Additionally, in rare cases, the emission of two protons and one neutron produces another isotope, $\mathrm{^{13}B}$. $\mathrm{^{16}N}$~(with a half-life of $7.13$~s) and $\mathrm{^{15}C}$~(with a half-life of $2.45$~s) emit electrons and $\gamma$-rays, releasing total energies of $10.42$~MeV for $\mathrm{^{16}N}$ and $9.77$~MeV for $\mathrm{^{15}C}$, respectively. Figure~\ref{fig:mc_expect-decay} shows the decay curves of $\mathrm{^{16}N}$ and $\mathrm{^{15}C}$, and Figure~\ref{fig:mc_expect-energy} shows the energy spectra of emitted particles. For $\mathrm{^{12}B}$ and $\mathrm{^{13}B}$, their half-lives and $Q$-values are $0.0202$~s and $13.37$~MeV, and $0.0172$~s and $13.44$~MeV, respectively. Figure~\ref{fig:mc_expect-decay-b12b13} shows the decay curves of $\mathrm{^{12}B}$ and $\mathrm{^{13}B}$. Comparing with the decay curves of $\mathrm{^{16}N}$ and $\mathrm{^{15}C}$ described in Fig.~\ref{fig:mc_expect-decay}, the decay of $\mathrm{^{12}B}$ quickly occurs and fully completes within $0.3$~s. Figure~\ref{fig:mc_expect-energy-b12b13} shows the energy spectra of emitted particles from $\mathrm{^{12}B}$ and $\mathrm{^{13}B}$ decays.

\subsection{Predicted production rates}

The measurement of production rates of $\mathrm{^{16}N}$ and $\mathrm{^{15}C}$ is important for validating simulation models of muon-induced products at underground experimental sites~\cite{Galbiati:2005ft, Li:2014sea, Li:2015kpa, Li:2015lxa}. For example, scintillation experiments such as Borexino, Double Chooz, and KamLAND have measured the production rates of light isotopes through muon captures on nuclei~\cite{KamLAND:2009zwo, Borexino:2013cke, DoubleChooz:2015jlf, KamLAND-Zen:2023spw}.

The production rates of unstable isotopes due to muon nuclear capture depend on the depth of the experimental site because the muon nuclear capture process occurs when the cosmic-ray muon stops inside the detector. The change of the number of stopping muons is expected due to the different propagation length of cosmic-ray muons at the underground site. Based on the MUSIC simulation~\cite{Tang:2006uu}, the estimated average energy of cosmic-ray muon is about $270$~GeV at the detector site. However, uncertainties in the energy spectrum of primary cosmic-rays, the atmospheric density structure, and the interaction model of QCD cause uncertainties in the intensity of cosmic-ray muons and their energy spectrum at the detector site. The production rates of unstable isotopes through muon capture on oxygen nuclei are estimated by simulation-based studies, and then uncertainties are taken into account. Table~\ref{tb:exp-rate} summarizes the expected production rate of $\mathrm{^{16}N}$ produced by the nuclear capture of negative muon on oxygen in the SK detector~\cite{Galbiati:2005ft, Li:2014sea}. 

\begin{table}[!h]
    \begin{center}
    \caption{Summary of the expected production rate of $\mathrm{^{16}N}$ by the nuclear capture of negative muons in water from the theoretical predictions~\cite{Galbiati:2005ft, Li:2014sea}. No theoretical uncertainty is mentioned in either publication.}
        \label{tb:exp-rate}
            \begin{tabular}{cc}
                \hline \hline
                Theoretical prediction & $\mathrm{^{16}N}$ production rate \\
                 & [event/kton/day]\\ \hline
                 C.~Galbiati~et al.~\cite{Galbiati:2005ft} & $3.2\phantom{0}$ \\       
                 S.~W.~Li~et al.~\cite{Li:2014sea}& $3.0$ \\
            \hline \hline
        \end{tabular}
    \end{center}
\end{table}

Based on the two theoretical predictions, the daily production rate of $\mathrm{^{16}N}$ is small, but the large fiducial volume and the long operation of the SK detector enable us to statistically measure the production rate of $\mathrm{^{16}N}$. Note that those predictions consider the production rate of $\mathrm{^{16}N}$ only through the nuclear muon capture process on oxygen in water, and other production methods such as spallation production due to through-going muon are not included.

\section{Analysis for $\bm{\mathrm{^{16}N}}$ and $\bm{\mathrm{^{15}C}}$ decays} \label{sec:analysis}

In this section, we present the analysis method and results for measuring the production rates of $\mathrm{^{16}N}$ and $\mathrm{^{15}C}$ in the timing window of $[0.5, 25.0]$~s after the stopping muon. The production rates of $\mathrm{^{12}B}$  and $\mathrm{^{13}B}$ are described in Sec.~\ref{sec:b12}.

\subsection{MC simulation} \label{sec:mc-intro}

In order to understand the detector response to cosmic-ray muons and the decays of unstable nuclei, a detector simulation based on the \texttt{GEANT3} toolkit~\cite{Brun:1994aa} was used for this study. This simulation has been tuned by comparing real calibration data with outputs from the MC simulation in SK~\cite{Abe:2013gga}.

The decay kinematics of unstable nuclei are computed using \texttt{Geant4}~\cite{agostinelli:2003250, Allison:2006, Allison:2016} version geant4.10.5p1, with \texttt{G4PreCompoundModel}~\cite{Quesada:2011} based on Evaluated Nuclear Structure Data Files~(ENSDF) version G4ENSDFSTATE2.2~\cite{ENSDF}. We simulated the decay times, the momenta of emitting electrons from $\beta$~decays, and the sequential de-excitation of progeny nuclei, and implemented the detector simulation to evaluate the response of decay products.

As detailed in the next subsections, the stopping-muon and $\mathrm{^{16}N}$~($\mathrm{^{15}C}$) decay events are separately recorded because of their relatively long half-lives. Therefore, we produced two different simulations: one simulates the stopping muon and the other simulates the $\mathrm{^{16}N}$ and $\mathrm{^{15}C}$ decay events. From the MC simulation for the stopping muon, we evaluate the selection efficiency of the stopping muon inside the fiducial volume. We generated $\mathrm{^{16}N}$ and $\mathrm{^{15}C}$ decay events whose generating vertex positions are taken from the stopping muon MC simulation. By combining two MC simulations, we then evaluate the total selection efficiency for measuring the production rate of $\mathrm{^{16}N}$ and $\mathrm{^{15}C}$ nuclei. Although we can not distinguish $\mathrm{^{16}N}$ decay events from $\mathrm{^{15}C}$ decay events, we can statistically measure the production rates of $\mathrm{^{16}N}$ and $\mathrm{^{15}C}$ by considering the difference in their half-lives and $Q$-values; this process will be explained in Sec.~\ref{sec:chi2}. For the event generation of the MC simulation, we used the decay time 
distribution shown in Fig.~\ref{fig:mc_expect-decay} and the energy spectra shown in Fig.~\ref{fig:mc_expect-energy}.

\subsection{Stopping muon selection}

\subsubsection{Reconstruction of cosmic-ray muon} \label{sec:mu-recon}

A decay event of $\mathrm{^{16}N}$~(or $\mathrm{^{15}C}$, $\mathrm{^{12}B}$ and $\mathrm{^{13}B}$) is separately recorded from a parent stopping muon since the recorded time of a single event is at most $+535~\mu$s as explained in Sec.~\ref{sec:daq}. We form a possible pair of stopping muon and decay events by considering the combination of timing and spatial information. 

\begin{figure}[h]
    \begin{center}
    \includegraphics[width=1.0\linewidth]{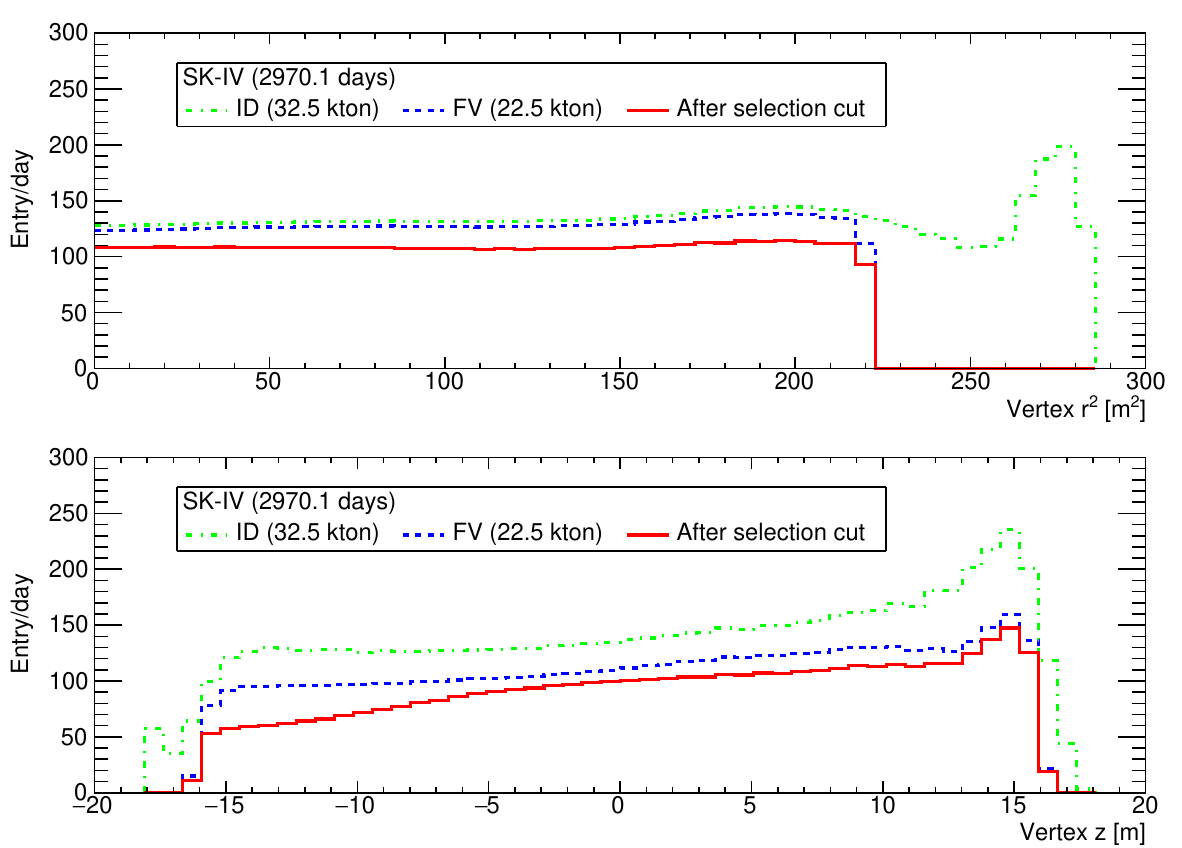}
    \end{center}
    \caption{The distributions of stopping muon positions before and after the muon selection using SK-IV real observed data. The top~(bottom) panel shows the distribution of the reconstructed vertex of radius~$r^{2}$~(height $z$). The light-green, blue, and red histograms show the number of stopping muons inside the inner detector, inside the fiducial volume, and after the muon selection. \label{fig:stop-vtx} }
\end{figure}

The first step of the measurement is selecting the stopping muon events inside the SK detector. The reconstruction of a cosmic-ray muon is performed by a dedicated muon fitter, called the MUBOY algorithm~\cite{Desai, Zoa}. MUBOY classifies reconstructed muons into four groups depending on the number of muon tracks and the event topology: (I)~Single through-going muons, (II)~Single stopping muons, (III)~Multiple muons, and (IV)~Corner-clipping muons. The procedure of reconstructing the stopping muon is detailed in Ref.~\cite{Kitagawa:2024ipa} and their entry position, direction of the muon track, and stopping position are reconstructed. Figure~\ref{fig:stop-vtx} shows the reconstructed stopping muon position in the SK detector.

Cosmic-ray muons with a short track inside the detector, as well as corner clipping muons, are sometimes mis-reconstructed as stopping muons. As detailed in Ref.~\cite{Kitagawa:2024ipa}, we optimized the selection criteria to maximize the selection efficiency for stopping muons and to minimize the mis-reconstruction of single-track muons without decays. The selection uses information on the topology of the observed muon such as track length, the total amount of observed charge, the maximum charge of a single PMT, and other reconstruction variables. When the stopping position is close to the detector structure, the MUBOY algorithm can not efficiently tag stopping muons. Based on the muon selection, stopping muons in the fiducial volume are selected with an efficiency of $(65.68\pm0.01)\%$ for SK-IV while a single penetrating muon is incorrectly recognized as a stopping muon at a rate less than $0.01\%$.

Table~\ref{tb:stop-eff} summarizes the selection efficiency for the stopping muon. After the muon selection, the resolution of the reconstructing stopping position inside the fiducial volume is about $50$~cm~\cite{Kitagawa:2024ipa}.

\begin{table}[!h]
    \begin{center}
    \caption{Summary of the selection efficiency for stopping muon in the fiducial volume. In this table, the statistical uncertainty of the selection efficiency is listed.}
        \label{tb:stop-eff}
            \begin{tabular}{lc}
                \hline \hline
                SK phase &  Selection efficiency \\
                 & in fiducial volume~[$\%$] \\ \hline
                SK-IV & $65.68 \pm 0.01$ \\
                SK-V\phantom{I} & $67.35\pm 0.04$ \\
                SK-VI & $67.26 \pm 0.03$ \\
                \hline \hline
        \end{tabular}
    \end{center}
\end{table}

\begin{figure}[h]
    \includegraphics[width=1.0\linewidth]{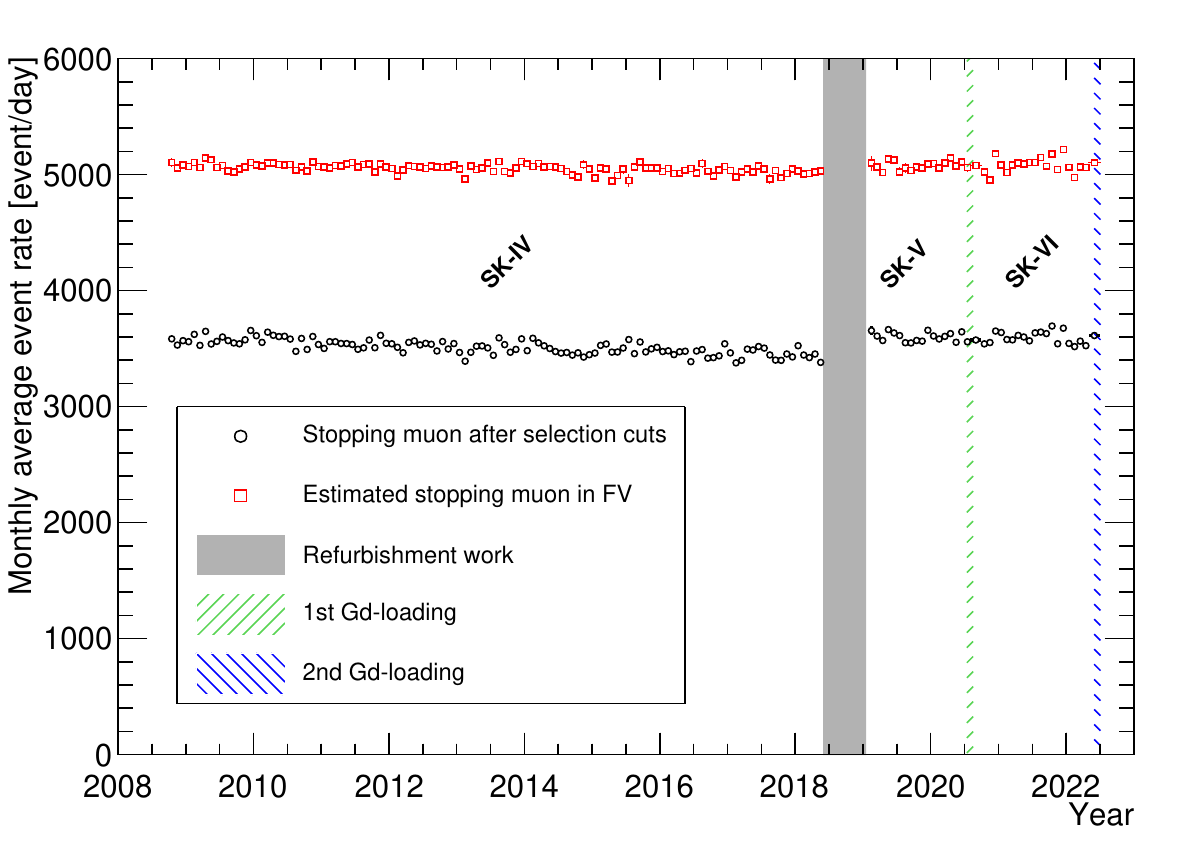}
    \caption{The monthly average of the number of stopping muons as a function of time. The black open circles show the event rate after the muon selection in units of event/day, and the red open squares show the estimated observed rate after correcting for selection efficiency. We should note that the statistical and systematic uncertainties are smaller than the symbols in this figure. The gray band shows the period of refurbishment work toward SK-Gd~\cite{Super-Kamiokande:2021the}. The green~(blue) slanting band shows the first~(second) Gd loading~\cite{Super-Kamiokande:2021the, Super-Kamiokande:2024kcb}. In the case of SK-IV, an increase of PMT gain was observed as explained in Ref.~\cite{Super-Kamiokande:2023jbt}, which resulted in a change to the cut criteria and a subsequent decrease of muon rate after the muon selection.}
    \label{fig:muon-rate}
\end{figure}

Figure~\ref{fig:muon-rate} shows the monthly stopping muon rate before and after the muon selection. The observed muon rate is basically stable during the SK operation after correcting the selection efficiency. 

Table~\ref{tb:stop-rate} summarizes the daily observed rates of stopping muons inside the fiducial volume. The observed rates after correcting the selection efficiency among the SK phases are consistent within their uncertainties.

\begin{table*}[]
    \begin{center}
    \caption{Summary of the daily observed event rate of the stopping muons inside the SK fiducial volume after the muon selection. The uncertainties described in this table are the combination of statistical uncertainties and systematic uncertainty described in Sect.~\ref{sec:sys_stopmu} and summarized in Table~\ref{tb:sys-stop-mu}.}
        \label{tb:stop-rate}
            \begin{tabular}{lcc}
                \hline \hline
                SK phase & \multicolumn{2}{c}{Observed event rate in fiducial volume~[event/day]} \\ 
                & Observed event rate & Estimated number of stopping muon \\
                & after muon selection & corrected by selection efficiency \\ \hline
                SK-IV & $3512 \pm 1\,(\mathrm{stat.})  {}^{+30}_{-34}\,(\mathrm{syst.})$ & $5347 \pm 2\,(\mathrm{stat.}) {}^{+46}_{-51}\,(\mathrm{syst.})$ \\
                SK-V\phantom{I} & $3602\pm 3\,(\mathrm{stat.})  {}^{+29}_{-32}\,(\mathrm{syst.})$ & $5347\pm 6\,(\mathrm{stat.})  {}^{+43}_{-47}\,(\mathrm{syst.})$ \\
                SK-VI & $3593 \pm 3\,(\mathrm{stat.})  {}^{+30}_{-33}\,(\mathrm{syst.})$ & $5341 \pm 4\,(\mathrm{stat.})  {}^{+45}_{-49}\,(\mathrm{syst.})$ \\
                \hline \hline
        \end{tabular}
    \end{center}
\end{table*}

\subsubsection{Search for decay electron}
 
\begin{figure}[h]
    \includegraphics[width=1.0\linewidth]{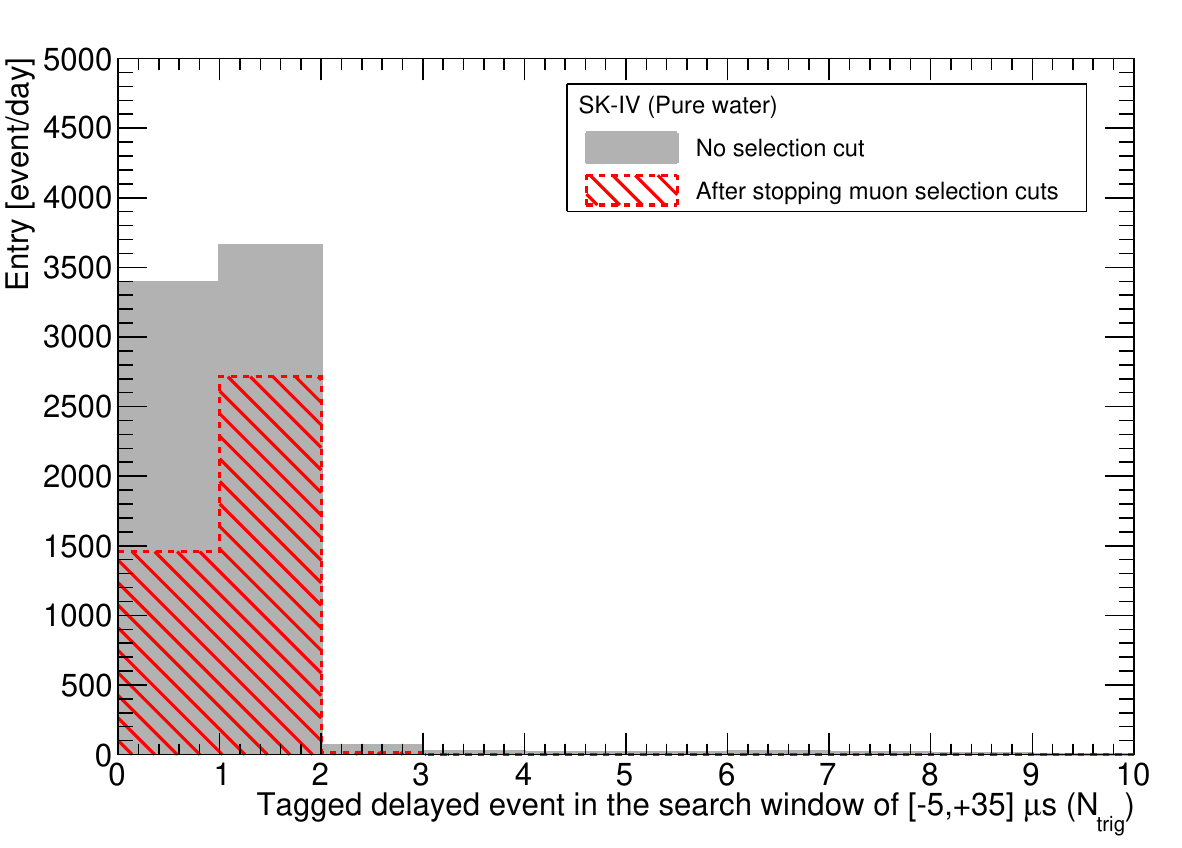}
    \caption{The distributions of delayed events~($N_{\mathrm{trig}}$) after stopping muons with and without the muon selection using the SK-IV data set. The vertical axis shows the daily number of events before~(illustrated in gray filled histogram) and after the muon selection~(illustrated in red slanting histogram). The horizontal axis shows the number of delayed events~($N_{\mathrm{trig}}$).}
    \label{fig:sub-trig-sk4}
\end{figure}

After installing the new electronics, the duration of the time window to record a single event is long enough to tag a decay electron event after a stopping muon. Hence, we search for a delayed event within the window of $+35~\mu$s after the stopping muon. The majority of non-captured stopping cosmic-ray muons produce their decay electrons and this results in expecting a single delayed event in that search window. Here, we defined the number of delayed events within $+35~\mu$s as $N_{\mathrm{trig}}$. Typically, the decay electron event is recognized as $N_{\mathrm{trig}}=1$. In contrast, no delayed event in the search window of $+35~\mu$s is expected when a nuclear muon capture occurs. In such a case, $N_{\mathrm{trig}}=0$ is expected. Figure~\ref{fig:sub-trig-sk4} shows the distribution of $N_{\mathrm{trig}}$ associated with observed stopping muons. We should note that $N_{\mathrm{trig}}\ge 2$ sometimes occurs when the accidental noise background event due to radioactive impurities, such as radon, is coincidentally observed within the timing window of $+35~\mu$s. However, the probability of $N_{\mathrm{trig}} \ge 2$ is quite small as demonstrated in Fig.~\ref{fig:sub-trig-sk4}. 

\begin{figure}[h]
    \includegraphics[width=1.0\linewidth]{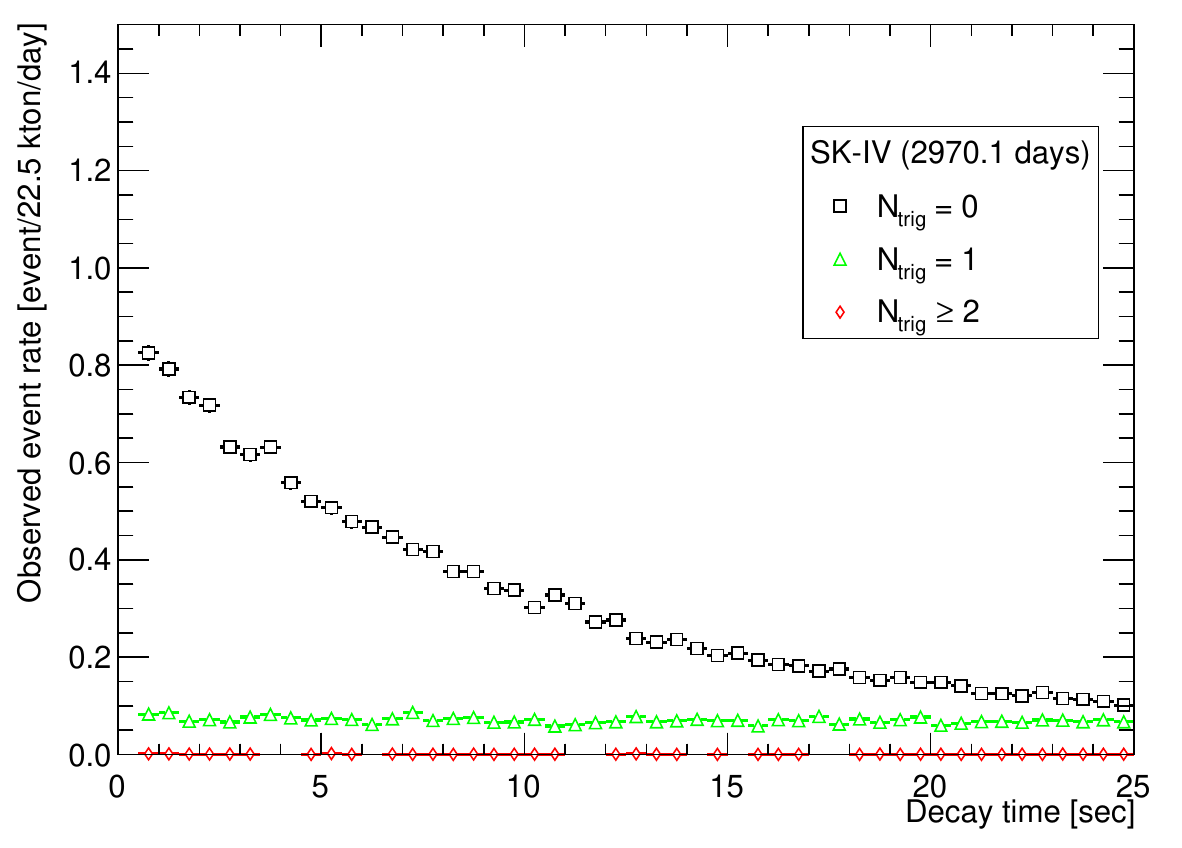}
    \caption{Decay time distributions in $[0.5,25.0]$~s with different $N_{\mathrm{trig}}$ samples using the SK-IV data set. The black open squares, light-green open triangles, and red open rhombuses show the number of decay candidate events around the stopping muon position with $N_{\mathrm{trig}}=0$~(without decay electron), $N_{\mathrm{trig}}=1$~(with decay electron), and $N_{\mathrm{trig}} \geq 2$~(with decay electron and accidental coincidence events), respectively. The sample of $N_{\mathrm{trig}}=0$ demonstrates that decay candidate events are observed.}
    \label{fig:data_decay_sub}
\end{figure}

Figure~\ref{fig:data_decay_sub} shows the decay time after the stopping muon in the window of $[0.5, 25.0]$~s. The sample of $N_{\mathrm{trig}}=0$ demonstrates that $\mathrm{^{16}N}$ decay candidate events are observed because its half-life is several seconds after stopping muon events are observed, as expected from $\mathrm{^{16}N}$ decays. On the other hand, no $\mathrm{^{16}N}$ decay events are observed when sampling the stopping muon with $N_{\mathrm{trig}} = 1$ and $N_{\mathrm{trig}} \ge 2$. Hence, the distribution in Fig.~\ref{fig:data_decay_sub} demonstrates that $\mathrm{^{16}N}$~(as well as $\mathrm{^{15}C}$) events are successfully tagged by sampling the stopping muons with $N_{\mathrm{trig}}=0$. Thus, selecting stopping muons with $N_{\mathrm{trig}}=0$ improves the signal-to-noise ratio of tagging $\mathrm{^{16}N}$ and $\mathrm{^{15}C}$ decay events.

We should note that contamination of $N_{\mathrm{trig}}$ sometimes occurs due to an accidental coincidence event just after the stopping muon. We evaluated the systematic uncertainty due to mis-counting the decay signals due to unstable isotopes in Sec.~\ref{sec:sys_subtrig}.

\subsection{Selection of $\bm{\mathrm{^{16}N}}$ and $\bm{\mathrm{^{15}C}}$ decay event} \label{sec:reduction}

\subsubsection{Pair construction with stopping muons}

Decay events of $\mathrm{^{16}N}$~(and $\mathrm{^{15}C}$) are recorded separately from stopping muon events due to their long half-lives; such decay events are triggered by SLE, LE, HE, and SHE triggers~(listed in Table~\ref{tb:trigger}) and reconstructed by a standard reconstruction fitter, called BONSAI~\cite{Smy:2007maa}.  Electrons from $\mathrm{^{16}N}$ decay or those scattered by $\gamma$-rays can travel only a few cm in water and the location of the Cherenkov emission is recognized as a point. Under this assumption, the vertex position is reconstructed with a maximum likelihood fit of the photon arrival time on the PMTs after subtracting the time of flight from the estimated interaction point~\cite{Super-Kamiokande:2024kcb}. 

The energy of those events is determined based on the number of hit PMTs with factors to account for delayed hits due to reflection and scattering in water, dark noise on nonhit PMTs, photo-cathode coverage, PMT gain, and water transparency. However, at the stage of the data acquisition, the energy threshold of the accumulated events changes depending on the detector conditions to avoid affecting the data transfer speed as mentioned in Sec.~\ref{sec:daq}. The QBEE electronics, installed at the SK-IV, allow the detector to accumulate more data with high-speed processing. This results in the lowering of the energy threshold at the stage of data acquisition. In this analysis, we set the lower energy threshold of selecting isotope decay events as $4.0$~MeV of the reconstructed total energy of electron.

To efficiently select $\mathrm{^{16}N}$ events, we first collected the stopping cosmic-ray muons and then selected the events that occurred within a certain distance from the stopping position and a relatively long time window after the stopping time. We set the lower limit of the time window as $0.5$~s to remove the background events from the decay electrons and the radioactive decays from spallation products, such as $\mathrm{^{12}B}$ and $\mathrm{^{12}N}$~\cite{Super-Kamiokande:2015xra, Super-Kamiokande:2021snn}. This selection will also include background events randomly occurring around the stopping muons. 

\subsubsection{Background estimation and subtraction}

For selecting the stopping muon, which is captured by oxygen in water, we require $N_{\mathrm{trig}}=0$ and therefore no delayed signal, such as a decay electron event as well as $\gamma$-rays from the excitation $\mathrm{^{14}N}$ and $\mathrm{^{15}N}$ decays within $[-5, +35]$~$\mu$s. 

\begin{figure}[h]
    \includegraphics[width=1.0\linewidth]{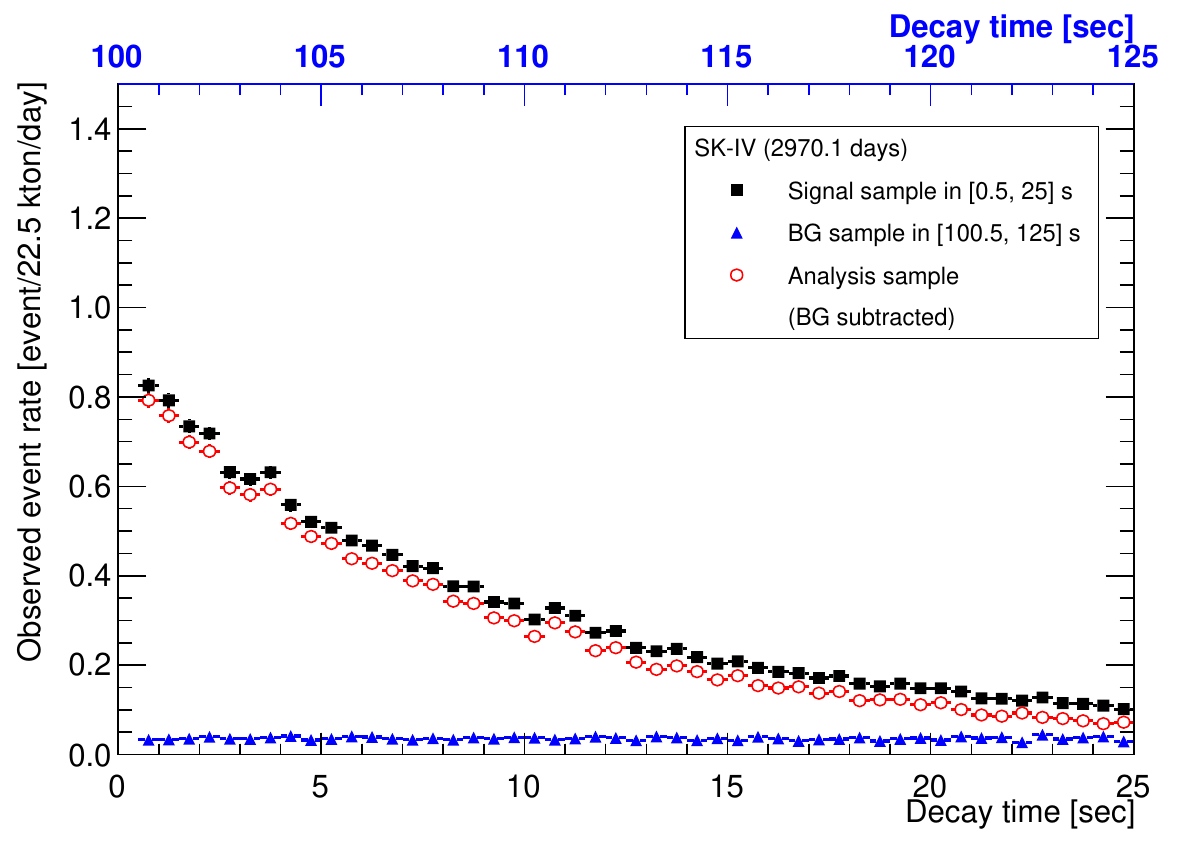}
    \caption{Decay time distribution of signal sample in $[0.5,25.0]$~s and background sample in $[100.5, 125.0]$~s after the selection cuts described in Sec.~\ref{sec:reduction}. After subtracting the background distribution from the signal distribution, the analysis sample is produced. The time difference from the stopping muon in the signal region~($[0.5, 25.0]$~s) drawn in black filled squares and the background region~($[100.5, 125.0]$~s with the upper horizontal axis) drawn in blue filled triangles. The analysis sample is drawn in red open circles.}
    \label{fig:data_decay}
\end{figure}

Since $\mathrm{^{16}N}$ and $\mathrm{^{15}C}$ decay with relatively long half-lives, the event selection described above can accidentally form an incorrect pair with a stopping muon. Even with the $N_{\mathrm{trig}}=0$ selection, some decay electrons can be missed due to being too close to the stopping muon in time, and these incorrectly labeled $N_{\mathrm{trig}}=0$ stopping muons can then form a random coincidence with a a delayed signal which is not a muon capture; this would mimic our signal and constitute our background. To estimate the possibility of forming a wrong pair due to the accidental background events, we also collected events within a $24.5$~s time window starting $100$~s after stopping muons, where this interval is the same length as the time window of the signal selection. Here, we refer to this $100$~s delayed window after the stopping muon as the background window. We should note that the selected events in the background window mainly originate from random background events due to radioactive impurities in the water~\cite{Nakano:2019bnr} and spallation products induced by through-going muons. Such events will not be correlated with stopping muons with $N_{\mathrm{trig}}=0$ because of the sufficiently delayed time. Figure~\ref{fig:data_decay} shows the decay time distribution after the stopping muon in both the signal window and the background window.

The events in the background window demonstrate that the analysis procedure described above can sometimes form an accidental coincidence event with a selected stopping muon. However, the event rate of the background window is flat and smaller than that in the signal window, as shown in Fig.~\ref{fig:data_decay}. Based on those distributions, the probability of forming a wrong pair due to an accidental background event is small. Hence, we subtracted the event rate in the background window from those in the signal window to reduce the contamination of non-$\mathrm{^{16}N}$ (and $\mathrm{^{15}C}$) isotope decays.

\begin{figure}[h]
    \includegraphics[width=1.0\linewidth]{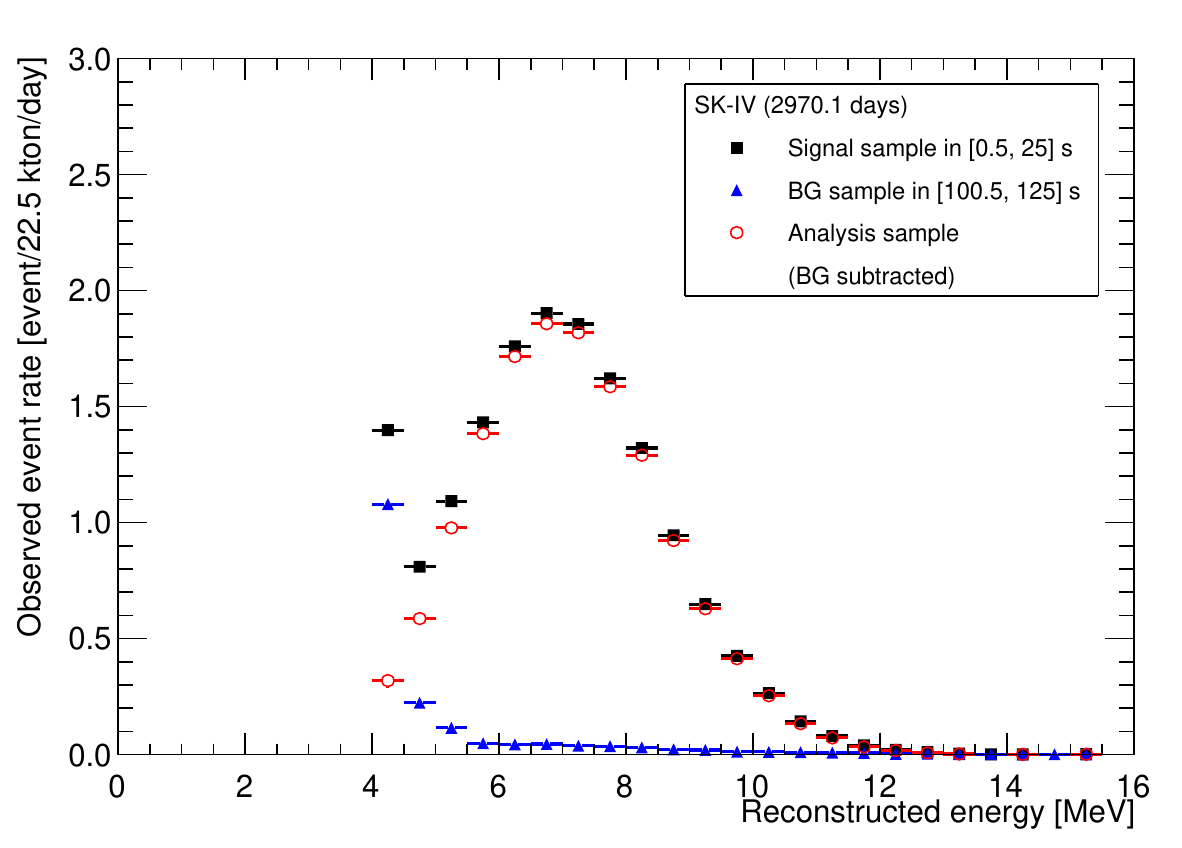}
    \caption{Reconstructed total energy distribution of signal sample in a $[0.5,25.0]$~s timing window and background sample in a $[100.5, 125.0]$~s timing window after the selection cuts described in Sec.~\ref{sec:reduction}. After subtracting the background distribution from the signal distribution, the analysis sample is produced. The definition of colors is the same as in Fig.~\ref{fig:data_decay}.}
    \label{fig:data_ene}
\end{figure}

Figure~\ref{fig:data_ene} shows the reconstructed energy distribution in both the signal window and the background window. After subtracting the background distribution, the peak is clearly visible. Hence, the energy distribution after subtracting the background distribution shown in Fig.~\ref{fig:data_ene} demonstrates that the $\mathrm{^{16}N}$ and $\mathrm{^{15}C}$ decay events are efficiently selected from the data samples.

\subsubsection{Optimization of selection cuts} \label{sec:opt}

After the initial event reconstruction by BONSAI, several cuts are applied to select the decay candidate events of $\mathrm{^{16}N}$ and $\mathrm{^{15}C}$. Backgrounds from radioactivity and poorly reconstructed events are seen close to the ID wall. To reduce these backgrounds, events with reconstructed vertices within $2$~m from the ID wall~(outside of the analysis fiducial volume) are rejected. We also define a backward-projected distance~(the distance from the reconstructed vertex to the wall opposite from the direction of travel~\cite{Super-Kamiokande:2005wtt}) and reject events for which this distance is less than $4$~m. 

In addition to the vertex and timing window cuts, we also apply an event quality cut, for which the quality of event reconstruction is quantified by two variables based on PMT hit timing~($g_{t}$) and hit pattern~($g_{p}$)~\cite{Super-Kamiokande:2016yck}. Some radioactive background events, originating mainly from the PMT enclosures, PMT glass, and detector wall structure, are mis-reconstructed inside the fiducial volume even though the true vertex lies outside the fiducial volume. To reject such backgrounds we select events whose ${g_{t}}^{2}-{g_{p}}^{2}$ is larger than $0.26$. Table~\ref{tb:eff-cut} summarizes the selection efficiencies of selection cuts in the SK-IV phase.

\begin{table*}[]
    \begin{center}
    \caption{Summary of selection efficiencies for $\mathrm{^{16}N}$, $\mathrm{^{15}C}$, $\mathrm{^{12}B}$, and $\mathrm{^{13}B}$ decay events using the SK-IV MC simulation with a timing window of $[0.5, 25.0]$~s. Here, efficiency of $100\%$ is the generated MC decay events inside the analysis fiducial volume. Note that both $\mathrm{^{12}B}$ and $\mathrm{^{13}B}$ decay events are completely rejected by the timing cut~($>0.5$~s) due to their short half-lives. Systematic uncertainties of other phases are summarized in Appendix~\ref{app:eff-sys}.}
        \label{tb:eff-cut}
            \begin{tabular}{p{0.20\textwidth}>{\centering}p{0.15\textwidth}>{\centering}p{0.15\textwidth}>{\centering}p{0.15\textwidth}>{\centering\arraybackslash}p{0.15\textwidth}}
                \hline \hline
                Selection cut & \multicolumn{4}{c}{Selection efficiency~[$\%$]}  \\ 
                 & $\mathrm{^{16}N}$ & $\mathrm{^{15}C}$ & $\mathrm{^{12}B}$ & $\mathrm{^{13}B}$\\ \hline
                $1$st reduction cut & $89.70 \pm 0.02$ & $84.12 \pm 0.02$ & $83.88\pm0.02$ & $84.52\pm0.02$\\
                Fiducial volume cut & $94.74 \pm 0.02$ & $94.16 \pm 0.02$ & $93.83\pm0.02$ & $94.01\pm0.02$\\
                Fit quality cut & $86.62 \pm 0.02$ & $81.34 \pm 0.01$ & $83.52\pm0.02$ & $84.02\pm0.01$\\
                Effective wall cut & $92.81 \pm 0.02$ & $92.10 \pm 0.02$ & $91.38\pm0.02$ & $91.57\pm0.02$\\
                Energy cut & $90.11 \pm 0.02$ & $84.95 \pm 0.02$ & $83.46\pm0.02$ & $84.24\pm0.02$\\
                Timing cut & $86.48\pm 0.02$ & $86.59 \pm 0.02$ & $0$ & $0$\\
                Distance cut & $63.58 \pm 0.01$ & $61.36 \pm 0.01$ & $60.29\pm0.01$ & $60.90\pm0.01$\\ \hline
                Total efficiency & $52.34 \pm 0.01$ & $46.56 \pm 0.01$ & $0$ & $0$\\ 
                \hline \hline
                \end{tabular}
    \end{center}
\end{table*}

\begin{figure}[h]
    \begin{center}
 
    \includegraphics[width=1.0\linewidth]{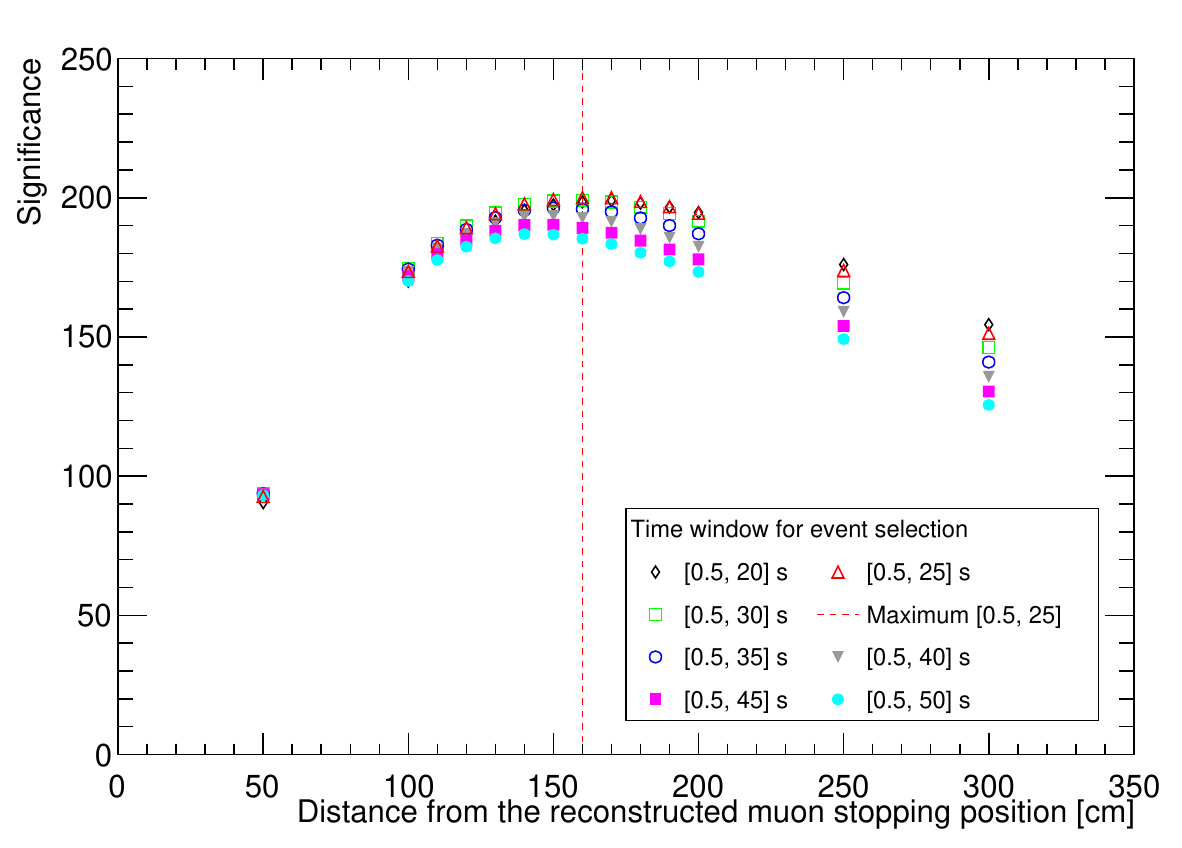}
    \end{center}
    \caption{Significance curves for the event selection of $\mathrm{^{16}N}$~(and $\mathrm{^{15}C}$) as a function of the distance between the stopping muon and the candidate event positions with different time windows. The plots with different colors and symbols show the significance values with different timing windows using the SK-IV data. The vertical red dashed line shows the maximum significance at $160$~cm with [$0.5, 25.0$]~s of the timing window. \label{fig:significance} }
\end{figure}

In addition to the standard event selection cuts described above, we also optimize the efficiency of forming the pair of the stopping muon and the decay candidate events. To maximize the signal-to-noise ratio to select both $\mathrm{^{16}N}$ and $\mathrm{^{15}C}$ events associated with the nuclear muon capture on oxygen in water, we calculated the significance of the event selection,

\begin{equation}
    \mathrm{Significance} = \frac{N_{\mathrm{Signal}}}{\sqrt{N_{\mathrm{Signal}}+N_{\mathrm{BG}}}} \label{eq:sig}
\end{equation}

\noindent where $N_{\mathrm{Signal}}$~($N_{\mathrm{BG}}$) is the number of selected events in the signal timing window~(background timing window). This selection considers the different durations of the timing window and the distance between the muon stopping position and the vertex position of decay candidate events. In the case of the signal timing window, both the decay of unstable isotopes and the accidental background events are selected. To remove the background events and improve the purity of the isotope in the analysis sample, we have subtracted the number of background events selected in the background timing window. Figure~\ref{fig:significance} shows the example of significance curves of the distance cut to select $\mathrm{^{16}N}$ and $\mathrm{^{15}C}$ events near the stopping muon position. Based on this evaluation, we set the distance cut to $160$~cm.

\subsection{$\bm{\chi^{2}}$ method} \label{sec:chi2}

To determine the production rates of $\mathrm{^{16}N}$ and $\mathrm{^{15}C}$ via a nuclear muon capture on oxygen in water, the decay times between the stopping muon and the selected event and the energy spectra of the selected events after subtracting the accidental backgrounds are simultaneously fit to the distributions derived from the MC simulation. The definition of total chi-square~($\chi^{2}_{\mathrm{Total}}$) is

\begin{equation}
    \chi^{2}_{\mathrm{Total}}~(R_{\mathrm{^{16}N}}, R_{\mathrm{^{15}C}}) = \chi^{2}_{\mathrm{Time}}+\chi^{2}_{\mathrm{Energy}} \label{eq:chi2}
\end{equation}

\noindent where $\chi^{2}$ is chi-square for each distribution, $R_{\mathrm{^{16}N}}$ and $R_{\mathrm{^{15}C}}$ are the given production rates for $\mathrm{^{16}N}$ and $\mathrm{^{15}C}$ in unit of event/kton/day, respectively. The $\chi^{2}$ for each distribution are defined as,

\begin{equation}
\left\{
\begin{array}{ll}
 \displaystyle \chi^{2}_{\mathrm{Time}} = \sum_{i}^{n_{\mathrm{Time}}} \frac{\left(N^{\mathrm{Data}}_{i}-N^{\mathrm{MC}}_{i} \right)^{2}}{(\sigma^{\mathrm{Data}}_{i})^{2}+(\sigma^{\mathrm{MC}}_{i})^{2}+(\sigma^{\mathrm{Syst.}}_{i})^{2}} \\
  \displaystyle \chi^{2}_{\mathrm{Energy}} = \sum_{i}^{n_{\mathrm{Energy}}} \frac{\left(N^{\mathrm{Data}}_{i}-N^{\mathrm{MC}}_{i} \right)^{2} }{(\sigma^{\mathrm{Data}}_{i})^{2}+(\sigma^{\mathrm{MC}}_{i})^{2}} + \left( \frac{1-p}{\sigma^{\mathrm{E\text{-}scale}}} \right)^{2}\\
\end{array}
\right. \label{eq:chi2-each}
\end{equation}

\noindent 
where $N_{i}^{\mathrm{Data}}$~($N_{i}^{\mathrm{MC}}$) is the number of selected events in $i$-th bin of the observed data~(MC) distribution, $n$ is the number of bins, $\sigma_{i}^{\mathrm{Data}}$~($\sigma_{i}^{\mathrm{MC}}$) is the statistical uncertainty on each bin of the observed data~(MC), and $\sigma_{i}^{\mathrm{Syst.}}$ is the systematic uncertainty on each bin which will be described in Sec.~\ref{sec:sys}, respectively. Since the energy scale of the detector response affects the value of $\chi^{2}_{\mathrm{Energy}}$, the pull term is introduced only for $\chi^{2}_{\mathrm{Energy}}$, where $\sigma^{\mathrm{E\text{-}scale}}$ is the systematic uncertainty of the energy scale determined from LINAC calibration~\cite{Super-Kamiokande:1998hbb}. The details are described in the next section. 

\subsection{Systematic uncertainties} \label{sec:sys}

In this section, we provide an overview of the systematic uncertainties for the measurement of production rates of $\mathrm{^{16}N}$ and $\mathrm{^{15}C}$. The systematic uncertainties listed in this section are also used in the analysis for the $\mathrm{^{12}B}$ and $\mathrm{^{13}B}$ measurements.

\subsubsection{Reconstruction and selection for the stopping muon events} \label{sec:sys_stopmu}

The systematic uncertainties on the number of selected stopping muons are estimated based on MC simulation. The accuracy of the reconstructed stopping muon direction as well as the track length is critical to select the stopping muon event. For direction and track length systematic uncertainties, we estimated the change in number of events which pass our selection based on truth and reconstruction differences in MC simulation and estimated the systematic uncertainty based on this. In addition to the track reconstruction, the observed charge is also critical to selecting the stopping muon. The energy scale of the stopping muon is estimated by its track length in the water tank and the variations are typically at the $\pm2\%$ level among SK phases~\cite{Super-Kamiokande:2023ahc}. By scaling the observed charge of PMTs in the MC simulation, we estimated the number of events after the muon selection. Table~\ref{tb:sys-stop-mu} summarizes the systematic uncertainties to select the stopping muon events. The systematic uncertainties originating from the stopping muon selection are consistent among the three SK phases.

\begin{table*}[]
    \begin{center}
    \caption{Summary of systematic uncertainties on the number of stopping muon events in SK-IV, SK-V, and SK-VI. The detail of the selection cut is described in Sect.~\ref{sec:mu-recon}~\cite{Kitagawa:2024ipa}.}
        \label{tb:sys-stop-mu}
            \begin{tabular}{lccc}
                \hline \hline
                Selection cut & \multicolumn{3}{c}{Systematic uncertainty~[$\%$]}  \\
                & SK-IV & SK-V & SK-VI \\ \hline
                Track length cut & $\pm0.4$ & $\pm0.4$ & $\pm 0.4$\\
                Stopping muon direction cut & $\pm0.5$ & $\pm0.5$ & $\pm0.5$\\
                Total charge cut & $+0.4/-0.6$ & $+0.4/-0.5$ & $+0.4/-0.5$\\
                Maximum charge cut & $\pm0.5$ & $\pm0.4$ & $\pm0.4$\\ \hline
                Total systematic uncertainty  & $+0.9/-1.0$ & $+0.8/-0.9$ & $+0.8/-0.9$\\
                \hline \hline
                \end{tabular}
    \end{center}
\end{table*}

\subsubsection{Reconstruction and selection cuts for the isotope decay events}

In general, the systematic uncertainties of the number of selected events are estimated using calibration data and simulation. To calculate uncertainty on the efficiency of the reduction steps and the accuracy of event reconstruction, calibration data and the MC simulated events are compared by using LINAC~\cite{Super-Kamiokande:1998hbb} and DT calibration devices~\cite{Super-Kamiokande:2000kzn}, and a Ni calibration source~\cite{Abe:2013gga} as used in the solar neutrino analysis~\cite{Super-Kamiokande:2023jbt}. Table~\ref{tb:sys-tab} summarizes the systematic uncertainties on selection cuts determined by comparing calibration data with MC simulation and by shifting the cut parameter values.

\begin{table}[]
    \begin{center}
    \caption{Summary of systematic uncertainties on the number of events for $\mathrm{^{16}N}$ and $\mathrm{^{15}C}$ decay events with the timing window of $[0.5, 25.0]$~s using the SK-IV MC simulation. Systematic uncertainties of other phases are summarized in Appendix~\ref{app:eff-sys}.}
        \label{tb:sys-tab}
            \begin{tabular}{lcc}
                \hline \hline
                Selection cut & \multicolumn{2}{c}{Systematic uncertainty~[$\%$]}  \\ 
                 & $\mathrm{^{16}N}$& $\mathrm{^{15}C}$  \\ \hline
                $1$st reduction cut & $<0.1$ & $<0.1$  \\
                Fiducial volume cut & $\pm0.2$ & $\pm0.2$ \\
                Fit quality cut & $\pm0.4$ & $\pm0.2$ \\
                Effective wall cut & $\pm0.1$ & $\pm0.1$ \\
                Energy cut & $\pm0.2$ & $\pm0.2$ \\
                Timing cut & $<0.1$ & $<0.1$  \\
                Distance cut & $\pm0.1$ & $\pm0.1$ \\ \hline
                Total  & $\pm0.5$ & $\pm0.4$ \\ 
                \hline \hline
                \end{tabular}
    \end{center}
\end{table}

\subsubsection{Mis-counting of $N_{\mathrm{trig}}$} \label{sec:sys_subtrig}

As demonstrated in Fig.~\ref{fig:data_decay_sub}, the event selection with $N_{\mathrm{trig}}=0$ clearly improves the signal-to-noise ratio of finding the unstable isotope decays after the stopping muon events. However, the mis-counting of $N_\mathrm{{trig}}$ sometimes occurs due to an accidental coincidence event or PMT after-pulse contamination just after stopping muon. To evaluate the mis-counting of $N_{\mathrm{trig}}$, we performed the same analysis procedure on the sample with $N_{\mathrm{trig}}\ge1$ and then evaluated the number of extra events in the signal sample after subtracting the number of events in the background sample. Table~\ref{tb:mis-subtrig} summarizes the number of selected events after subtracting the background sample with different $N_\mathrm{trig}$.

\begin{table*}[]
    \begin{center}
    \caption{Summary of the number of selected events with the different $N_{\mathrm{trig}}$. In this table, the number of events after subtracting the background sample distribution is listed.}
        \label{tb:mis-subtrig}
            \begin{tabular}{l|ccc|ccc}
                \hline \hline
                SK phase & \multicolumn{6}{c}{Timing window} \\ \hline
                & \multicolumn{3}{c|}{$[0.5,25.0]$~s}  & \multicolumn{3}{c}{$[0.001, 0.5]$~s} \\
                & $N_{\mathrm{trig}}=0$ & $N_{\mathrm{trig}} \ge 1$  & Assigned systematics & $N_{\mathrm{trig}}=0$ & $N_{\mathrm{trig}} \ge 1$ & Assigned systematics \\ \hline
                SK-IV & $41624$ & $533$ & $\pm1.3\%$ & $7504$ & $113$ & $\pm1.5\%$\\
                SK-V & $4755$ & $237$ & $\pm5.0\%$ & $932$ & $38$ & $\pm4.1\%$\\
                SK-VI & $6664$ & $269$ & $\pm4.0\%$ & $1267$ & $60$ & $\pm4.7\%$\\ 
                \hline \hline
        \end{tabular}
    \end{center}
\end{table*}

Even though we selected events with $N_{\mathrm{trig}} \ge 1$, a small excess of the selected events exists in the analysis sample after subtracting the background sample. To compensate for those events, we assigned the systematic uncertainty on the number of selected events in the analysis sample. We should note that the different relative fractions among the three phases originate from the statistical uncertainties of the selected events after subtracting the background sample from the signal sample. We finally assign the systematic uncertainties on the number of selected events with $N_{\mathrm{trig}}=0$ as $\pm1.3\%$, $\pm5.0\%$, and $\pm4.0\%$ for the SK-IV, SK-V, and SK-VI samples, respectively.

\subsubsection{Absolute energy scale}

The energy reconstruction~(energy scale) is critical in this study to analyze the shape of the energy distribution. The energy reconstruction is tuned by comparing calibration data against MC simulation with LINAC~\cite{Super-Kamiokande:1998hbb} and deuterium-tritium neutron~(DT) generator~\cite{Super-Kamiokande:2000kzn} sources. The former determines the absolute energy scale by injecting mono-energetic electron beams and the latter evaluates the directional dependence of the energy scale as well as the stability of the energy scale in time with high statistics radioactive $\beta$~decays. The energy calibration is explained in detail in Appendix.~\ref{sec:energy}.

Table~\ref{tb:escale} summarizes the systematic uncertainties on the energy scale determined by two calibration sources, for each of the SK phases analysed~\cite{Super-Kamiokande:2023jbt}. The relatively large errors for SK-V and SK-VI are a result of the limited number of LINAC calibrations compared to SK-IV, due to their short running times.

\begin{table}[!h]
    \begin{center}
    \caption{Summary of systematic uncertainties on the energy scale, determined by the LINAC and DT calibrations~\cite{Super-Kamiokande:1998hbb, Super-Kamiokande:2000kzn, Super-Kamiokande:2023jbt}.}
        \label{tb:escale}
            \begin{tabular}{lc}
                \hline \hline
                SK phase & Systematic \\ 
                & uncertainty~[$\%$] \\ \hline 
                SK-IV~\cite{Super-Kamiokande:2023jbt} & $\pm0.48$ \\ 
                SK-V~\cite{Kitagawa:2024ipa} & $\pm0.87$ \\ 
                SK-VI~\cite{Kitagawa:2024ipa} & $\pm1.32$ \\ 
            \hline \hline
        \end{tabular}
    \end{center}
\end{table}

\subsubsection{Charge ratio of cosmic-ray muons}

Since a negatively charged muon undergoes nuclear capture by oxygen in water, the fraction of negative muons in the cosmic-ray muon sample affects the measurement of the branching ratio. The charge ratio~($R_{\mu}$) of cosmic-ray muons is defined as $R_{\mu} = N_{\mu^{+}}/N_{\mu^{-}}$, where $N_{\mu^{+}}~(N_{\mu^{-}})$ is the number of positively~(negatively) charged cosmic-ray muons.

The SK detector cannot determine the charge of cosmic-ray muons. However, the SK detector can statistically measure the charge ratio of stopping cosmic-ray muons by evaluating the decay time of paired muons and decay electrons because of the different decay times of positive and negative muons. Table~\ref{tb:charge_ratio} summarizes the charge ratio of cosmic-ray muons measured by the SK detector after installing the new front-end electronics~\cite{Kitagawa:2024ipa}.

\begin{table}[!h]
    \begin{center}
    \caption{The summary of the charge ratio~$(R_{\mu}=N_{^{+}}/N_{\mu^{-}})$ measured by the SK detector~\cite{Kitagawa:2024ipa}. As explained in Sec.~\ref{sec:daq123}, such a measurement could not be performed up to and including SK-III due to electronics performance.}
        \label{tb:charge_ratio}
            \begin{tabular}{cc}
                \hline \hline
                Phase & Charge ratio~($R_{\mu}$)  \\ \hline
                SK-IV & $1.32\pm0.03$ \\
                SK-V\phantom{I}& $1.26\pm 0.07$ \\
                SK-VI & $1.33\pm0.06$ \\ \hline
                Combined & $1.32 \pm 0.02$ \\
            \hline \hline
        \end{tabular}
    \end{center}
\end{table}

We used the uncertainties of the charge ratio measurement as the systematic uncertainty of the production rates measurement.

\subsubsection{Natural abundance of oxygen}

Three stable isotopes of oxygen~($\mathrm{^{16}O}$, $\mathrm{^{17}O}$, and $\mathrm{^{18}O}$) exist, and their natural abundance should be considered when measuring the production rate of isotopes~(as well as the branching ratio). The ratio of formation of muonic atom due to Coulomb force on each element can be expressed by the number of nuclei and the relative probabilities $P(Z)$, where $Z$ is the atomic number, and the parameter $P(Z)$ is experimentally measured~\cite{Daniel:1979ay}. Since they have the same number of protons, $P(Z)$ is the same among the three isotopes. Hence, the systematic uncertainty of the number of nuclear muon captures on oxygen is evaluated based on their natural abundance. According to the NuDat database~\cite{meija2016isotopic}, the dominant component is $\mathrm{^{16}O}$, whose natural abundance is $99.76\%$ while the small contributions are $\mathrm{^{17}O}$ and $\mathrm{^{18}O}$, whose natural abundances are $0.04\%$ and $0.20\%$, respectively.

As listed in Table~\ref{tb:stop-rate}, the event rate of stopping muons is $5347\pm2$~event/day in the fiducial volume after correcting for the selection efficiency of stopping muons in SK-IV. Then, the number of the nuclear muon captures on oxygen is estimated as $424.1\pm6.0$~event/day after additionally considering the charge ratio of stopping muons~\cite{Kitagawa:2024ipa} and the probability of nuclear muon capture by oxygen, which is expressed as $P_{\mathrm{cap}}=(18.44\pm0.01)\%$~\cite{Suzuki:1987jf}. Considering the natural abundances, we estimated the number of nuclear muon captures as $423.0$~event/day on $\mathrm{^{16}O}$, $0.2$~event/day on $\mathrm{^{17}O}$, and $0.9$~event/day on $\mathrm{^{18}O}$, respectively. Table~\ref{tb:oxygen} summarizes the estimated number of nuclear muon captures on $\mathrm{^{16}O}$, $\mathrm{^{17}O}$, and $\mathrm{^{18}O}$.

\begin{table}[!h]
    \begin{center}
    \caption{Summary of the estimated number of nuclear muon capture process on $\mathrm{^{16}O}$, $\mathrm{^{17}O}$, and $\mathrm{^{18}O}$ in the analysis fiducial volume~($22.5$~kton) in a single day. Based on this calculation, the contributions of $\mathrm{^{17}O}$ and $\mathrm{^{18}O}$ are negligible due to their low natural abundance.}
        \label{tb:oxygen}
            \begin{tabular}{l|c|ccc}
                \hline \hline
                SK phase & Nuclear muon capture & \multicolumn{3}{c}{Estimated number}\\
                & [event/day]  & $\mathrm{^{16}O}$ & $\mathrm{^{17}O}$ & $\mathrm{^{18}O}$  \\ \hline 
                SK-IV & $424.1\pm \phantom{0}6.0$ & $423.1$ & $0.2$ & $0.9$ \\
                SK-V & $435.3\pm13.7$ & $434.3$ & $0.2$ & $0.9$ \\
                SK-VI & $421.8\pm11.1$ & $420.8$ & $0.2$ & $0.9$ \\
            \hline \hline
        \end{tabular}
    \end{center}
\end{table}

Based on the calculation above, the sum of nuclear muon captures by $\mathrm{^{17}O}$ and $\mathrm{^{18}O}$ is smaller than the total uncertainty of nuclear muon capture on $\mathrm{^{16}O}$. Therefore, we ignored their contributions in this study.

\subsubsection{Nuclear capture after gadolinium loading}

As briefly mentioned in Sec.~\ref{sec:dataset}, the SK-VI phase started after the first gadolinium loading in July 2020. During this loading work, $13$~tons of $\mathrm{Gd_{2}(SO_{4})_{3} \cdot 8 H_{2}O}$ was dissolved, resulting in $0.021\%$ of $\mathrm{Gd_{2}(SO_{4})_{3}}$ concentration in the SK tank~\cite{Super-Kamiokande:2021the}. These additional elements may produce additional muonic atoms instead of oxygen in water.

As detailed in Ref.~\cite{Kitagawa:2024ipa}, the fraction of such additional muonic atoms is small, and their production rate is smaller than that of nuclear muon capture by oxygen in water. We ignored their contribution in this analysis. 

\subsection{Measurement of production rates of $\bf{\mathrm{^{16}N}}$ and $\bm{\mathrm{^{15}C}}$} \label{sec:production}

\begin{figure}[h]
    \begin{center}
    \includegraphics[width=1.0\linewidth]{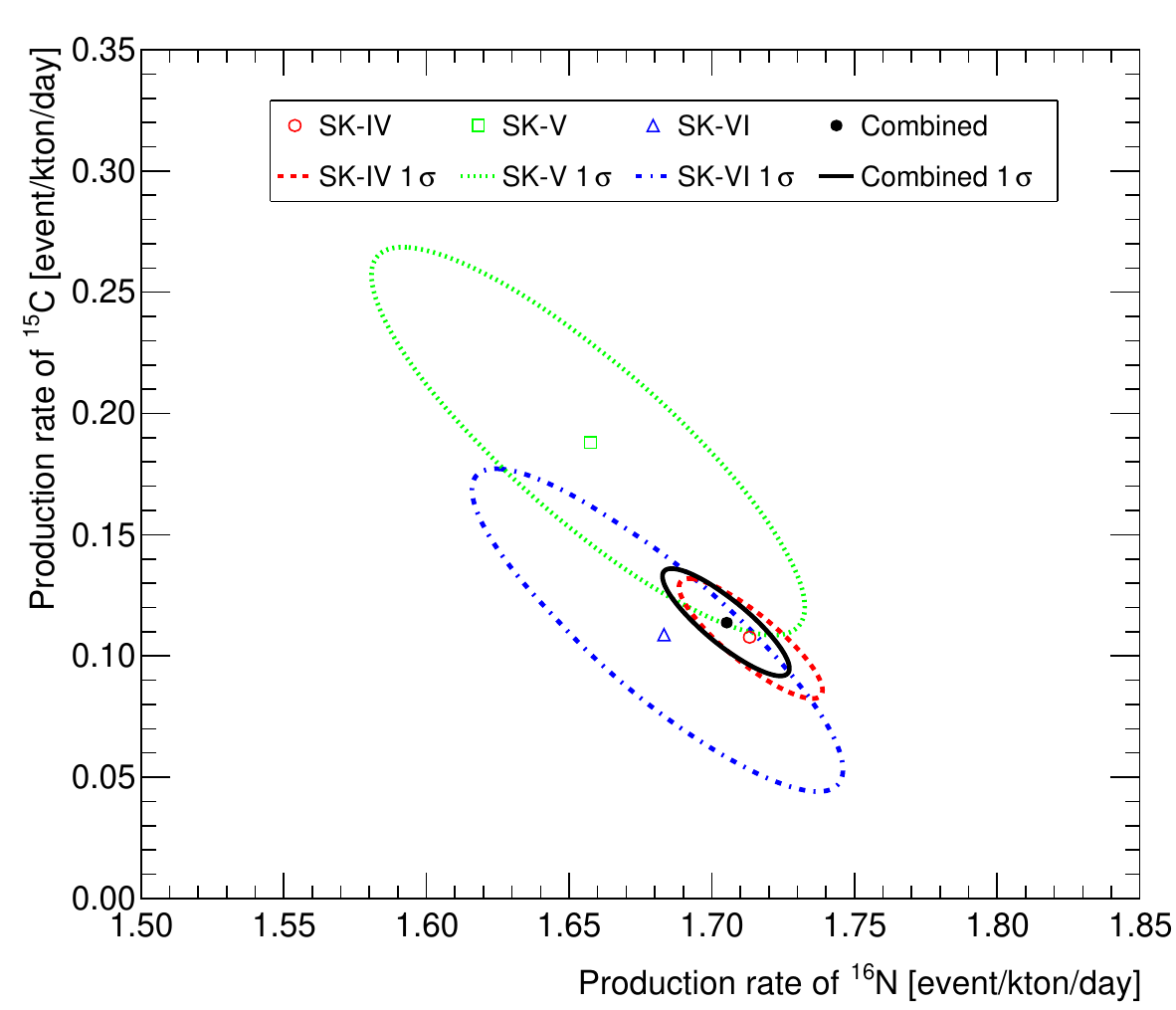}
    \end{center}
    \caption{The 1$\sigma$ allowed regions of the production rates of $\mathrm{^{16}N}$ and $\mathrm{^{15}C}$ using SK-IV, SK-V, and SK-VI data sets. The red open square (dashed line), green open square~(dotted line), and blue open upward-triangle~(dashed-dotted line) show the best-fit values~(their $1\sigma$ allowed regions determined by $\Delta \chi^{2}~(R_{\mathrm{^{16}N}}, R_{\mathrm{^{15}C}})=2.30$) of SK-IV, SK-V, and SK-VI, respectively. The black filled circle~(solid line) shows the combined value~($1\sigma$ allowed region). \label{fig:chi2-map}}
\end{figure}

To determine the production rates of naturally produced $\mathrm{^{16}N}$ and $\mathrm{^{15}C}$ through the nuclear muon capture on oxygen in water, we calculated the $\chi^{2}$ defined in Eq.~(\ref{eq:chi2}) and then extracted the difference between each value and the minimum value of $\chi^{2}$, which is expressed as $\Delta \chi^{2}(R_{\mathrm{^{16}N}}, R_{\mathrm{^{15}C}}) = \chi^{2} (R_{\mathrm{^{16}N}}, R_{\mathrm{^{15}C}}) - \chi^{2}_{\mathrm{Min}}$, where $\chi^{2}_{\mathrm{Min}}$ is the minimum value of $\chi^{2}$. Figure~\ref{fig:chi2-map} shows the result of $\Delta \chi^{2}$ calculation using SK-IV, SK-V, and SK-VI data sets. Based on the $\chi^{2}$ method, we determined the production rates of $\mathrm{^{16}N}$ and $\mathrm{^{15}C}$. Because non-zero $\mathrm{^{15}C}$ contribution does not explain the decay time and energy distributions, the observed data demonstrate the contribution from $\mathrm{^{15}C}$ decay events.
The measured production rates of $\mathrm{^{16}N}$ and $\mathrm{^{15}C}$ are consistent within their estimated uncertainties. 

Table~\ref{tb:event-per-day} summarizes the measured production rates of $\mathrm{^{16}N}$ and $\mathrm{^{15}C}$ among the SK phases. The combined values are $R_{\mathrm{^{16}N}}= 1.71 \pm 0.01\,(\mathrm{stat.+syst.})$~event/kton/day and $R_{\mathrm{^{15}C}} = 0.11 \pm 0.01\,(\mathrm{stat.+syst.})$~event/kton/day, respectively.

\begin{table}[]
    \begin{center}
    \caption{Summary of the production rates of $\mathrm{^{16}N}$ and $\mathrm{^{15}C}$ in water measured by the SK detector. The uncertainties shown in this table include both statistical and systematic uncertainties. 
    }
        \label{tb:event-per-day}
            \begin{tabular}{lcc}
                \hline \hline
                 SK phase &  \multicolumn{2}{c}{Production rate} \\
                 & \multicolumn{2}{c}{[event/kton/day]} \\ 
                 & $\mathrm{^{16}N}$  & $\mathrm{^{15}C}$ \\ \hline   
                 SK-IV & $1.71 \pm 0.02$ & $0.11 \pm 0.02$ \\ 
                 SK-V & $1.66 \pm 0.05$ & $0.19\pm 0.05$ \\ 
                 SK-VI & $1.68 \pm 0.04$ & $0.11 \pm 0.04$ \\ \hline
                 Combined & $1.71 \pm 0.01$ & $0.11\pm0.01$ \\
                 \hline \hline
        \end{tabular}
    \end{center}
\end{table}

\begin{figure}[]
    \includegraphics[width=1.0\linewidth]{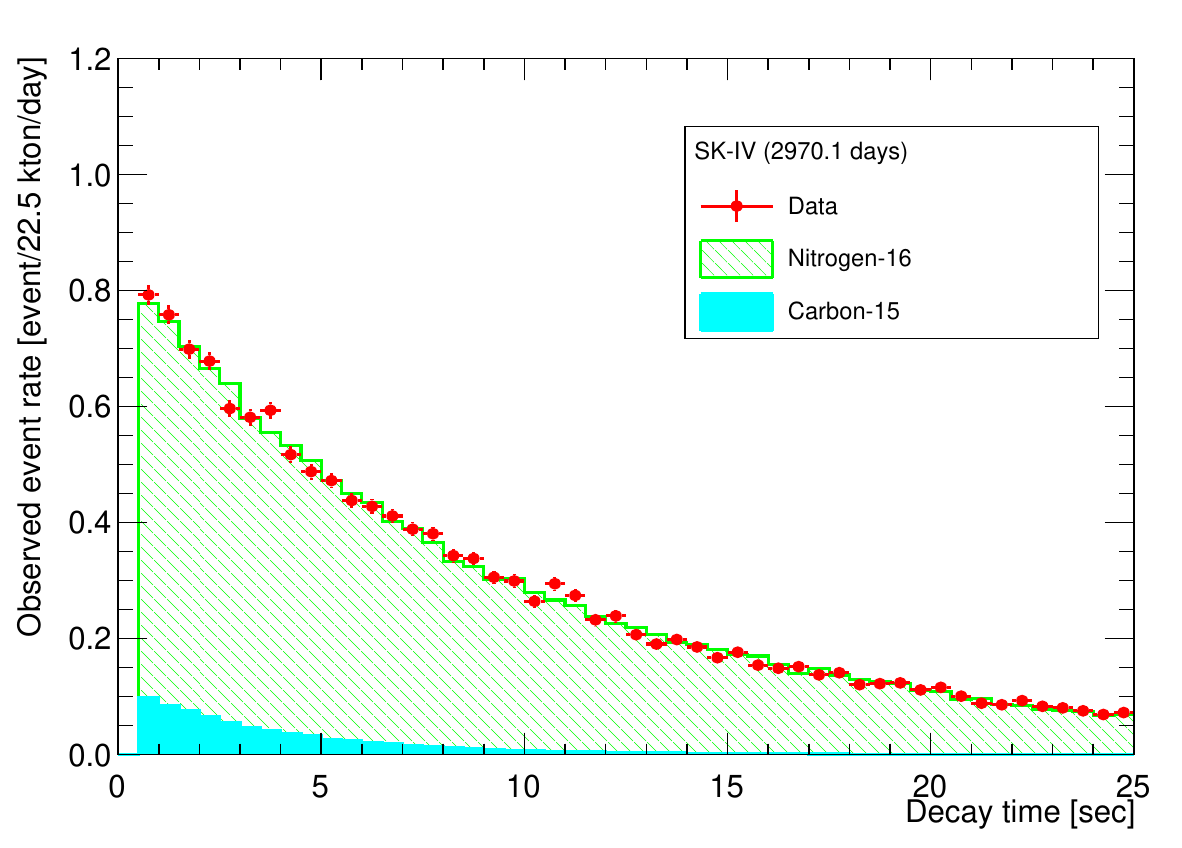} \\
    \includegraphics[width=1.0\linewidth]{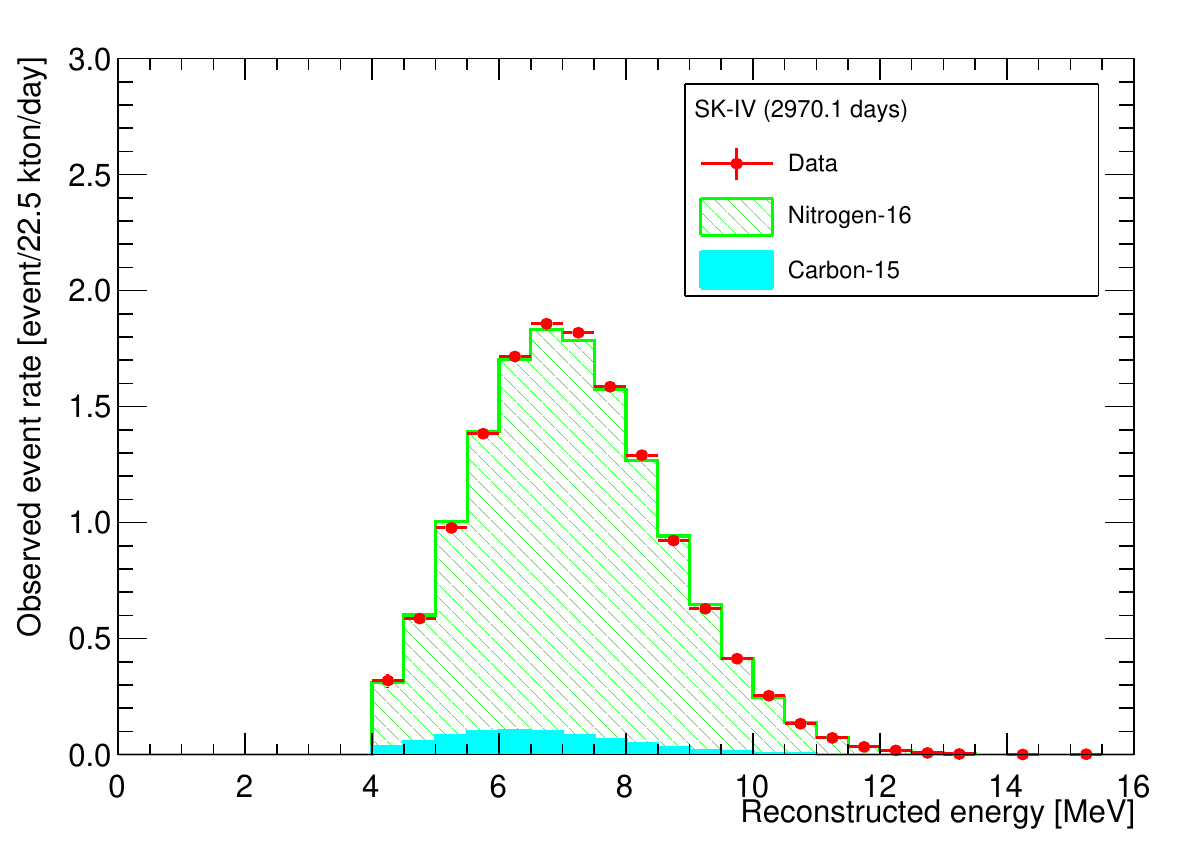}
    \caption{Distributions of the time difference~(top) in the timing window of $[0.5, 25.0]$~s and the reconstructed total energy~(bottom) together with the best fit of the components from $\mathrm{^{16}N}$ and $\mathrm{^{15}C}$ using the SK-IV data. Red-filled circles show the observed data after the selection cuts. The light-green right slanting histogram and the light-blue filled histogram show the best fit of $\mathrm{^{16}N}$, and $\mathrm{^{15}C}$ components, respectively. The distributions of other phases are shown in Appendix~\ref{sec:app-other-dist}.}
    \label{fig:best-sk4}
\end{figure}

Figure~\ref{fig:best-sk4} shows examples of decay time and energy distributions using the SK-IV data together with their best-fit distributions. The distributions using other SK phases are shown in the Appendix~\ref{sec:app-other-dist}. 

\subsection{Comparison with theoretical predictions} \label{sec:n16-exp-theor}

\begin{figure}[h]
    \begin{center}
    \includegraphics[width=1.0\linewidth]{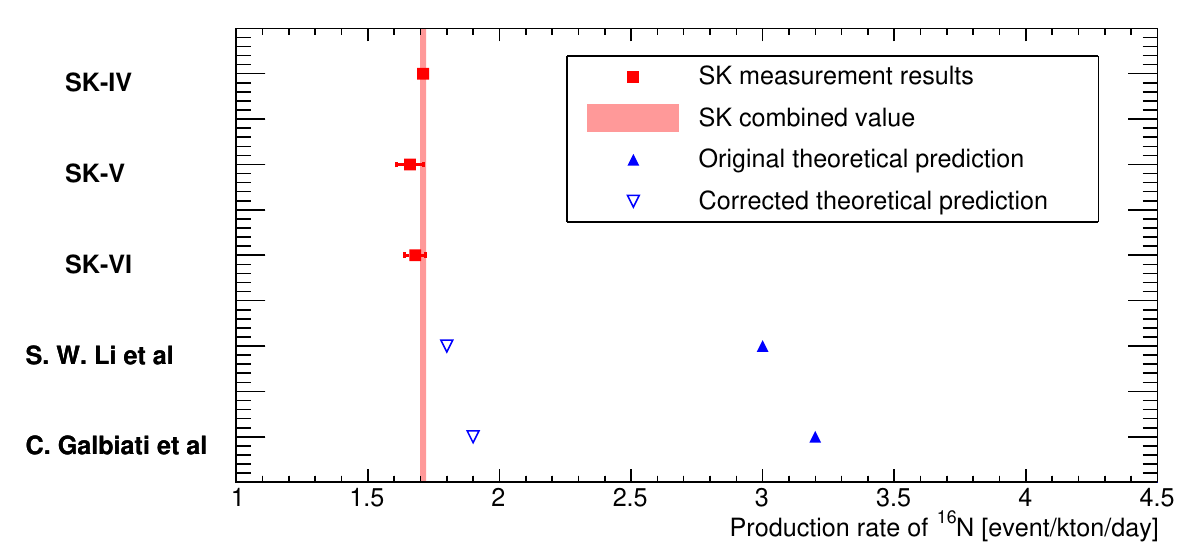}
    \end{center}
    \caption{Comparison of the measurement of the production rate of $\mathrm{^{16}N}$ through nuclear muon capture in water with the predictions from two theoretical models~\cite{Galbiati:2005ft, Li:2014sea}. The red square points show the measured production rate of $\mathrm{^{16}N}$, the red band shows the combined value of the SK measurement~(range of $1\sigma$), the blue filled upward-triangle points show the original theoretical predictions, and the blue open downward-triangle points show the corrected theoretical predictions by considering the topology of stopping muon track at the SK experimental site. \label{fig:comp_exp_theor} }
\end{figure}

In order to test the theoretical predictions as introduced in Sec~\ref{sec:overview}, we compared the measured production rates with them. Figure~\ref{fig:comp_exp_theor} shows the comparison of the production rate of $\mathrm{^{16}N}$ measured by the SK detector with theoretical predictions. The combined production rate of $\mathrm{^{16}N}$ is $1.71 \pm 0.01\,(\mathrm{stat.+syst.})$~event/kton/day, where we merged the measurement results from SK-IV, SK-V, and SK-VI. Comparing the measured production rates with the theoretical predictions~\cite{Galbiati:2005ft, Li:2014sea}, the estimated production rate is roughly a factor of two larger than that measured in this study.

To properly compare theoretical and experimental production rates, differences in stopping muon rates must be understood. Theoretical predictions consider only vertically down-going muons in their calculation. For the theoretical predictions, the average track length is about $32.2~\mathrm{m}$, and muons with energies below $6.4~\mathrm{GeV}$ at the experimental site are recognized as stopping muons. This corresponds to $5\%$ of cosmic-ray muons reaching the SK detector. In a realistic situation, muons with non-vertical angles can also be detected, and the measured average track length is about $24.3~\mathrm{m}$~\cite{Super-Kamiokande:2022cvw}. Based on this value, muons with energies below approximately $4.8~\mathrm{GeV}$ will stop inside the detector, which corresponds to $3\%$ of cosmic-ray muons. Taking into account the differences in the stopping muon rates, the theroetical predictions for the production rates of $\mathrm{^{16}N}$ are corrected to about $1.9~\mathrm{event/kton/day}$~\cite{Galbiati:2005ft} and $1.8~\mathrm{event/kton/day}$~\cite{Li:2014sea}, respectively. This consideration of stopping muon rate significantly reduces the discrepancy between the predictions and measurements, as shown in Fig.~\ref{fig:comp_exp_theor}. However, there still exists a difference between the measurement and predictions. This remaining discrepancy suggests that the simulation of the production of radioactive isotopes be updated by including this measurement result.

\section{Analysis for $\bm{\mathrm{^{12}B}}$ and $\bm{\mathrm{^{13}B}}$ decays} \label{sec:b12}

In this section, we present the analysis method and the result of measuring the production rates of $\mathrm{^{12}B}$ and $\mathrm{^{13}B}$ in the timing window of $[0.001, 0.5]$~s after the stopping muon. We basically follow the procedure of the decay event selection for $\mathrm{^{16}N}$ and $\mathrm{^{15}C}$ measurements as described in Sect.~\ref{sec:reduction}.

\subsection{Analysis for $\bm{\mathrm{^{12}B}}$ and $\bm{\mathrm{^{13}B}}$ measurement}

As demonstrated in Fig.~\ref{fig:mc_expect-decay-b12b13}, both the $\mathrm{^{12}B}$ and $\mathrm{^{13}B}$ decay events are expected to be observed within $0.5$~s after stopping muons due to their short half-lives. To evaluate the detector response of $\mathrm{^{12}B}$ and $\mathrm{^{13}B}$ decay events using the MC simulation, we generated their decays based on distributions shown in Fig.~\ref{fig:mc_expect-decay-b12b13} and Fig.~\ref{fig:mc_expect-energy-b12b13}. The selection of both decays is performed by sampling the events within a relatively short time window after stopping muons. For the range of the timing window, we set the upper limit of $0.5$~s so that the time interval of the $\mathrm{^{16}N}$ and $\mathrm{^{15}C}$ measurement does not overlap with it. We also set the lower limit of $0.001$~s because of the following reasons: After installing the new electronics in SK-IV onwards, the DAQ system generates the additional trigger~(AFT trigger), whose range of recording delayed events is at most $535~\mu$s. We use the time window of $[0.001, 0.5]$~s for measuring the production rates of $\mathrm{^{12}B}$ and $\mathrm{^{13}B}$ isotopes. Because the measured four isotopes have similar $Q$-values, the vertex resolutions of selected decay events do not change even in the different timing windows. Hence, the distance cut~($160$~cm) is the same as that for $\mathrm{^{16}N}$ and $\mathrm{^{15}C}$ measurements. Other selection criteria are also optimized by following the procedure described in Sect.~\ref{sec:opt}.

Table~\ref{tb:eff-cut-b12} summarizes the selection efficiencies for the timing window of $[0.001,0.5]$~s. Although the duration of the timing window is short, background events from $\mathrm{^{16}N}$ and $\mathrm{^{15}C}$ exist in the window.

\begin{table*}[]
    \begin{center}
    \caption{Summary of selection efficiencies for $\mathrm{^{16}N}$, $\mathrm{^{15}C}$, $\mathrm{^{12}B}$, and $\mathrm{^{13}B}$ decay events using the SK-IV MC simulation with a timing window of $[0.001, 0.5]$~s. Except for the timing cut, the efficiency of selection cuts is the same as Table~\ref{tb:eff-cut}. Selection efficiencies and their uncertainties of other phases are summarized in Appendix~\ref{app:eff-sys}.}
        \label{tb:eff-cut-b12}
            \begin{tabular}{p{0.12\textwidth}>{\centering}p{0.10\textwidth}>{\centering}p{0.10\textwidth}>{\centering}p{0.10\textwidth}>{\centering\arraybackslash}p{0.10\textwidth}}
                \hline \hline
                Selection cut &\multicolumn{4}{c}{Selection efficiency~[$\%$]}  \\ 
                 & $\mathrm{^{16}N}$ & $\mathrm{^{15}C}$ & $\mathrm{^{12}B}$ & $\mathrm{^{13}B}$\\ \hline
                Timing cut& $\phantom{0}4.75 \pm 0.01$ & $13.30 \pm 0.01$ & $96.66\pm0.02$  & $96.04\pm0.02$\\ 
                ([$0.001,0.5$]~s) & & & & \\ 
                Selection cuts &\multicolumn{4}{c}{Same as Table~\ref{tb:eff-cut}} \\ \hline
                Total efficiency & $\phantom{0}2.88 \pm 0.01$ & $\phantom{0}7.14 \pm 0.01$ & $53.78\pm0.01$ & $54.19\pm0.01$\\
                \hline \hline
                \end{tabular}
    \end{center}
\end{table*}

We also estimated the systematic uncertainties for $\mathrm{^{12}B}$ and $\mathrm{^{13}B}$ measurements. Table~\ref{tb:sys-tab-b12b13} summarizes the systematic uncertainty of the selection cuts. Other systematic uncertainties are the same as those described in Sec.~\ref{sec:sys}.

\begin{table}[]
    \begin{center}
    \caption{Summary of systematic uncertainties on the number of events for $\mathrm{^{12}B}$ and $\mathrm{^{13}B}$ decay events with the timing window of $[0.001, 0.5]$~s using the SK-IV MC simulation. Systematic uncertainties of other phases are summarized in Appendix~\ref{app:eff-sys}.}
        \label{tb:sys-tab-b12b13}
            \begin{tabular}{lcccc}
                \hline \hline
                Selection cut &\multicolumn{2}{c}{Systematic uncertainty~[$\%$]}  \\ 
                 &  $\mathrm{^{12}B}$ & $\mathrm{^{13}B}$\\ \hline
                $1$st reduction cut & $<0.1$& $<0.1$ \\
                Fiducial volume cut & $\pm0.1$& $\pm0.1$ \\
                Fit quality cut & $\pm0.4$ & $\pm0.4$ \\
                Effective wall cut & $\pm0.1$ & $\pm0.1$ \\
                Energy cut & $\pm0.1$ & $\pm0.1$ \\
                Timing cut & $<0.1$ & $<0.1$  \\
                Distance cut & $\pm0.2$ & $\pm0.1$ \\ \hline
                Total  & $\pm0.4$ & $\pm0.4$ \\
                \hline \hline
                \end{tabular}
    \end{center}
\end{table}

\subsection{$\chi^{2}$ method for $\mathrm{^{12}B}$ and $\mathrm{^{13}B}$ measurement} 

To determine the production rates of $\mathrm{^{12}B}$ and $\mathrm{^{13}B}$ isotopes, the $\chi^{2}$ calculation described in Eq.~(\ref{eq:chi2}) is modified by introducing the contribution of $\mathrm{^{12}B}$ and $\mathrm{^{13}B}$ decays as following;

\begin{align}
    & \hspace{-7mm}  \chi^{2}_{\mathrm{Total}}~(R_{\mathrm{^{16}N}}^{\mathrm{Best}}, R_{\mathrm{^{15}C}}^{\mathrm{Best}}, R_{\mathrm{^{12}B}}, R_{\mathrm{^{13}B}}) \notag \\ 
     & =  \chi^{2}_{\mathrm{Time}}+\chi^{2}_{\mathrm{Energy}} \notag \\
     & +   \left( \frac{(1-\alpha)~N_{\mathrm{^{16}N}}^{\mathrm{Best}}}{\sigma^{\mathrm{Best}}_{\mathrm{^{16}N}}} \right)^{2} 
      + \left( \frac{(1-\beta)~N_{\mathrm{^{15}C}}^{\mathrm{Best}}}{\sigma^{\mathrm{Best}}_{\mathrm{^{15}C}}} \right)^{2} \label{eq:chi2-b12-b13}
\end{align}

\noindent where $R_{\mathrm{^{12}B}}$~($R_{\mathrm{^{13}B}}$) is the production rate of $\mathrm{^{12}B}$~($\mathrm{^{13}B}$) to be measured by this method, $R_{\mathrm{^{16}N}}^{\mathrm{Best}}$~($R_{\mathrm{^{15}C}}^{\mathrm{{Best}}}$) is the measured production rate in $\mathrm{^{16}N}$~($\mathrm{^{15}C}$) by the analysis described in Sec.~\ref{sec:analysis} above, $\chi^{2}_{\mathrm{Time}}$ and $\chi^{2}_{\mathrm{Energy}}$ are the same ones defined in Eq.~(\ref{eq:chi2-each}), $\sigma_{\mathrm{^{16}N}}^{\mathrm{Best}}$~($\sigma_{\mathrm{^{15}C}}^{\mathrm{Best}}$) is the total uncertainty of production rate as summarized in Table~\ref{tb:event-per-day}, $N_{\mathrm{^{16}N}}^{\mathrm{Best}}$~($N_{\mathrm{^{15}C}}^{\mathrm{Best}}$) is the estimated number of decay events in the analysis sample after the selection cuts for $\mathrm{^{16}N}$~($\mathrm{^{15}C}$), and a parameter $\alpha$~($\beta$) is a scaling factor for the fluctuation of the number of $N_{\mathrm{^{16}N}}^{\mathrm{Best}}$~($N_{\mathrm{^{15}C}}^{\mathrm{Best}}$) decay events in the signal region.

\subsection{Measurement of production rates of $\mathrm{^{12}B}$ and $\mathrm{^{13}B}$}

\begin{figure}[h]
    \begin{center}
    \includegraphics[width=1.0\linewidth]{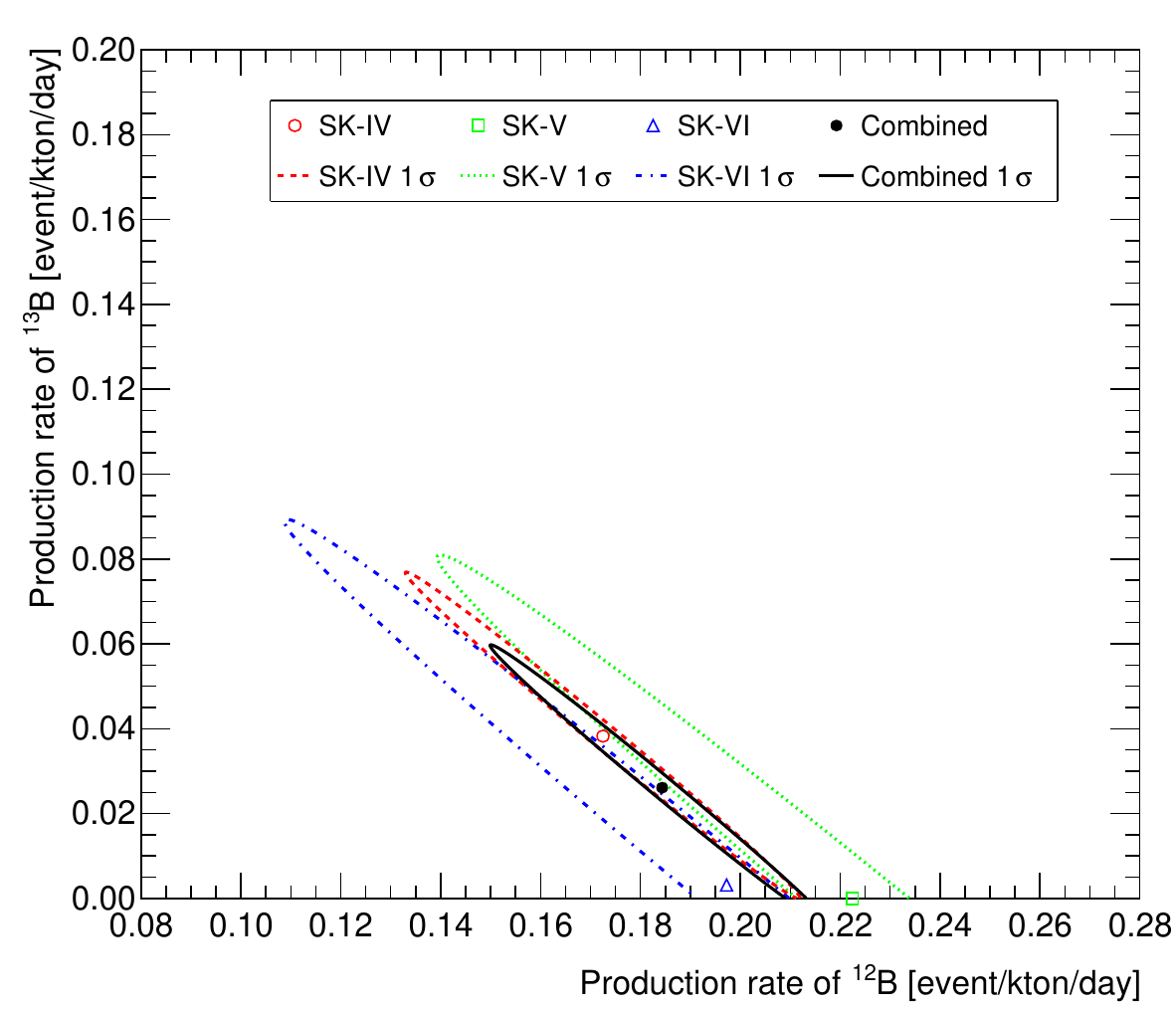}
    \end{center}
    \caption{The allowed regions of the production rates of $\mathrm{^{12}B}$ and $\mathrm{^{13}B}$ using SK-IV, SK-V, and SK-VI data sets calculated by Eq.~(\ref{eq:chi2-b12-b13}). The red open square~(dashed line), green open square~(dotted line), and blue open upward-triangle~(dashed-dotted line) show the best-fit values~(their $1\sigma$ allowed regions determined by $\Delta \chi^{2}~(R_{\mathrm{^{12}B}}, R_{\mathrm{^{13}B}})=2.30$) of SK-IV, SK-V, and SK-VI, respectively. The black filled circle~(solid line) shows the combined value~($1\sigma$ allowed region). \label{fig:chi2-map_b12b13}}
\end{figure}

To determine the production rates of naturally produced $\mathrm{^{12}B}$ and $\mathrm{^{13}B}$ through the nuclear muon capture on oxygen in water, we calculated the $\chi^{2}$ defined in Eq.~(\ref{eq:chi2-b12-b13}) and followed procedure outlined in Sec.~\ref{sec:production}. Figure~\ref{fig:chi2-map_b12b13} shows the result of the $\Delta \chi^{2}$ calculation using SK-IV, SK-V, and SK-VI data sets.

Table~\ref{tb:event-per-day-b12b13} summarizes the measured production rates of $\mathrm{^{12}B}$ and $\mathrm{^{13}B}$ isotopes among the SK phases. The values obtained when combining the SK phases are $R_{\mathrm{^{12}B}}= 0.18\pm 0.02$~event/kton/day, and $R_{\mathrm{^{13}B}}= 0.03\pm 0.02$~event/kton/day, where uncertainty is both statistical and systematic. 

\begin{table}[h]
    \begin{center}
    \caption{Summary of the production rates of $\mathrm{^{12}B}$ and $\mathrm{^{13}B}$ measured by the SK detector. The uncertainty shown in this table includes both statistical and systematic uncertainties. For the production rate of $\mathrm{^{13}B}$, we list both the 1$\sigma$ and 90\% C.L. values when the 1$\sigma$ uncertainty is lower than the mean value, and only the 90\% C.L. value otherwise.}
        \label{tb:event-per-day-b12b13}
            \begin{tabular}{lccc}
                \hline \hline
                SK phase &  \multicolumn{3}{c}{Production rate} \\
                & \multicolumn{3}{c}{[event/kton/day]} \\ 
                & $\mathrm{^{12}B}$  & $\mathrm{^{13}B}~(1\sigma)$ & $\mathrm{^{13}B}~(90\%~\mathrm{C.L.})$ \\ \hline   
                SK-IV & $0.17\pm 0.03$ & $0.04 \pm 0.03$ & $<0.09$\\ 
                SK-V & $0.22^{+0.01}_{-0.05}$ & -- & $<0.12$  \\ 
                SK-VI & $0.20^{+0.01}_{-0.06}$ &-- & $<0.13$ \\ \hline
                Combined & $0.18\pm0.02$& $0.03\pm0.02$ & $<0.07$  \\
                \hline \hline
        \end{tabular}
    \end{center}
\end{table}

\begin{figure}[h]
       \includegraphics[width=1.0\linewidth]{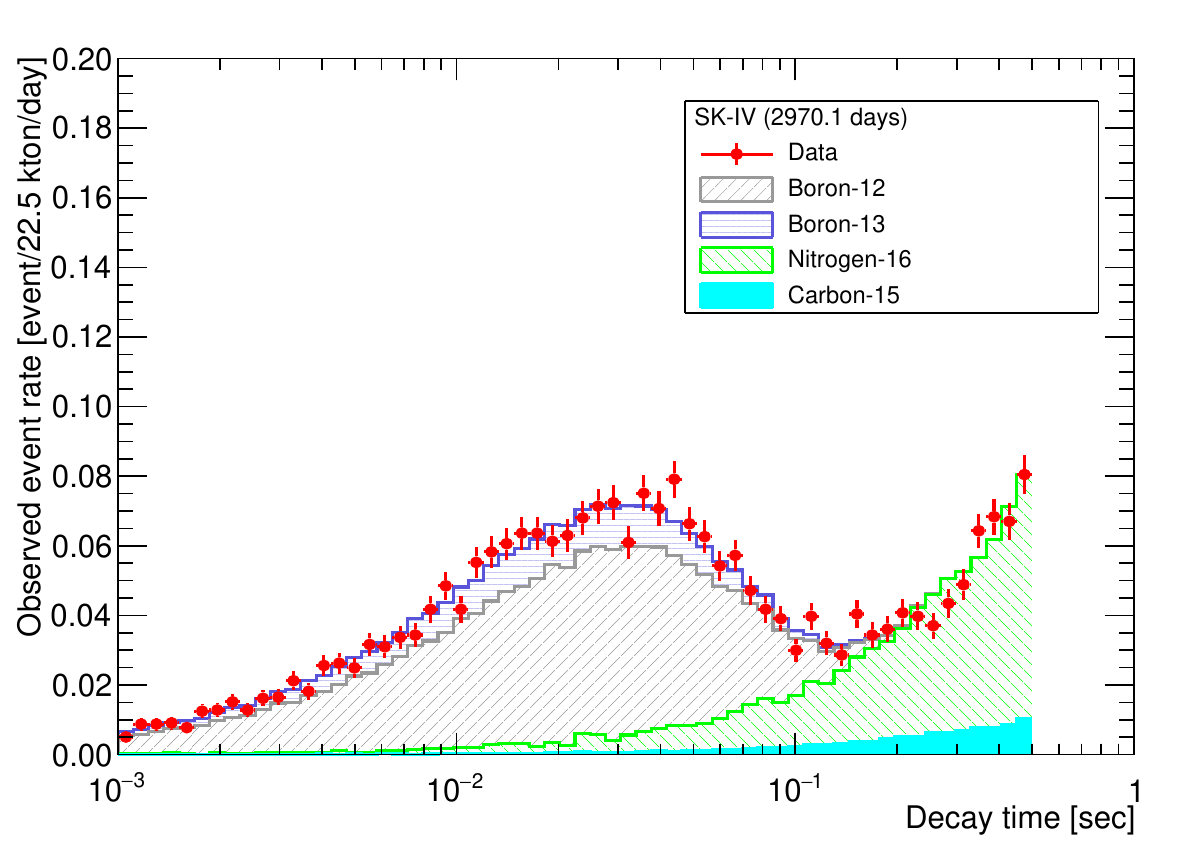} \\
       \includegraphics[width=1.0\linewidth]{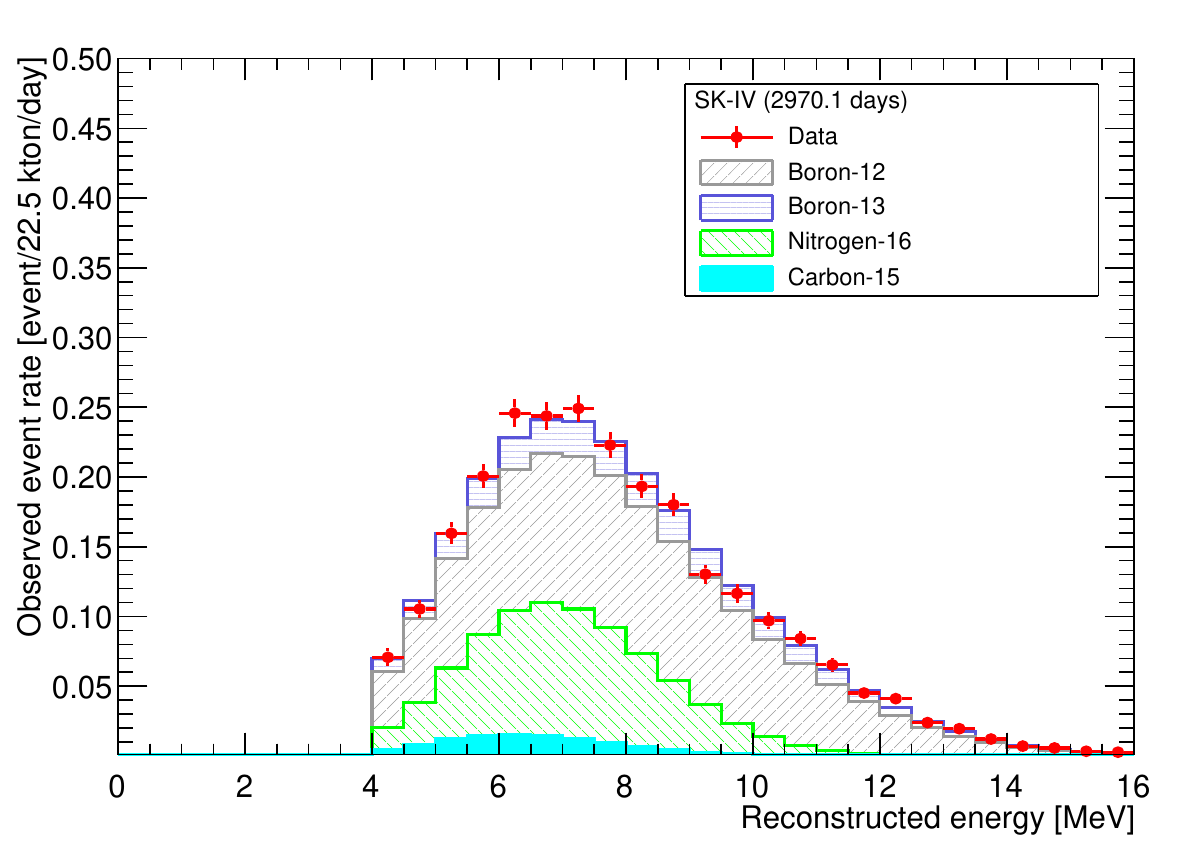}
       \caption{Distributions of the time difference~(top) in the timing window of $[0.001, 0.5]$~s and the reconstructed energy~(bottom) together with the best-fit of the components from $\mathrm{^{12}B}$,$\mathrm{^{13}B}$,$\mathrm{^{16}N}$, and $\mathrm{^{15}C}$ using the SK-IV data. Red filled circle shows the observed data after the selection cuts. Light-gray left slanting histogram, light-purple horizontal lines histogram, light-green right slanting histogram, and light-blue filled histogram show the best-fit of $\mathrm{^{12}B}$,$\mathrm{^{13}B}$,$\mathrm{^{16}N}$, and $\mathrm{^{15}C}$ components, respectively. The distributions for other SK phases are shown in Appendix~\ref{sec:app-other-dist}.}
    \label{fig:best-sk4-b12b13}
\end{figure}

Figure~\ref{fig:best-sk4-b12b13} shows examples of decay time and energy distributions using the SK-IV data together with their best-fit distributions. The distributions using other SK phases are shown in Appendix~\ref{sec:app-other-dist}. Based on the analysis of the SK-IV data sample, we statistically extracted the excess of $\mathrm{^{13}B}$ events over the distributions of $\mathrm{^{16}N}$, $\mathrm{^{15}C}$, and $\mathrm{^{12}B}$, as shown in Fig.~\ref{fig:best-sk4-b12b13}. However, SK phases other than SK-IV do not have enough statistics to extract $\mathrm{^{13}B}$ events due to their short operation period and the low production rate of $\mathrm{^{13}B}$. 

Due to the low production rate of $\mathrm{^{13}B}$, the sensitivity of the measurement is still limited, where the excess of $\mathrm{^{13}B}$ at the $1\sigma$ level as summarized in Table~\ref{tb:event-per-day-b12b13}. Hence, further data accumulation is required to precisely measure the production rate of $\mathrm{^{13}B}$ by improving energy and timing resolutions.

\section{Measurement of branching ratio} \label{sec:branch}

\subsection{Measurement of branching ratios}

By considering the charge ratio of stopping muons, the number of nuclear muon captures can be estimated. The branching ratio of unstable isotopes can be determined by dividing the production ratio of isotopes by the total number of stopping negative muons with the correction of selection efficiencies. The branching ratio~$Br(i)$ of $i$-th isotope can be expressed as,

\begin{equation}
   Br(i) = \frac{R(i)\times V}{N^{\mu}_{\mathrm{stop}}\left(\frac{1}{1+R_{\mu}}\right)P_{\mathrm{cap}}}, \label{eq:branch}
\end{equation}

\noindent where $R(i)$ is the production rate of $i$-th isotope, $V$ is the fiducial volume of the analysis sample~($22.5$~kton), $N^{\mu}_{\mathrm{stop}}$ is the total number of stopping muon inside the fiducial volume, $R_{\mu}$ is the charge ratio of stopping muon listed in Table~\ref{tb:charge_ratio}, and $P_{\mathrm{cap}}$ is the fraction of nuclear capture by oxygen~\cite{Suzuki:1987jf}.

Based on the measured production rates described in Sect.~\ref{sec:analysis}, we can calculate the branching ratio from Eq.~(\ref{eq:branch}). Table~\ref{tb:branch_skresult} summarizes the branching ratio measured by the SK detector.  By combining the SK measurements, we obtained $Br(\mathrm{^{16}N})= (9.0 \pm 0.1\,(\mathrm{stat.+syst.}))\%$, $Br(\mathrm{^{15}C})=(0.6\pm 0.1\,(\mathrm{stat.+syst.}))\%$, $Br(\mathrm{^{12}B})=(0.98 \pm 0.18\,(\mathrm{stat.+syst.}))\%$, and $Br(\mathrm{^{13}B})=(0.14 \pm 0.12\,(\mathrm{stat.+syst.}))\%$, respectively, where the range of their uncertainty is $1\sigma$.

\begin{table*}[]
    \begin{center}
    \caption{Summary of branching ratio of negative muon capture on oxygen nuclei measured by the SK detector. Owing to the low production rate of $\mathrm{^{13}B}$, the upper limits of the branching ratio are set for SK-IV, SK-V, and SK-VI.}
        \label{tb:branch_skresult}
            \begin{tabular}{p{0.10\textwidth}>{\centering}p{0.10\textwidth}>{\centering}p{0.10\textwidth}>{\centering}p{0.10\textwidth}>{\centering}p{0.10\textwidth}>{\centering\arraybackslash}p{0.12\textwidth}}
                \hline \hline
                SK phase &  \multicolumn{5}{c}{Branching ratio [\%]} \\            
                & $\mathrm{^{16}N}$  & $\mathrm{^{15}C}$ & $\mathrm{^{12}B}$ & $\mathrm{^{13}B}~(1\sigma)$ & $\mathrm{^{13}B}~(90\%~\mathrm{C.L.})$ \\ \hline   
                SK-IV & $9.1 \pm 0.2$ & $0.6 \pm 0.1$ & $0.92\pm 0.14$ & $0.20 ^{+0.14}_{-0.13} $ & $<0.49$ \\ 
                SK-V & $8.6 \pm 0.4$ & $1.0\pm0.3$ &$1.1^{+0.1}_{-0.3}$ & -- & $<0.6$\\ 
                SK-VI & $9.0 \pm 0.3$ & $0.6 \pm 0.2$&$1.1^{+0.1}_{-0.3}$ & -- & $<0.7$  \\ \hline
                Combined & $9.0 \pm 0.1$ & $0.6\pm0.1$&$0.98\pm0.18$ & $0.14\pm 0.12$& $<0.39$ \\
                \hline \hline
        \end{tabular}
    \end{center}
\end{table*}

\subsection{Comparison with theoretical predictions and experimental measurements} \label{sec:comp}

In this section, we compare the measurement results described in the former sections with the measurements by other groups and the predictions by simulations.

\begin{figure}[h]
    \begin{center}
    \includegraphics[width=1.0\linewidth]{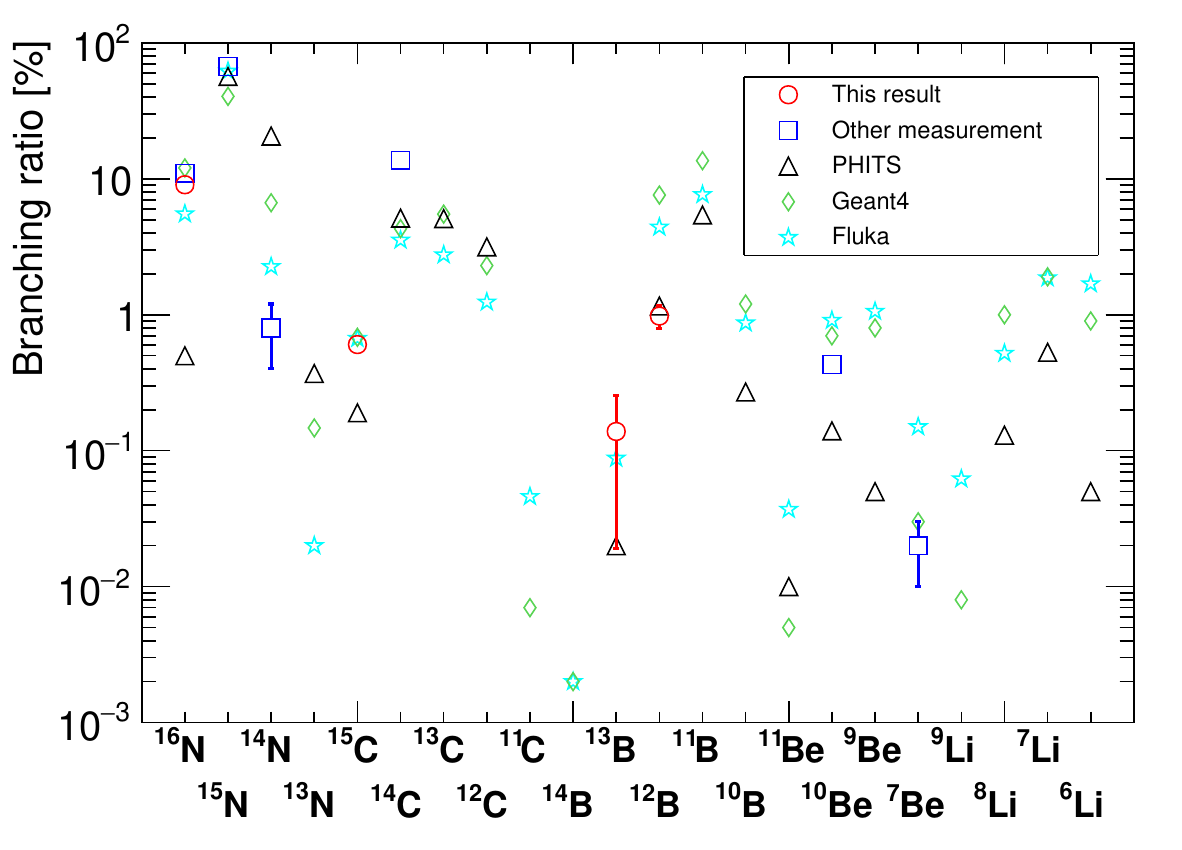}
    \end{center}
    \caption{Comparison of measured branching ratio through nuclear muon capture in water. Red circles show the result measured by the SK detector, blue squares show the other measurement results, and other symbols show simulated values from \texttt{PHITS}, \texttt{Geant4}, and \texttt{Fluka}, as summarized in Table~\ref{tb:branch}. \label{fig:comp_branching_ratio} }
\end{figure}

As mentioned in Sec.~\ref{sec:intro}, the branching ratio is not fully understood because only a limited number of unstable isotopes have been measured. Figure~\ref{fig:comp_branching_ratio} shows the branching ratios measured by the SK detector together with past measurements by other groups~\cite{2002heisinger, Measday:2001yr} and three simulations.

For the $\mathrm{^{16}N}$ isotope, the branching ratio measured by the SK detector is consistent with that measured by Ref.~\cite{Measday:2001yr, Suzuki:1987jf, 1360580234906296832, Guichon:1979ga} within at most $1.9\sigma$. Considering the total uncertainty, the branching ratio measured by the SK detector is three times more accurate than the previous measurement. This measurement of the branching ratio of $\mathrm{^{16}N}$ is currently the most accurate in the world. 

For $\mathrm{^{15}C}$, $\mathrm{^{12}B}$, and $\mathrm{^{13}B}$ isotopes, these are the first branching ratio measurements and provide new inputs for updating simulations of the nuclear muon capture process. 

\begin{figure}[h]
    \begin{center}
    \includegraphics[width=1.0\linewidth]{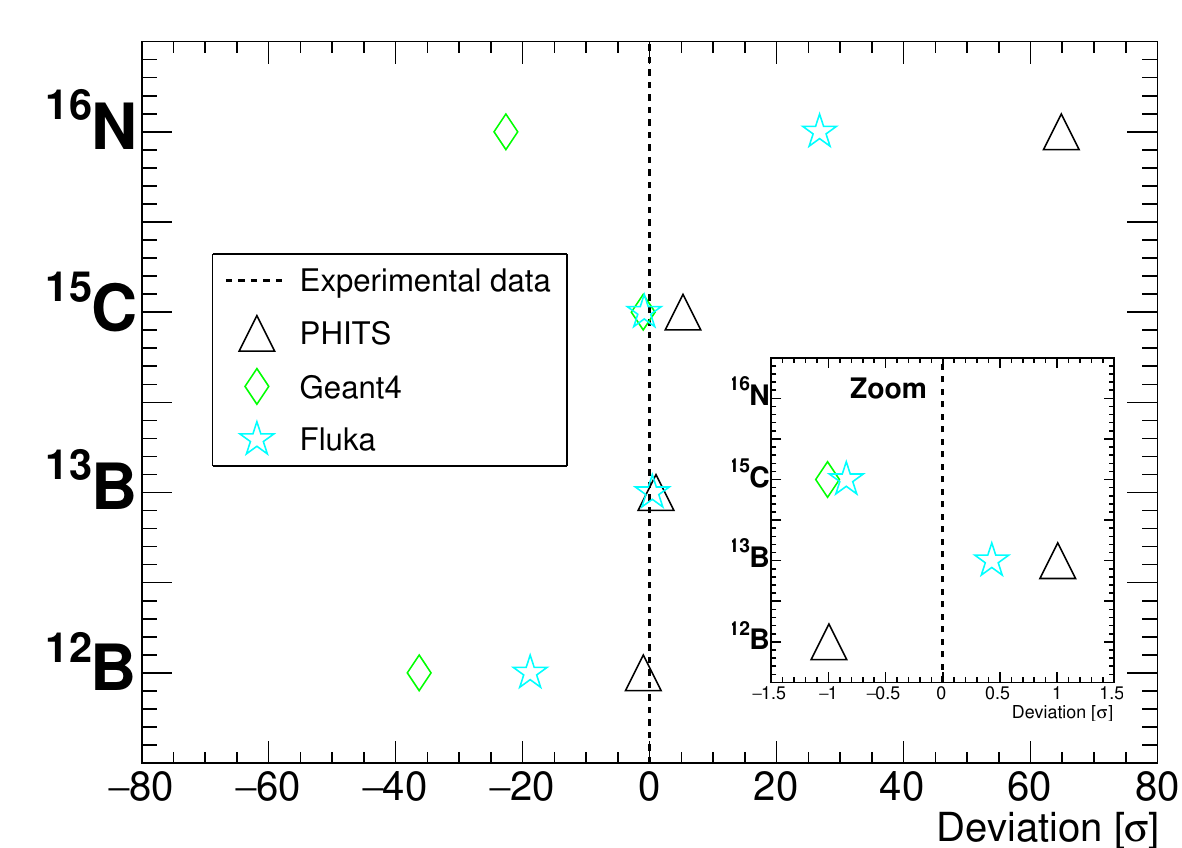}
    \end{center}
    \caption{Deviations between the experimental data and the predicted branching ratios of $\mathrm{^{16}N}$, $\mathrm{^{15}C}$, $\mathrm{^{13}B}$ and $\mathrm{^{12}B}$ from three MC simulations, such as \texttt{PHITS}, \texttt{Geant4}, and \texttt{Fluka}. The definition of symbols and color is the same as Fig.~\ref{fig:comp_branching_ratio}. The inset panel shows a zoomed-in figure around zero deviation. Note that \texttt{Geant4} does not produce $\mathrm{^{13}B}$ as listed in Table~\ref{tb:branch}. Hence, no point is plotted. \label{fig:comp_simulation} }
\end{figure}

Figure~\ref{fig:comp_simulation} shows the deviation of the predicted branching ratio from the SK measurement result. Because of high precision measurements, the SK data can test the predicted values extracted from the MC simulation.

In the case of $\mathrm{^{16}N}$ measurement, the SK measurement data completely exclude the predicted values from three simulations, \texttt{PHITS}, \texttt{Geant4}, and \texttt{Fluka}. The $\mathrm{^{15}C}$ measurement is consistent with \texttt{Geant4} and \texttt{Fluka} simulations within $1\sigma$ while excluding the \texttt{PHITS} prediction by $5.3\sigma$. Due to the current accuracy of the measurement, the branching ratio of $\mathrm{^{13}B}$ is consistent with \texttt{PHITS} and \texttt{Fluka} within $1\sigma$. Finally, the $\mathrm{^{12}B}$ measurement is consistent with the \texttt{PHITS} simulation within $1\sigma$ while completely excludes the \texttt{Geant4} and \texttt{Fluka} predictions.

We suggest that further improvement of simulations for muonic capture processes is required for future studies based on those measurement results.

\subsection{Comparison with the neutron multiplicity measurement in SK-VI}

In an independent analysis, we measured the neutron multiplicity in the nuclear muon capture reaction using the SK-VI data set ~\cite{Super-Kamiokande:2025vvn}. This measurement utilizes $\gamma$-rays emitted from $\mathrm{Gd}(n, \gamma)$ reactions to detect free neutrons. In the stopping muon sample, we counted the number of detected neutrons per event, and the observed distribution is corrected with the neutron detection efficiency and the false-positive rate by a $\chi^{2}$~method. The corrected distribution is scaled by the expected number of nuclear muon capture events in the sample to obtain branching ratios for each neutron multiplicity. As a result, the probability $P(i)$ of $i$~neutrons emitted in the nuclear muon capture process on oxygen is obtained to be $P(0)=(24\pm3)\%$, $P(1)=(70^{+3}_{-2})\%$, $P(2)=(6.1\pm0.5)\%$, and $P(3)=(0.38\pm0.09)\%$, respectively. 

When the isotopes measured in the main text are produced in the nuclear muon capture, no neutron is expected to be emitted for $\mathrm{^{16}N}$, $\mathrm{^{15}C}$, and $\mathrm{^{12}B}$ isotopes, while one neutron is expected for the $\mathrm{^{13}B}$ isotope, as summarized in Table~\ref{tb:nucleus}. The total branching ratio for $\mathrm{^{16}N}$, $\mathrm{^{15}C}$, and $\mathrm{^{12}B}$ in this measurement is $10.6\%$ and is around half of $P(0)$ from the neutron measurement. That indicates other branches contribute to $P(0)$ with a large fraction.

The measurement of isotope production, together with the neutron multiplicity measurement, can provide inputs for theoretical studies of nuclear structure on light nuclei.

\section{Summary and prospects}\label{sec:summary}

The measurement of production rates of isotopes produced by a nuclear muon capture can validate models of background due to muons in underground detectors and provide input for the simulation of nuclear structure. We selected stopping muons in the SK detector and successfully measured the production rates of $\mathrm{^{16}N}$, $\mathrm{^{15}C}$, $\mathrm{^{12}B}$, and $\mathrm{^{13}B}$ as $R_{\mathrm{^{16}N}}=1.71 \pm 0.01\,(\mathrm{stat.+syst.})$~event/kton/day, $R_{\mathrm{^{15}C}} = 0.11 \pm 0.01\,(\mathrm{stat.+syst.})$~event/kton/day, $R_{\mathrm{^{12}B}} = 0.18 \pm 0.02\,(\mathrm{stat.+syst.})$~event/kton/day, and $R_{\mathrm{^{13}B}} = 0.03 \pm 0.02\,(\mathrm{stat.+syst.})$, respectively. Those measurements are useful in the SK experiment for modeling the realistic background due to muons overlapped with solar neutrinos, diffuse supernova neutrino background, and reactor neutrinos.

Based on the production rates of the unstable isotopes, we measured the branching ratio as $Br(\mathrm{^{16}N})=(9.0\pm 0.1\,(\mathrm{stat.+syst.}))\%$, $Br(\mathrm{^{15}C})=(0.6\pm 0.1\,(\mathrm{stat.+syst.}))\%$, $Br(\mathrm{^{12}B})=(0.98\pm 0.18\,(\mathrm{stat.+syst.}))\%$, and $Br(\mathrm{^{13}B})=(0.14\pm 0.12\,(\mathrm{stat.+syst.}))\%$, respectively. The result for $\mathrm{^{16}N}$ is the most precise measurement at present and consistent with past results measured by other groups within $1.9\sigma$. The results for $\mathrm{^{15}C}$,  $\mathrm{^{12}B}$, and  $\mathrm{^{13}B}$ are the first measurements of branching ratios of those isotopes.

Based on those measurement results, we suggest that further improvement of simulations for muonic capture processes is required for future studies. The measurement results are also helpful to test theoretical models, which calculate muon capture rates with the momentum transfer near the muon mass, and to further study low-energy neutrino-nucleus reactions~\cite{OConnell:1972edu}.

We also suggest that the modeling of background due to muons in underground experiments be tuned and updated according to the measured production rates of unstable isotopes resulting from nuclear muon capture. Such updates are helpful to improve the sensitivities for astrophysical neutrinos, such as solar neutrinos, diffuse supernova neutrino background, etc.

\section*{Acknowledgment}

We would like to thank M.~Niikura from RIKEN Nishina Center for Accelerator-Based Science for suggestions about the decay modes of boron associated with nuclear muon captures on oxygen. We also thank S.~Abe from Nuclear Science Research Institute, Japan Atomic Energy Agency for providing their calculation results of the branching ratio by using \texttt{PHITS}. We also thank J.F.~Beacom and O.~Nairat from Ohio State University for providing their calculation results of the branching ratio by using \texttt{Fluka}, and their helpful comments and suggestions relating to the production rate of $\mathrm{^{16}N}$ described in Sec.~\ref{sec:n16-exp-theor}.

We gratefully acknowledge the cooperation of the Kamioka Mining and Smelting Company. The Super-Kamiokande experiment has been built and operated from funding by the Japanese Ministry of Education, Culture, Sports, Science and Technology; the U.S. Department of Energy; and the U.S. National Science Foundation. Some of us have been supported by funds from the National Research Foundation of Korea~(NRF-2009-0083526, NRF-2022R1A5A1030700, NRF-2022R1A3B1078756, RS-2025-00514948) funded by the Ministry of Science, Information and Communication Technology~(ICT); the Institute for Basic Science~(IBS-R016-Y2); and the Ministry of Education~(2018R1D1A1B07049158, 2021R1I1A1A01042256, RS-2024-00442775); the Japan Society for the Promotion of Science; the National Natural Science Foundation of China under Grants No.~12375100; the Spanish Ministry of Science, Universities and Innovation~(grant PID2021-124050NB-C31); the Natural Sciences and Engineering Research Council~(NSERC) of Canada; the Scinet and Digital Research of Alliance Canada; the National Science Centre~(UMO-2018/30/E/ST2/00441 and UMO-2022/46/E/ST2/00336) and the Ministry of  Science and Higher Education~(2023/WK/04), Poland; the Science and Technology Facilities Council~(STFC) and Grid for Particle Physics~(GridPP), UK; the European Union’s Horizon 2020 Research and Innovation Programme H2020-MSCA-RISE-2018 JENNIFER2 grant agreement no.~822070, H2020-MSCA-RISE-2019 SK2HK grant agreement no.~872549; and European Union's Next Generation EU/PRTR  grant CA3/RSUE2021-00559; the National Institute for Nuclear Physics~(INFN), Italy. This work is partially supported by MEXT KAKENHI Grant Numbers 24K00654 and 24H02243.

\appendix

\section{Selection efficiencies among the SK phases} \label{app:eff-sys}

In the main text, the cut criteria and the selection efficiencies for isotope decay events are described using the SK-IV data because of that phase's long and stable operation. In this appendix, we summarize the cut criteria and selection efficiencies for each SK phase.

Based on the significance value defined in Eq.~(\ref{eq:sig}), we determined the selection criteria for both the distance cut and the time window. Among SK phases, the selection criteria are the same, which are explained in Sec.~\ref{sec:opt}. We also optimized the selection cuts in the same manner. Table~\ref{tb:eff-all} shows the selection efficiencies among the three SK phases for the search window of $[0.5,25.0]$~s. The selection efficiencies in both SK-V and SK-VI are lower than that in SK-IV because of the worse ID PMT response under single electron yield. In particular, the timing resolution of the PMT was changed in SK-V onwards due to the change of the high-voltage setting of the ID PMTs.

\begin{table}[!h]
    \begin{center}
    \caption{Summary of the selection efficiency for the timing window of $[0.5, 25.0]$~s among the SK phases.  As explained in Sec.~\ref{sec:analysis}, the decay events of $\mathrm{^{12}B}$ and $\mathrm{^{13}B}$ are completely rejected by the timing cut because of their relatively short lifetimes.}
        \label{tb:eff-all}
            \begin{tabular}{lcc}
                \hline \hline
                SK phase & \multicolumn{2}{c}{Selection efficiency~[$\%$]} \\ 
                & $\mathrm{^{16}N}$ & $\mathrm{^{15}C}$ \\ \hline
                SK-IV & $52.34\pm0.01$ & $46.56\pm0.01$ \\
                SK-V & $45.43\pm0.05$ & $38.57\pm0.05$ \\
                SK-VI & $44.76\pm0.04$ & $37.81\pm0.04$ \\
                \hline \hline
        \end{tabular}
    \end{center}
\end{table}

Table~\ref{tb:eff-all-b12} shows the selection efficiencies among the three SK phases for the search window of $[0.001,0.5]$~s. The differences in efficiency between periods can be explained by the same reason mentioned above.

\begin{table*}[]
    \begin{center}
    \caption{Summary of the total selection efficiency for the timing window of $[0.001, 0.5]$~s among the three SK phases. The details of selection cuts are summarized in Table~\ref{tb:eff-cut}. As explained in Sec.~\ref{sec:b12}, some of $\mathrm{^{16}N}$ and $\mathrm{^{15}C}$ decays enter into this timing window due to their relatively long lifetimes.}
        \label{tb:eff-all-b12}
            \begin{tabular}{p{0.12\textwidth}>{\centering}p{0.10\textwidth}>{\centering}p{0.10\textwidth}>{\centering}p{0.10\textwidth}>{\centering\arraybackslash}p{0.10\textwidth}}
                \hline \hline
                SK phase & \multicolumn{4}{c}{Selection efficiency~[$\%$]} \\ 
                & $\mathrm{^{16}N}$  & $\mathrm{^{15}C}$ & $\mathrm{^{12}B}$ & $\mathrm{^{13}B}$ \\ \hline
                SK-IV & $2.88\pm0.01$ & $7.14\pm0.01$ & $53.78\pm0.01$ & $54.19\pm0.01$\\
                SK-V & $2.47\pm0.01$ & $5.90\pm0.02$ & $47.95\pm0.06$ & $48.23\pm0.06$\\
                SK-VI & $2.43\pm0.01$ & $5.81\pm0.01$ & $47.39\pm0.05$ & $47.74\pm0.05$\\
                \hline \hline
        \end{tabular}
    \end{center}
\end{table*}

\section{Systematic uncertainties for the event selection cuts}
 
In Sect.~\ref{sec:sys} we show the list of systematic uncertainties caused by the selection cuts in SK-IV data set. Here, we list the estimated systematic uncertainties in other SK phases. Table~\ref{tb:eff-sys-phase} summarizes the total systematic uncertainties originating from the selection cuts of isotope decay events.

\begin{table}[!h]
    \begin{center}
    \caption{Summary of total systematic uncertainties originating from the selection cuts of the isotope decay events. The list of the selection cuts is described in Table~\ref{tb:sys-tab} and Table~\ref{tb:sys-tab-b12b13}.}
        \label{tb:eff-sys-phase}
            \begin{tabular}{lcccc}
                \hline \hline
                SK phase & \multicolumn{4}{c}{Total systematic uncertainties~$[\%]$} \\
                & $\mathrm{^{16}N}$  & $\mathrm{^{15}C}$ & $\mathrm{^{12}B}$ & $\mathrm{^{13}B}$ \\ \hline
                SK-IV & $\pm0.5$ & $\pm0.4$ & $\pm0.4$ & $\pm0.4$\\
                SK-V & $\pm1.4$ & $\pm1.9$ & $\pm1.0$ & $\pm1.0$\\
                SK-VI & $\pm1.5$ & $\pm1.8$ & $\pm1.1$ & $\pm1.1$\\
                \hline \hline
        \end{tabular}
    \end{center}
\end{table}

\section{$\bm{N_{\mathrm{trig}}}$ distributions in SK-V and SK-VI data sets}

\begin{figure}[h]
    \includegraphics[width=1.0\linewidth]{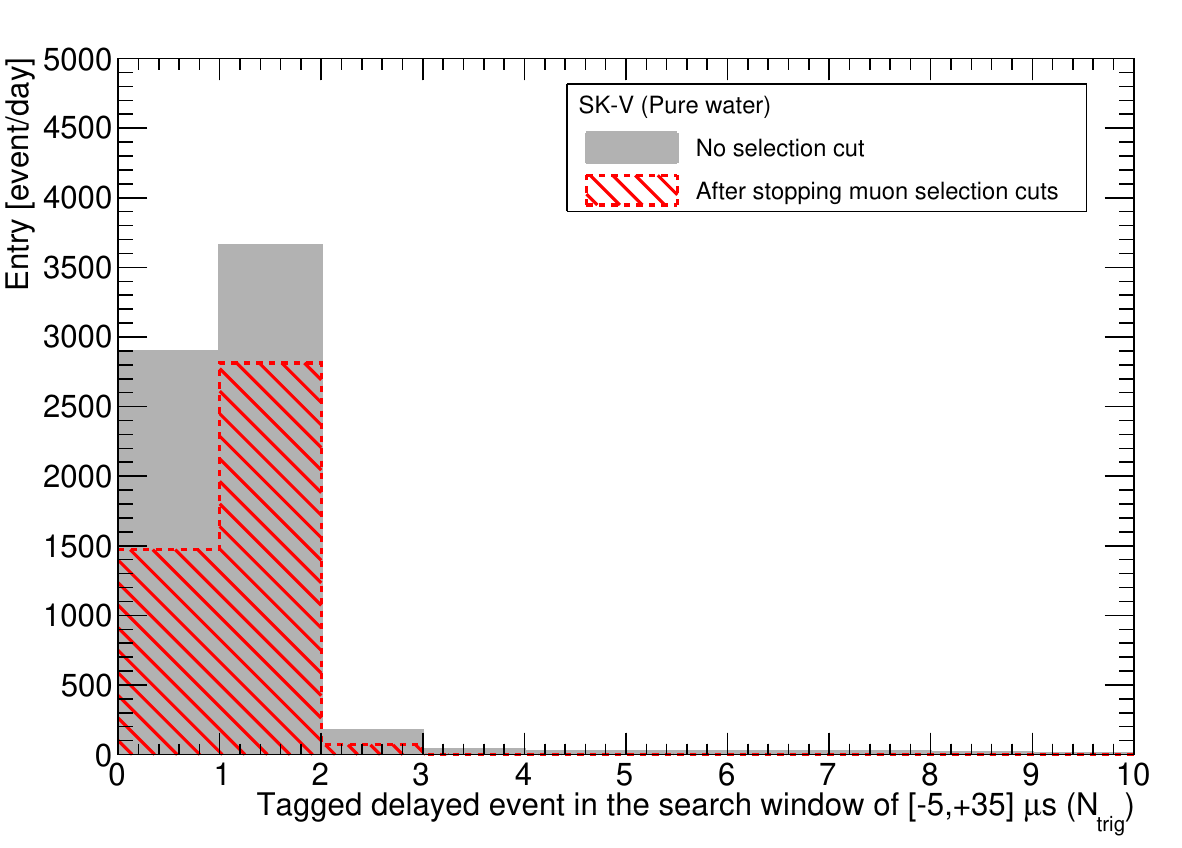} \\
    \includegraphics[width=1.0\linewidth]{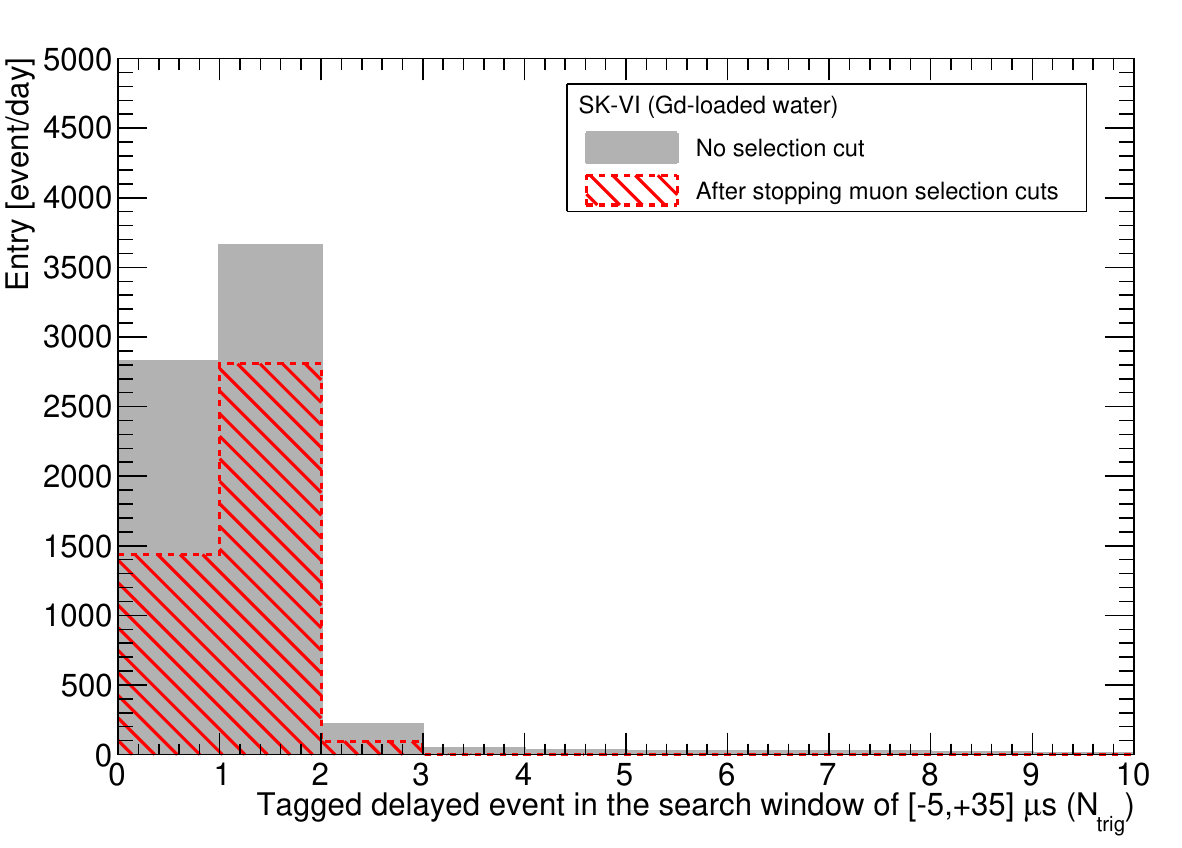}
    \caption{The distributions of delayed events~($N_{\mathrm{trig}}$) after stopping muons with and without the selection cuts using the SK-V~(top) and SK-VI~(bottom) data sets. The daily observed rates of stopping muons without the delayed signal~($N_{\mathrm{trig}}=0$) are consistent within their uncertainty.}
    \label{fig:sub-trig-sk56}
\end{figure}

In the main text, we show the distributions using SK-IV data set since the livetime is the longest among the three SK phases. Figure~\ref{fig:sub-trig-sk56} shows the $N_{\mathrm{trig}}$ distributions using the SK-V and SK-VI data sets. Comparing the distributions with those in the SK-IV phase, no clear change was observed. In particular, gadolinium loading from SK-VI onwards does not affect the distribution~\cite{Super-Kamiokande:2024kcb}. 

\section{Distributions of other SK data samples} \label{sec:app-other-dist} 

In this section, the energy and the decay time distributions measured by the SK phases are shown except for the SK-IV phase which was already shown in the main text. 

\begin{figure*}[]
        \begin{minipage}{0.49\textwidth}
            \centering
            \includegraphics[width=\textwidth]{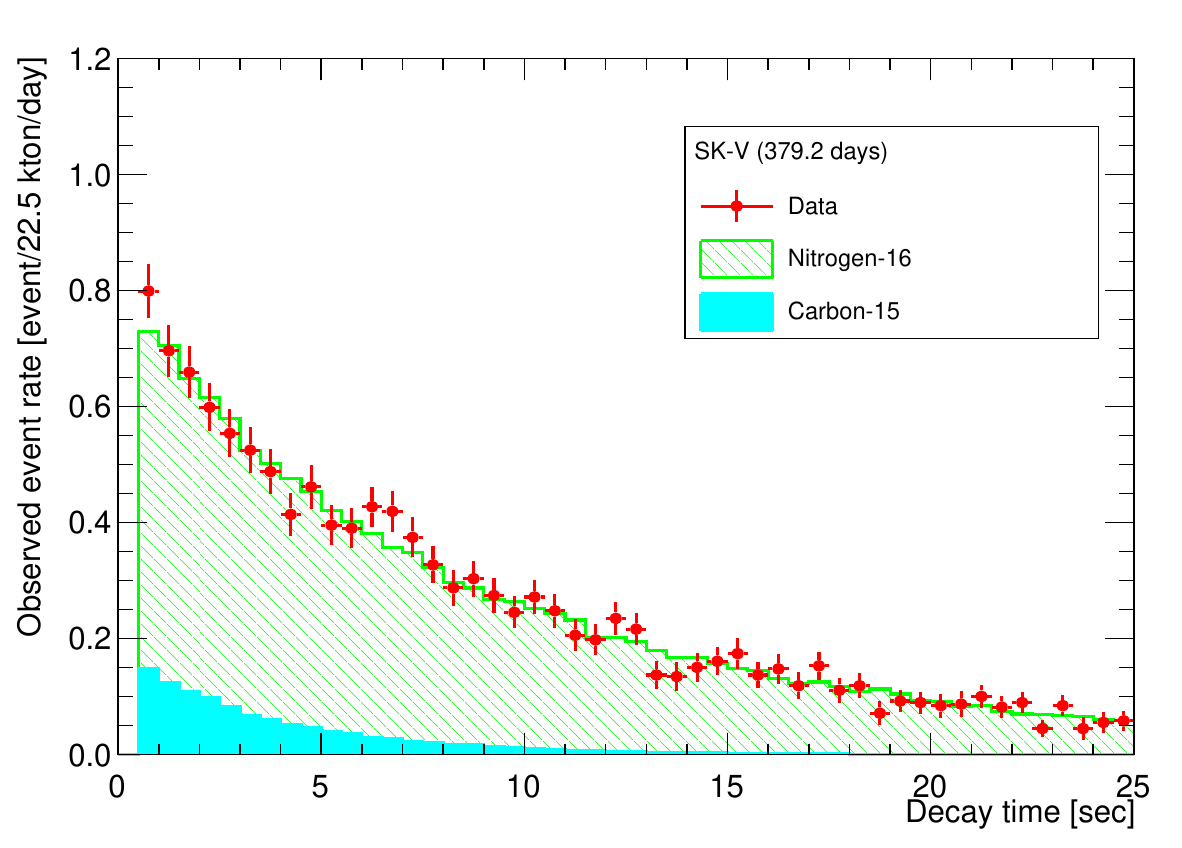}
        \end{minipage} 
        \begin{minipage}{0.49\textwidth}
            \centering           
            \includegraphics[width=\textwidth]{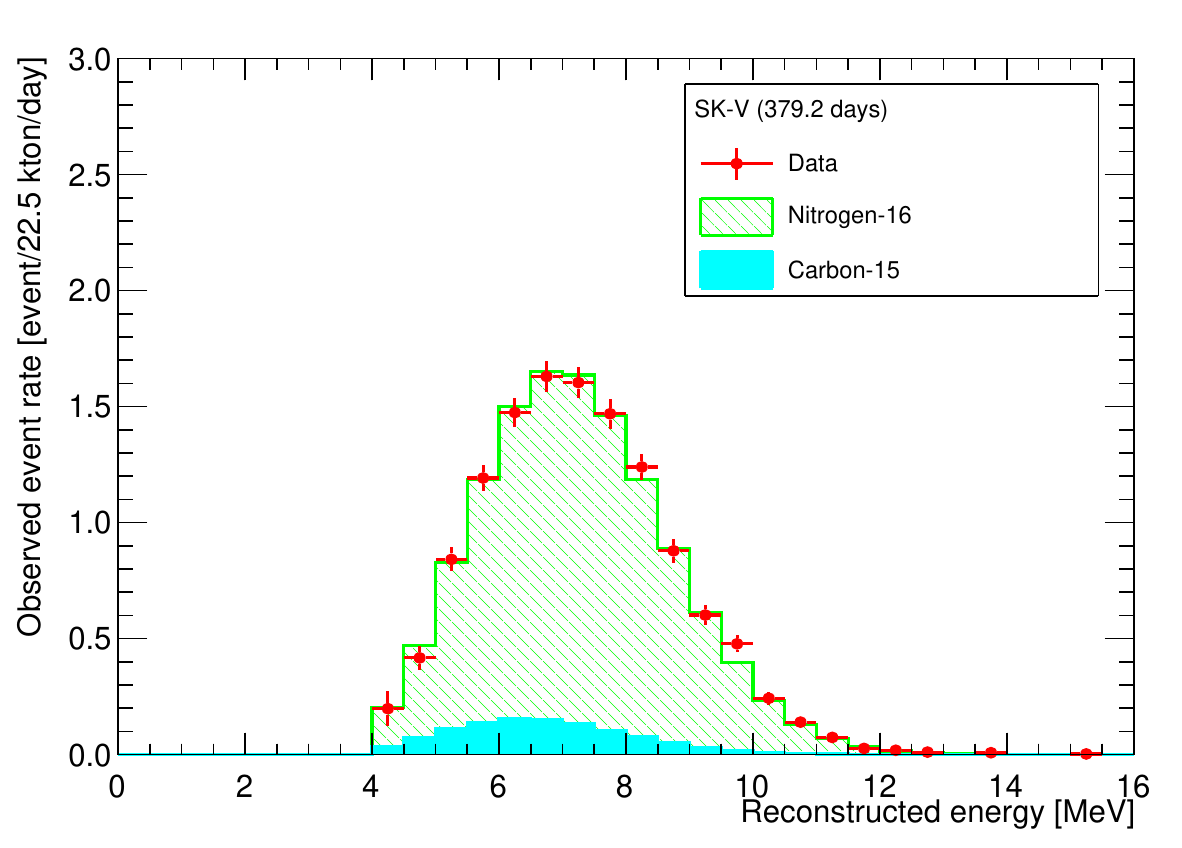}
	\end{minipage} 
    \caption{Distributions of the time difference~(left) in the timing window of $[0.5, 25.0]$~s and the reconstructed total energy~(right) together with the best-fit of the components from $\mathrm{^{16}N}$ and $\mathrm{^{15}C}$ using the SK-V data The definition of colors is the same as Fig.~\ref{fig:best-sk4}.}
    \label{fig:best-sk5}
\end{figure*}

Figure~\ref{fig:best-sk5} shows the distributions of the decay time and the energy in the time window of $[0.5,25.0]$~s using the SK-V sample.

\begin{figure*}[]
        \begin{minipage}{0.49\textwidth}
            \centering
            \includegraphics[width=\textwidth]{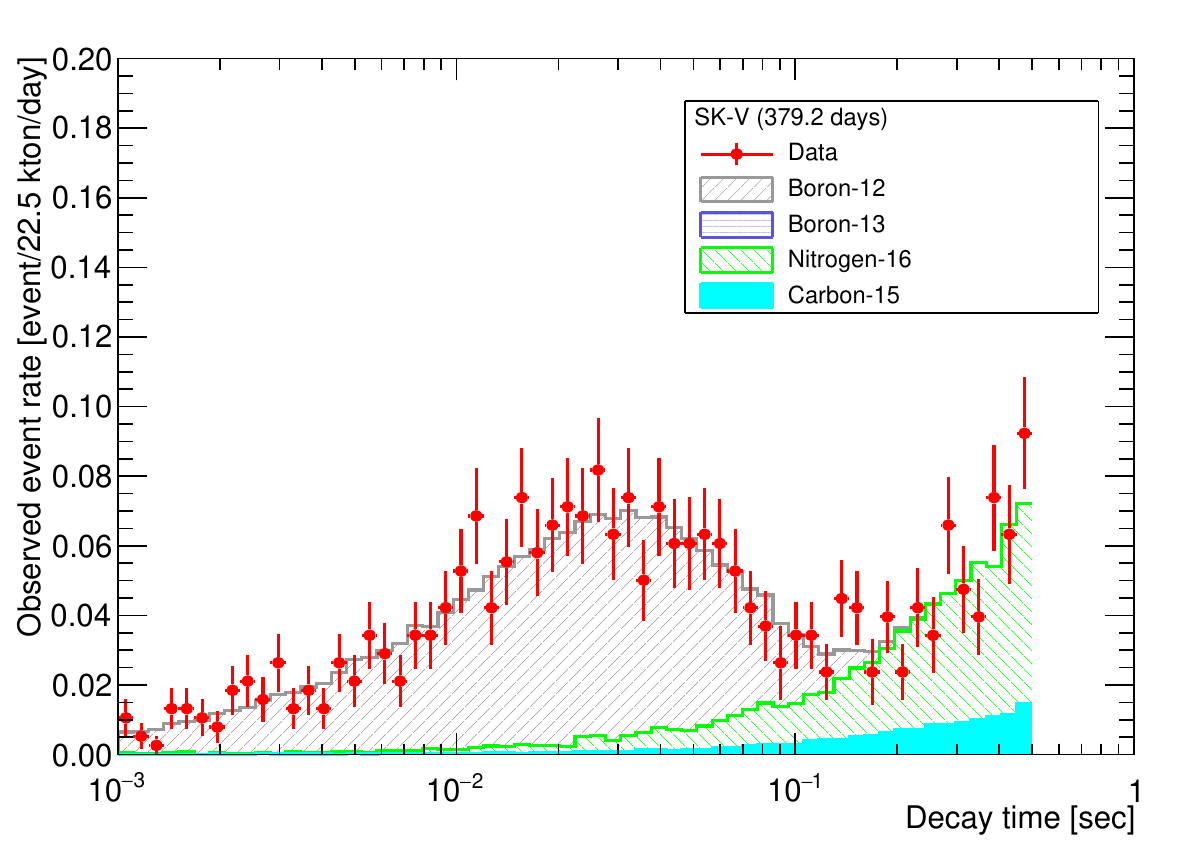}  
        \end{minipage} 
        \begin{minipage}{0.49\textwidth}
            \centering
            \includegraphics[width=\textwidth]{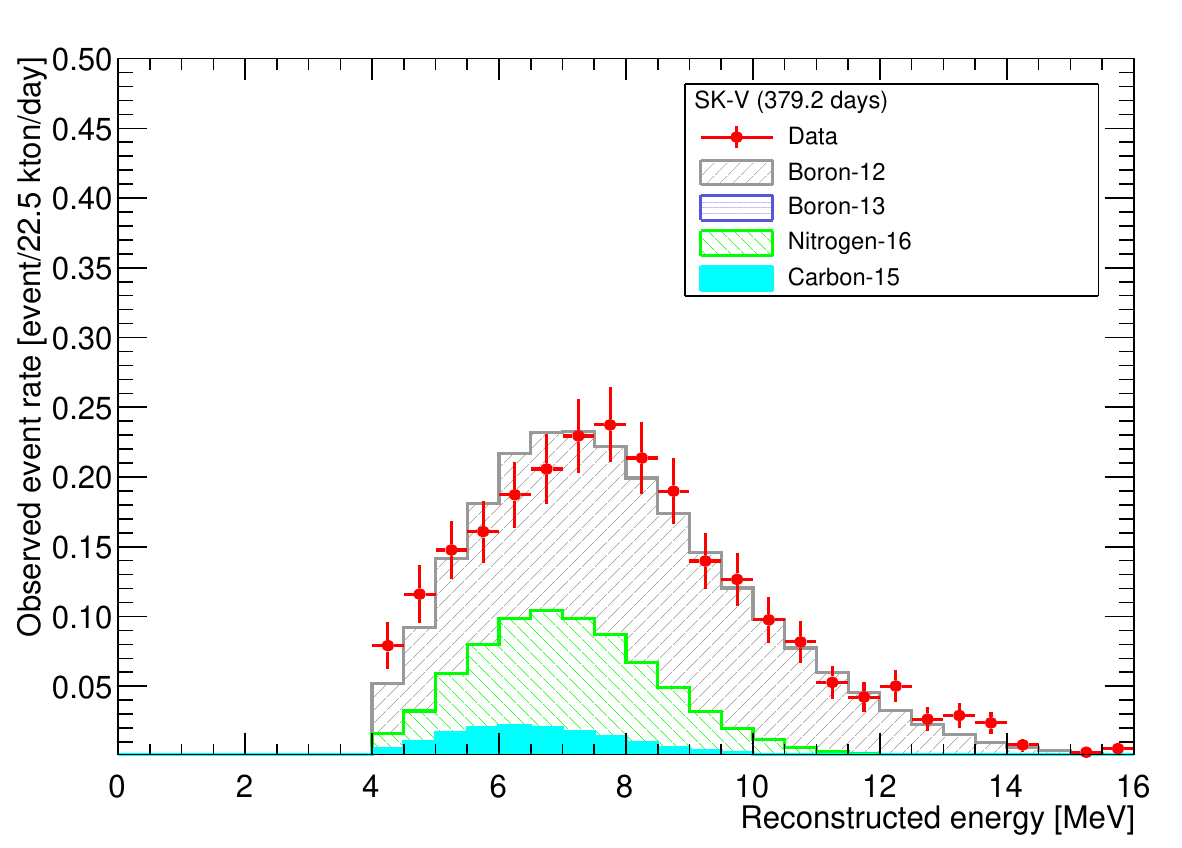}
	\end{minipage} 
    \caption{Distributions of the time difference~(left) in the timing window of $[0.001, 0.5]$~s and the reconstructed total energy~(right) together with the best-fit of the components from $\mathrm{^{12}B}$, $\mathrm{^{13}B}$, $\mathrm{^{16}N}$, and $\mathrm{^{15}C}$ using the SK-V data.  The definition of colors is the same as Fig.~\ref{fig:best-sk4-b12b13}. As shown in Fig.~\ref{fig:chi2-map_b12b13}, the best-fit result of the SK-V demonstrates no component of $\mathrm{^{13}B}$ decays in the analysis sample. As such, there is no contribution from  $\mathrm{^{13}B}$ decay in both panels.}
    \label{fig:best-sk5-b12b13}
\end{figure*}

Figure~\ref{fig:best-sk5-b12b13} shows the distributions of the decay time and the energy in the $[0.001,0.5]$~s time window using SK-V sample.

\begin{figure*}[]
        \begin{minipage}{0.49\textwidth}
            \centering
            \includegraphics[width=\textwidth]{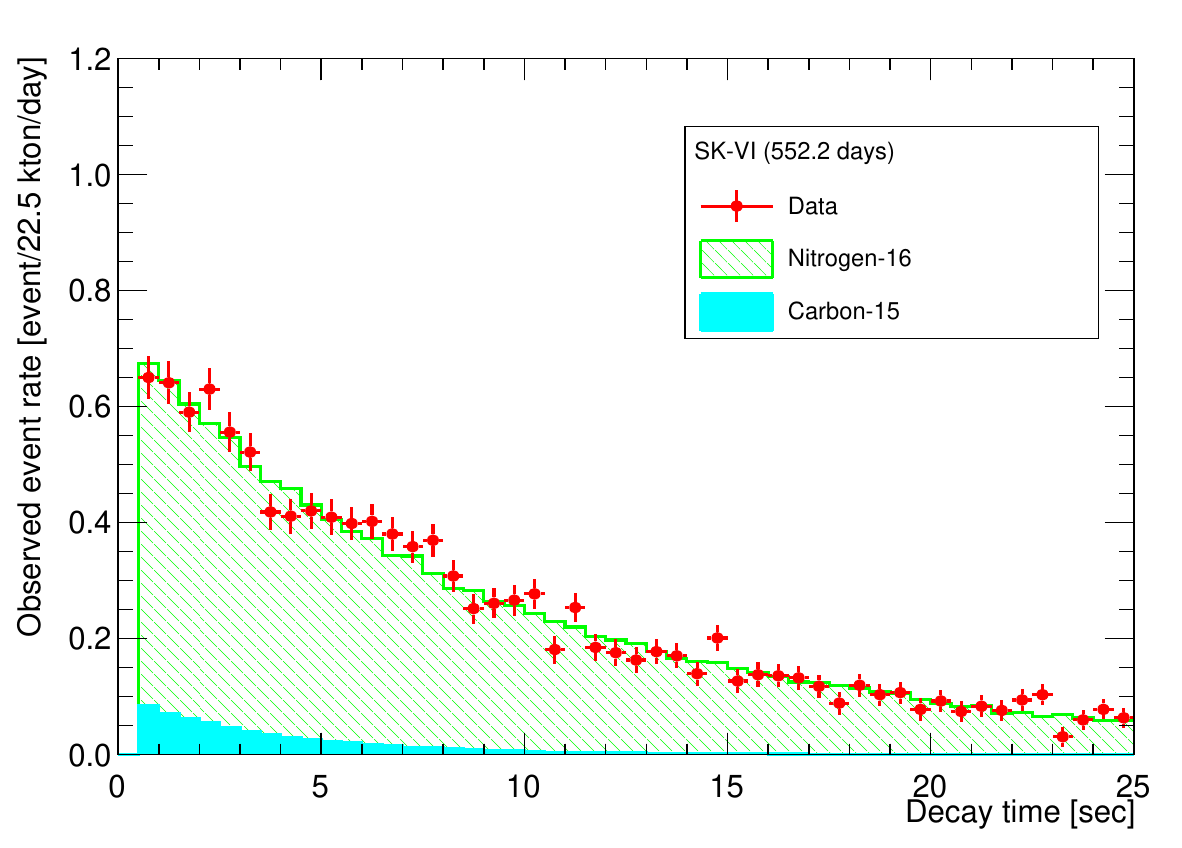}            
        \end{minipage} 
        \begin{minipage}{0.49\textwidth}
            \centering
            \includegraphics[width=\textwidth]{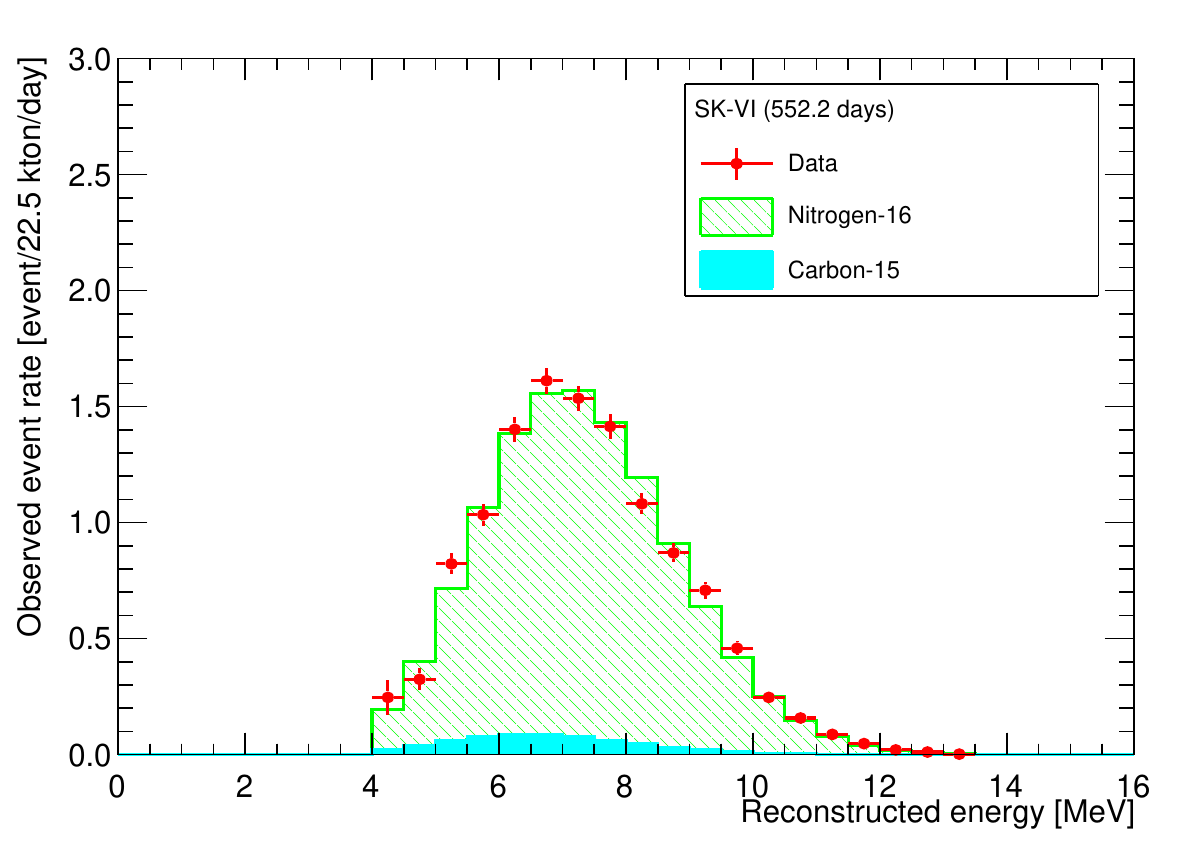}            
	\end{minipage} 
    \caption{Distributions of the time difference~(left) in the timing window of $[0.5, 25.0]$~s and the reconstructed total energy~(right) together with the best-fit of the components from $\mathrm{^{16}N}$ and $\mathrm{^{15}C}$ using the SK-VI data The definition of colors is the same as Fig.~\ref{fig:best-sk4}.}
    \label{fig:best-sk6}
\end{figure*}

Figure~\ref{fig:best-sk6} shows the distributions of the decay time and the energy in the time window of $[0.5,25.0]$~s using SK-VI sample.

\begin{figure*}[]
        \begin{minipage}{0.49\textwidth}
            \centering
            \includegraphics[width=\textwidth]{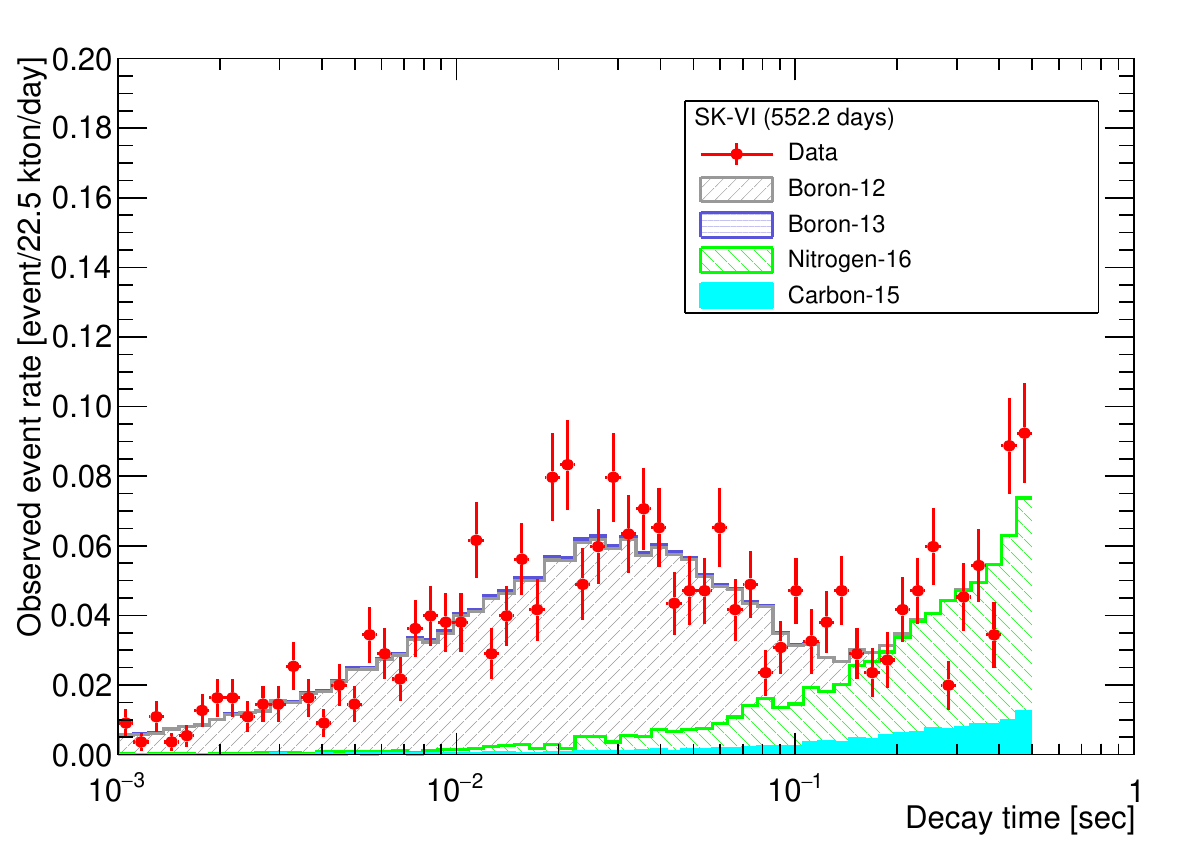}
        \end{minipage} 
        \begin{minipage}{0.49\textwidth}
            \centering
            \includegraphics[width=\textwidth]{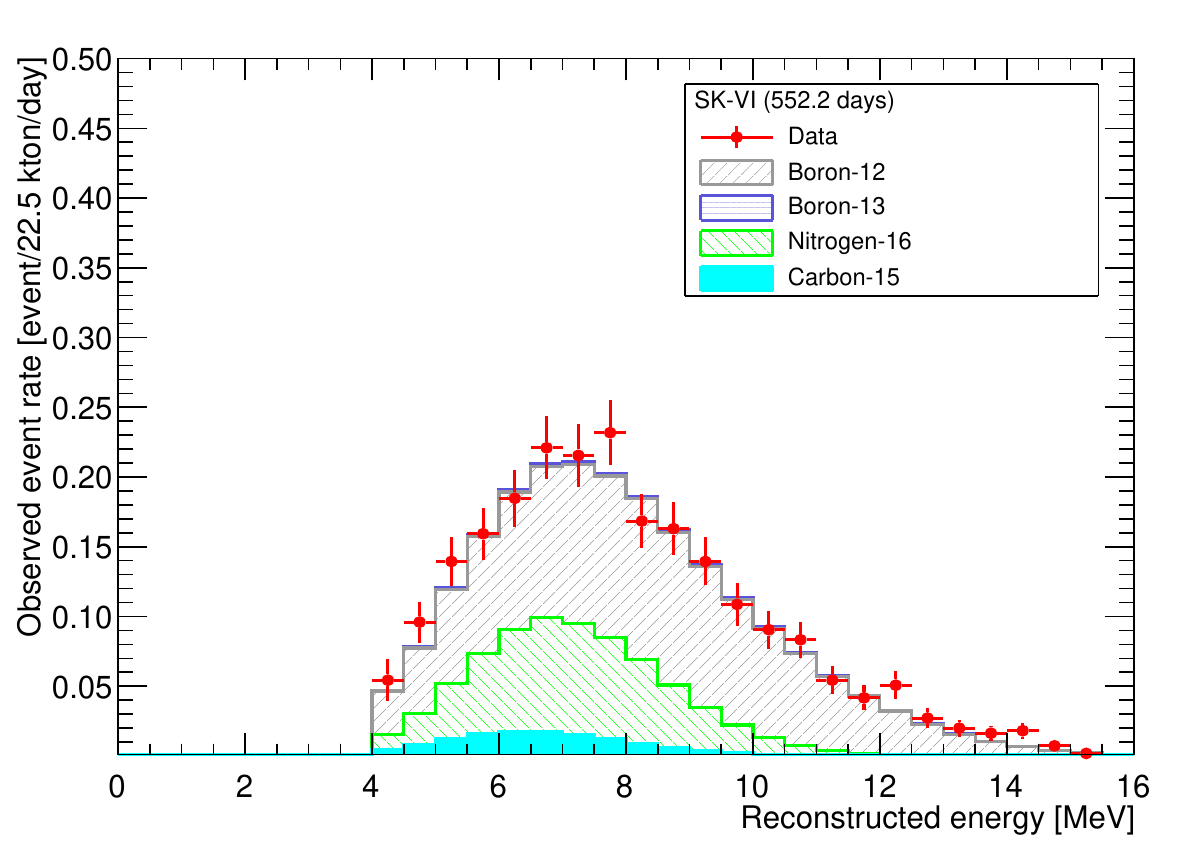}
	\end{minipage} 
    \caption{Distributions of the time difference~(left) in the timing window of $[0.001, 0.5]$~s and the reconstructed total energy~(right) together with the best-fit of the components from $\mathrm{^{12}B}$, $\mathrm{^{13}B}$, $\mathrm{^{16}N}$, and $\mathrm{^{15}C}$ using the SK-VI data. The definition of colors is the same as Fig.~\ref{fig:best-sk4-b12b13}.}
    \label{fig:best-sk6-b12b13}
\end{figure*}

Figure~\ref{fig:best-sk6-b12b13} shows the distributions of the decay time and the energy in the time window of $[0.001,0.5]$~s using SK-VI sample.

\section{Energy scale evaluation using $\bm{\mathrm{^{16}N}}$ and $\bm{\mathrm{^{15}C}}$ sample} \label{sec:energy}

Since $\mathrm{^{16}N}$~(and $\mathrm{^{15}C}$) isotope is naturally produced via a negative muon capture on oxygen nuclei in the water during the usual operation of the SK detector, their decays produce background events when their energies overlap those of solar neutrinos, diffuse supernova neutrinos, and reactor neutrinos. However, selecting such pure samples of isotopes with long lifetimes allows us to monitor the stability of the energy scale in the MeV region of the detector. Furthermore, the position and direction dependence of the energy scale can be evaluated by analyzing these isotopes because the nuclear capture process occurs throughout the detector volume, as shown in Fig.~\ref{fig:stop-vtx}, and the decay particles are isotropically emitted~\cite{Super-Kamiokande:2000kzn, Dragowsky:2001ax}. For that purpose, we collected $\mathrm{^{16}N}$~(and $\mathrm{^{15}C}$) decay events as the natural source sample, and we evaluated the energy scale during the usual operation of the SK detector. Then, we demonstrate the consistency of the energy scale with another calibration source, e.g., the DT generator~\cite{Super-Kamiokande:2000kzn}, in this Appendix.

\subsection{Monitoring the energy scale}

For monitoring purposes, we first selected $\mathrm{^{16}N}$ and $\mathrm{^{15}C}$ decay events in the timing window of $[0.5, 25.0]$~s according to the analysis method described in Sec.~\ref{sec:overview}. In parallel, we also generated realistic MC simulations of both $\mathrm{^{16}N}$ and $\mathrm{^{15}C}$ whose fraction is fixed with the measured values described in Sec.~\ref{sec:overview}.

For the evaluation of the energy scale, we used the effective hit parameter~$N_{\mathrm{eff}}$, which is defined as the number of PMT hits after applying a correction for the relative difference of PMT performance, the contributions of dark rate and late arrival hits, the effective surface area of PMTs, and light attenuation in the water~\cite{Super-Kamiokande:2023jbt}.

\begin{figure}[h]
    \begin{center}
    
    \includegraphics[width=1.0\linewidth]{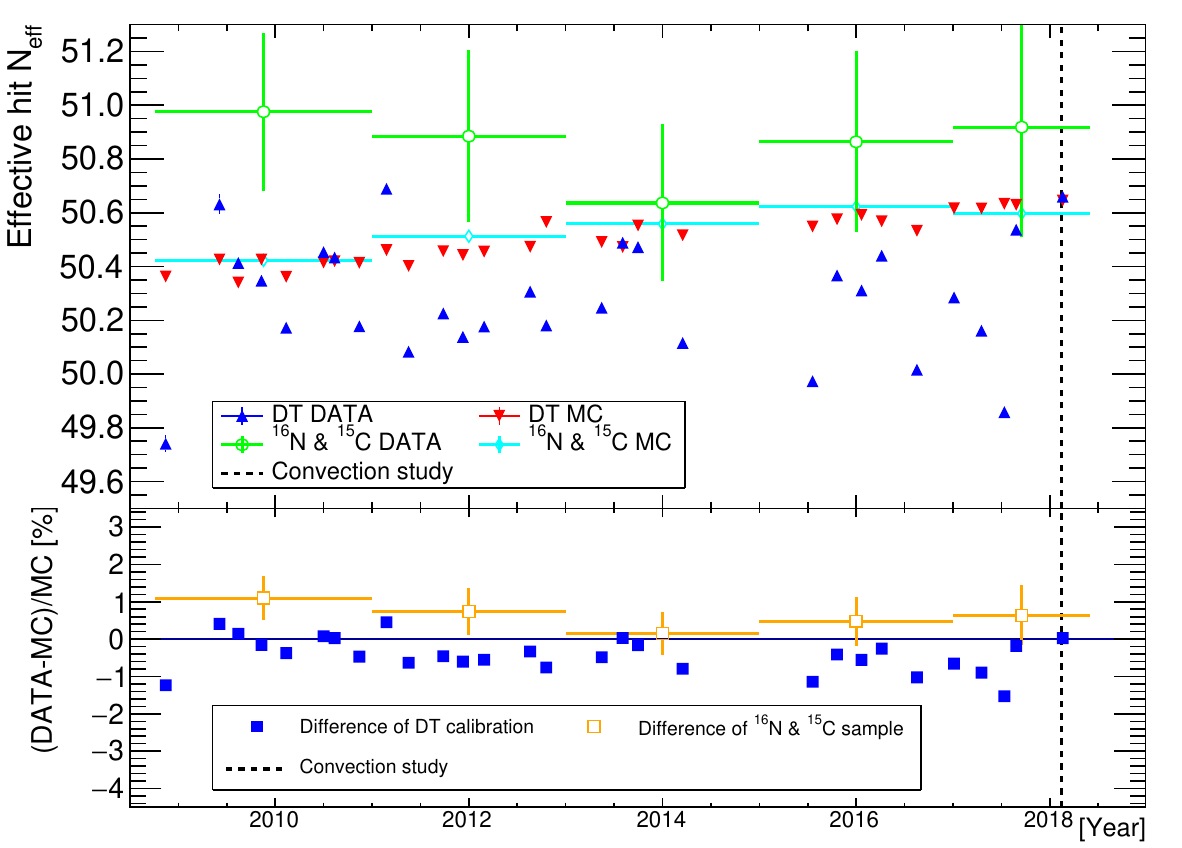}
    \end{center}
    \caption{Top panel: The energy scale using the effective hit~$N_{\mathrm{eff}}$ as a function of time estimated by the $\mathrm{^{16}N}$ sample as well as the DT calibration~\cite{Super-Kamiokande:2000kzn, Super-Kamiokande:2023jbt} during the SK-IV period from 2008 to 2018. Due to low statistics, two years of natural source sample are merged into one dataset. Bottom panel: Their relative difference between the data and the MC simulations. \label{fig:year} }
\end{figure}

Figure~\ref{fig:year} shows the time variation of $N_{\mathrm{eff}}$ and compares the energy scale by the natural source sample with that evaluated by the DT calibration~\cite{Super-Kamiokande:2000kzn}. The energy scale is evaluated by the natural source sample and it was found that the estimated energy scale is stable within a $\pm1\%$ level when we ignore the offset from the zero. This demonstrates that we can monitor the energy scale by using the natural source sample during the usual operation of the SK detector.

As detailed in Ref.~\cite{Super-Kamiokande:2023jbt}, we carried out water convection throughout the SK detector by controlling the supply water temperature to achieve the uniformity of the water quality. During that period, the energy scale evaluated by the DT calibration is almost zero~(right most point in the bottom panel of Fig.~\ref{fig:year}) and this demonstrates that the difference between the DT calibration data and the MC simulation in the SK-IV data sample can be explained by the incomplete modeling of the water properties. Although completeness exists in the MC simulation, the energy scale estimated by the DT calibration keeps within a $\pm0.5\%$ level. 

Due to low statistics of natural source samples, there exist larger uncertainties compared with those of DT calibration. But, we demonstrated the consistency of two calibration results within a $\pm2\%$ level as shown in Fig.~\ref{fig:year}.

\subsection{Position dependence of energy scale}

\begin{figure}[h]
    \begin{center}
    
    \includegraphics[width=1.0\linewidth]{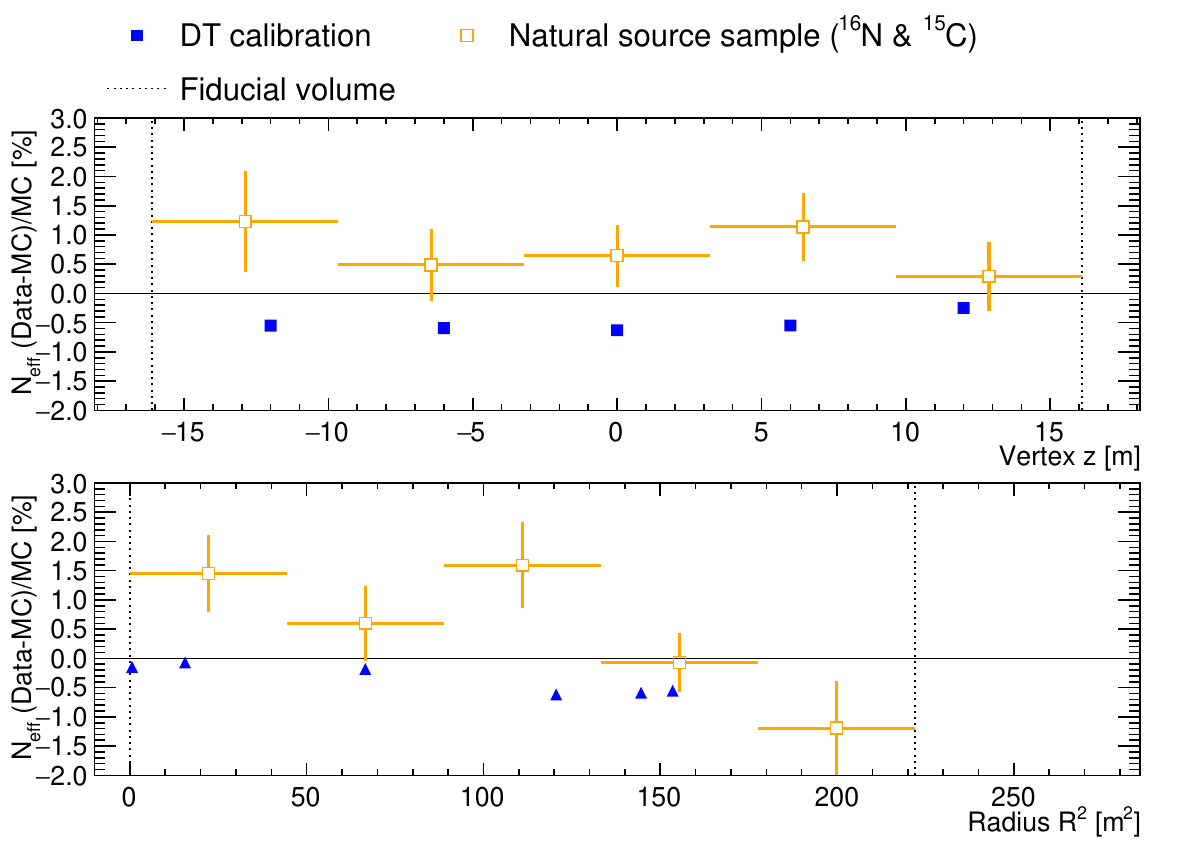}
    \end{center}
    \caption{The position dependence of the relative difference between the calibration data and the MC simulation by the natural source sample as well as the DT calibration~\cite{Super-Kamiokande:2000kzn, Super-Kamiokande:2023jbt}. The top panel shows the position dependence on the vertex~$Z$~(height of the SK detector) and the bottom panel shows that on the detector's radius. \label{fig:position} }
\end{figure}

In general, the solar neutrino interactions uniformly occur throughout the detector volume. The uniformity of the water pattern inside the SK detector affects the energy scale because the propagation of photons reflects the situation of absorption, reflection, and scattering in water. Hence, the position dependence of the energy scale is critical for the precise measurement of the solar neutrino spectrum in the SK detector.
Figure~\ref{fig:position} shows the position dependence of the energy scale. Although the statistical fluctuation can be seen, the natural source sample demonstrates the position dependence of the energy scale within a $\pm2\%$ level and is consistent with that evaluated by the DT calibration source.

\subsection{Directional dependence of energy scale}

\begin{figure}[h]
    \begin{center}
   
    \includegraphics[width=1.0\linewidth]{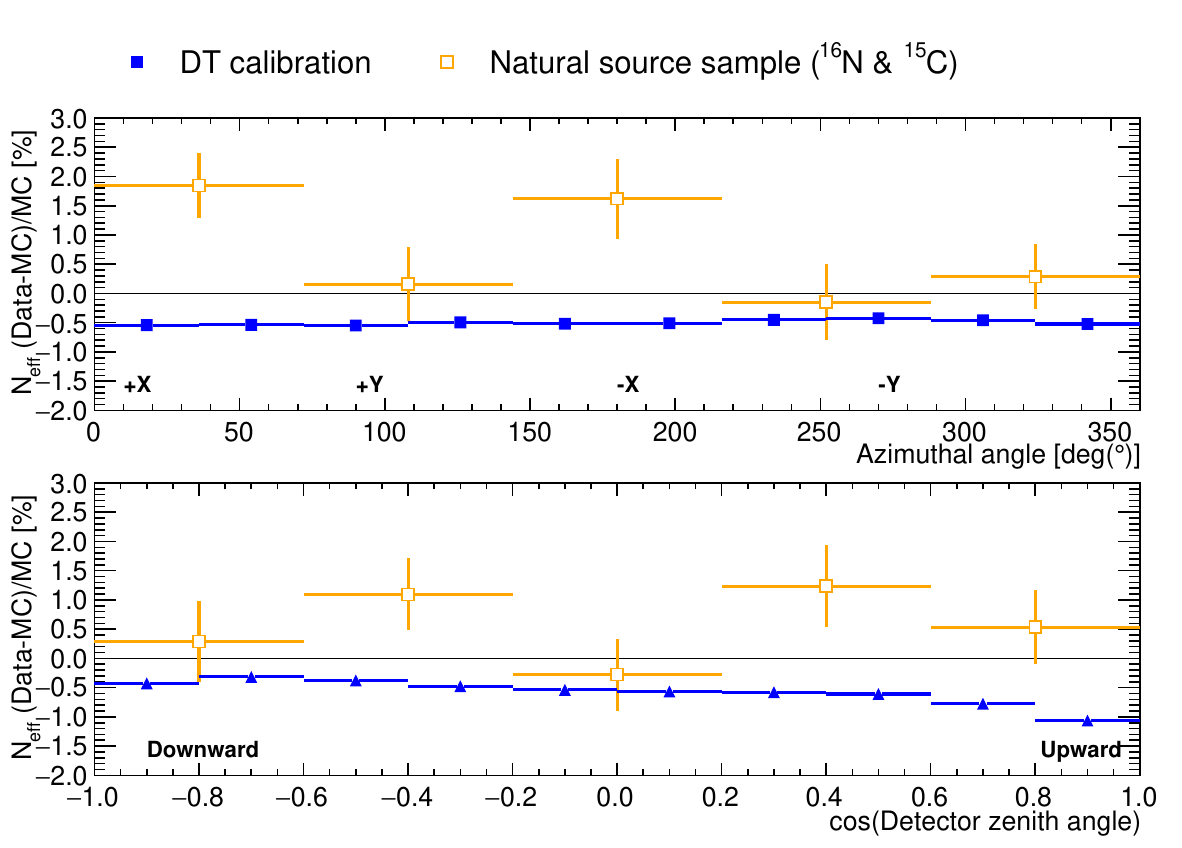}
    \end{center}
    \caption{The directional dependence of the relative difference between the calibration data and the MC simulation by the natural source sample as well as the DT calibration~\cite{Super-Kamiokande:2000kzn, Super-Kamiokande:2023jbt}. The top panel shows the angular dependence on the azimuthal angle and the bottom panel shows that on the zenith angle. As mentioned in Ref.~\cite{Super-Kamiokande:2023jbt}, the upward events are affected by the shadow due to the DT device above the target position. Hence, in the most upward bin $(0.8 < \cos \theta_{\mathrm{Zenith}}< 1.0)$ is noticeably reduced due to the actual shadow effect in the calibration data while such effect is not included in the MC simulation. \label{fig:direction} }
\end{figure}

The direct test of matter effects on neutrino oscillations is the day/night cycle of solar neutrino interaction rates~\cite{Super-Kamiokande:2003yed, Borexino:2011bhn}. The solar zenith angle of the scattered electron signal is used to approximate the neutrino travelling length through the Earth. To measure the day-night flux asymmetry of solar neutrinos in the SK detector, the directional dependence of the energy scale is important~\cite{Super-Kamiokande:2013mie}. 
Figure~\ref{fig:direction} shows the directional dependence of the energy scale by natural source samples together with the DT calibration. As was seen for position dependence, the natural source sample also demonstrates the directional dependence of the energy scale within a $\pm2\%$ level and is consistent with that evaluated by the DT calibration source.

\subsection{Summary of this section}
As demonstrated in this article, naturally produced $\mathrm{^{16}N}$ via negative muon capture on nuclei can statistically monitor the energy scale in the energy region of MeV during the usual operation of the detector even though the production rate is low. Once the detector volume becomes larger, such as will be the case for the Hyper-Kamiokande detector~\cite{Hyper-Kamiokande:2018ofw}, this analysis technique using natural source samples is quite helpful for monitoring the energy scale in the MeV range during usual operation without calibration devices.

\section{Correction of SK-I result} \label{app:sk1}

In the previous study~\cite{Super-Kamiokande:2000kzn}, we explained that the production rate of $\mathrm{^{16}N}$ is $11.9\pm1.0\,(\mathrm{stat.+syst.})~\mathrm{event/11.5 \,  kton/day}$ by analyzing the SK-I data set assuming about $4\%$ of the contamination of $\mathrm{^{15}C}$ isotope decay events. This value corresponds to a $\mathrm{^{16}N}$ production rate of $1.0 \pm 0.1~\mathrm{event/kton/day}$. However, this value is completely different from that measured in this analysis, listed in Table~\ref{tb:event-per-day}. This discrepancy happens because of the improper treatment of the selection efficiency of the stopping muon in the detector in the previous analysis. In fact, the selection efficiency of stopping muons in the previous analysis was estimated as $100\%$. Owing to this improper selection efficiency of stopping muon,  we wrongly reported the low production rate of $\mathrm{^{16}N}$ in the previous publication.

\bibliography{main}

\end{document}